\documentclass{psp-book9x6-modified}

\usepackage{psp-book-van-modified} % Citation -- Vancouver system
\usepackage{cite,graphicx,color,amsmath,booktabs,microtype,afterpage,subfigure,fixmath,truncate}
\usepackage{mathpazo} % Setting up Palatino math fonts as requested by the publisher
\usepackage[scaled]{helvet} % Sans-serif font for headings is Helvetica
\usepackage{bibunits} % Needed for multiple bibliographies (one per chapter).
\usepackage{perpage} % Needed to restart numbering footnotes on every page
\usepackage[colorlinks,plainpages=false,linkcolor=black,urlcolor=black,citecolor=black,pdfpagemode=UseNone,pdfstartview=FitBH]{hyperref}
\usepackage[dvipsnames]{xcolor}
\usepackage[T1]{fontenc}\usepackage[latin1]{inputenc}

\let\Gamma\varGamma
\let\Delta\varDelta
\let\Theta\varTheta
\let\Lambda\varLambda
\let\Xi\varXi
\let\Pi\varPi
\let\Sigma\varSigma
\let\Upsilon\varUpsilon
\let\Phi\varPhi
\let\Psi\varPsi

\makeatletter
\renewcommand\thefigure{\thechapter.\@arabic\c@figure}
\makeatother

\usepackage{perpage}\MakePerPage{footnote}

\defaultbibliographystyle{inosov-book}

\copyline{Rare-Earth Borides}{Dmytro S. Inosov (ed.)}{2020}
\title{Rare-Earth Borides} % set your book title here

\makeindex

\graphicspath{{Figures/}} %Put all figure files in the Figures folder

\begin{document}

%\titlepages                        % pls. do not remove this line

%\begin{dedication}
%\large Dedication Page \\[13pt]    % pls. input your dedication data here
%\large (optional)
%\end{dedication}

%\input{preface.tex}                %\begin{preface}...\end{preface}

%\input{foreword.tex}               %\begin{foreword}...\end{foreword}

%\input{acknowledgments.tex}        %\begin{acknowledgments}...\end{acknowledgments}

%\tableofcontents

%\listoffigures                     % optional list of figures

%\listoftables                      % optional list of tables

\cleardoublepage
\setcounter{page}{1}
\setcounter{chapter}{3}

%\input{Chapter_Shitsevalova.tex}    % every chapter should have a separate file named "Chapter_[AuthorName].tex" that is included here
                                    % and a bibliography file with the same name

%\input{Chapter_Takeuchi.tex}

%\input{Chapter_Bolotina.tex}

\chapter[\bf \mbox{Magnetism, quantum criticality, and metal-insulator\hspace{-3em}} \hspace{3em}transitions in \textit{R}B$_{\text{12}}$]{\bf Magnetism, quantum criticality, and metal-insulator transitions in \textit{R}B$_{\text{12}}$ \label{Chapter:Sluchanko}}
\addtocontents{toc}{\vspace{-2pt}\hspace{2.7em}\textit{by Nikolay E.~Sluchanko}\smallskip}

\chapauth{Nikolay E. Sluchanko$^{\ast}$
\chapaff{\noindent
Prokhorov General Physics Institute, Russian Academy of Sciences, Vavilova str. 38, 119991 Moscow, Russia\\
$^\ast$E-mail address: \href{mailto:nes@lt.gpi.ru}{nes@lt.gpi.ru}}}

\begin{bibunit}

\section*{Abstract}

This chapter presents a review of the physical properties of rare-earth (RE) dodecaborides $R$B$_{12}$ that are characterized by a cage-glass crystal structure with loosely bound RE ions. The analysis of available literature leads to a conclusion that the RE dodecaborides are strongly correlated electron systems with simultaneously active charge, spin, orbital, and lattice degrees of freedom. This explains the complexity of all $R$B$_{12}$ compounds including antiferromagnetic (TbB$_{12}$\,--\,TmB$_{12}$) and nonmagnetic (LuB$_{12}$) metals, from one side, and the so-called Kondo insulator compound YbB$_{12}$ and Yb-based Yb$_xR_{1-x}$B$_{12}$ solid solutions, from the other. It is shown that the reason for the complexity is based on the development of the cooperative dynamic Jahn-Teller instability of the covalent boron network, which produces trigonal and tetragonal distortions of the rigid cage and results in the symmetry lowering of the fcc lattice in the dodecaborides. The ferrodistortive dynamics in the boron sub-lattice generates both the collective modes and quasilocal vibrations (rattling modes) of the heavy RE ions, causing a modulation in the density of conduction electrons and the emergence of dynamic charge stripes in these strongly correlated compounds. We consider the manifestation of the charge stripes both in the properties of the nonmagnetic reference compound LuB$_{12}$ and in the phase diagrams of the $R$B$_{12}$ antiferromagnets that exhibit multiple magnetic phases with anisotropic field-angular phase diagrams in the form of the Maltese cross. We will also discuss the metal-insulator transitions in YbB$_{12}$ and in Yb-based dodecaborides in terms of the instability of the Yb~4$f$-electron configuration, which appears in addition to the Jahn-Teller instability of the boron cage, providing one more mechanism of the charge and spin fluctuations. The experimental results lead to conclusions that challenge the established Kondo-insulator scenario in YbB$_{12}$, providing arguments in favor of the appearance of Yb-Yb vibrationally coupled pairs which should be considered as the main factor responsible for the charge- and spin-gap formation.\vspace{-3pt}

\section{Introduction}

Rare-earth (RE) dodecaborides $R$B$_{12}$ with a ``cage-glass'' structure\index{cage glass} \cite{Slu_SluchankoAzarevich11} attract considerable attention of researchers due to a unique combination of their physical properties, including high melting temperature, microhardness, high chemical stability, etc. In these antiferromagnetic (AFM) metals, the N\'{e}el temperature decreases monotonically from TbB$_{12}$ ($T_{\rm N} \approx 22$~K) to TmB$_{12}$ ($T_{\rm N} \approx 3.2$~K) in the $R$B$_{12}$ series, while the conduction band remains similar, consisting of $5d$ ($R$) and $2p$ (B) atomic orbitals and changing only the filling of the $4f$ shell of the RE ion (\mbox{$8\leq n_{4f} \leq 14$}) \cite{Slu_SluchankoBogach09, Slu_CzopnikShitsevalova04}. In these RE antiferromagnets, the principal interaction which couples the magnetic moments of $4f$ orbitals is the indirect RKKY-type exchange.\index{RKKY interaction} The long-range and oscillatory character of this coupling in the presence of other interactions (for example, the crystal-electric-field (CEF) anisotropy,\index{crystal electric field} magnetoelastic coupling, dipole-dipole, or two-ion quadrupolar interactions and many-body effects) may lead to a competition (frustration)\index{magnetic frustration}\index{frustration} among the interionic interactions, resulting in complicated magnetic structures. So, these AFM metals with an unfilled $4f$ shell and incommensurate helical or amplitude-modulated magnetic structures,\index{amplitude-modulated magnetic structure} having the simple-cubic crystal structure with a single type of magnetic ions located in high-symmetry positions at large enough distances ($\sim$5.3~\r{A}) between them, look, at the first glance, like very suitable systems for testing theories and models developed for the magnetic properties of RE intermetallics.

The $4f$ filling between $n_{4f} = 12$ and $n_{4f} \approx 13$ leads to dramatic changes both in the magnetic and charge transport characteristics \cite{Slu_IgaShimizu98, Slu_IgaTakakuwa84, Slu_AltshulerAltshuler01}, demonstrating the transition from an AFM metal (TmB$_{12}$) to a paramagnetic semiconductor with the intermediate valence of Yb ions (YbB$_{12}$). The metal-insulator transition (MIT)\index{metal-insulator transition} brings a considerable growth at helium temperatures of the DC resistivity, from $4~\mu\Omega\cdot$cm in TmB$_{12}$ to $10~\Omega\cdot$cm in YbB$_{12}$, changing its behavior from metal-like to semiconductor-like \cite{Slu_SluchankoBogach09}. In spite of intensive investigations, the nature of this nonmagnetic semiconducting state in YbB$_{12}$ (so-called Kondo insulator)\index{Kondo insulator} remains a subject of debate \cite{Slu_Riseborough03, Slu_OtsukiKusunose07, Slu_AkbariThalmeier09, Slu_BarabanovMaksimov09}. There are several models proposed to explain the insulating nature of YbB$_{12}$. One of them is based on the coherent band picture where the energy gap is formed due to the strong hybridization of the $f$ electrons with the conduction band \cite{Slu_GreweSteglich91, Slu_SasoHarima03}. The inversion between $4f$ and $5d$ bands that accompanies this process is considered as an essential characteristic of the topological Kondo insulator~\cite{Slu_LuZhao13, Slu_WengZhao14, Slu_HagiwaraOhtsubo16}.\index{Kondo insulator!topological} Other models are based on a local picture, where the electrons contributing to the Kondo screening are captured by the local magnetic moments of Yb ions, resulting in an excitonic local Kondo bound state \cite{Slu_Kasuya96, Slu_Kasuya94}. The single-site scenarios were proposed also by Liu~\cite{Slu_Liu01} and by Barabanov and Maksimov~\cite{Slu_BarabanovMaksimov09}.

To shed more light on the nature of both the charge and spin gaps in the quasiparticle spectra of YbB$_{12}$, a number of studies have been undertaken during the last decade since the previous reviews devoted to the physics and chemistry of RE dodecaborides\footnote{\nocite{Slu_FlachbartAlekseev08, Slu_Mori08}During the preparation of this chapter, one more review article devoted to the studies of RE dodecaborides has been published \cite{Slu_GabaniFlachbart20}.} were published \cite{Slu_FlachbartAlekseev08, Slu_Mori08}. Some of these results are presented in this book in the chapters devoted to the details of fine crystal structure (Chapter~\ref{Chapter:Bolotina}), Raman scattering spectra of $R$B$_{12}$ (Chapter~\ref{Chapter:Ponosov}), and magnetic excitations (Chapter~\ref{Chapter:Alekseev}). In studies of the charge transport, magnetic and thermal properties of the nonmagnetic LuB$_{12}$, antiferromagnetic HoB$_{12}$\,--\,TmB$_{12}$ compounds, and solid solutions Tm$_{1-x}$Yb$_x$B$_{12}$, it was recently established that the cooperative Jahn-Teller (JT)\index{Jahn-Teller effect!cooperative} dynamics of the B$_{12}$ clusters\index{B$_{12}$ cluster} should be considered as the main factor responsible for a strong renormalization of the quasiparticle spectra, electron phase separation, and the symmetry breaking in the RE dodecaborides with the face-centered cubic (fcc) crystal structure. In this chapter, I present the analysis of both well-known and recent results in this area, arguing in favor of a spatial inhomogeneity and a non-Kondo physics which should be used to interpret the unusual magnetic ground states and the MIT in these exemplary strongly correlated electron systems.\index{strongly correlated electrons}

The chapter is organized as follows. In Section~\ref{Sec:Slu2} we discuss the results on the electronic band structure of the RE dodecaborides, Sections~\ref{Sec:Slu3} and \ref{Sec:Slu4} are devoted to the nonmagnetic reference compound LuB$_{12}$, magnetic dodecaborides $R$B$_{12}$ ($R$~=~Tb, Dy, Ho, Er, Tm), and the solid solutions $R_x$Lu$_{1-x}$B$_{12}$. Section~\ref{Sec:Slu5} collected the numerous results accumulated in the studies of the MIT in YbB$_{12}$ and Yb$_xR_{1-x}$B$_{12}$ ($R$~=~Lu, Tm) solid solutions, and Section~\ref{Sec:Slu_Conclusions} concludes the chapter.

\vspace{-3pt}\section{Electronic band structure of dodecaborides}\vspace{-2pt}
\label{Sec:Slu2}
\index{rare-earth dodecaborides!electronic structure}

\subsection{Rough estimations}

A very rough description of the electronic structure of the RE dodecaborides is produced if all valence electrons of the elements are taken into account. Some of these electrons fill the bonding orbitals, the other are delocalized in the conduction band. Thus, according, for instance, to the model of Lipscomb and Britton~\cite{Slu_LipscombBritton60},\index{Lipscomb-Britton model} the stabilization of the boron network in $R$B$_{12}$ requires the addition of two electrons from each RE atom. Indeed, the cuboctahedral arrangement of the boron atoms leads to six different bonding states (molecular orbitals) which can accommodate 26 electrons when we take their degeneracy into account. Since three valence electrons are provided by each boron atom, there are altogether 36 valence electrons coming from one B$_{12}$ unit, of which 12 are required for the external bonds with a neighboring B$_{12}$ unit. Therefore, the remaining 24 electrons plus two additional valence electrons of the RE metal are required to fill up all bonding orbitals of the cuboctahedral cluster, leaving one remaining electron per metal ion to occupy the conduction band. Consequently, the conduction band is half-filled, and the dodecaborides are metals \cite{Slu_JaegerPaluch06}. Thus, even if this model does not account for the boron-metal and metal-metal bonding~\cite{Slu_EtourneauHagenmuller85}, it can qualitatively predict the metallic properties of RE dodecaborides with a trivalent state of $R$ ions and one electron per unit cell in the conduction band. This electron transfer generates ionic bonding between RE atoms and B$_{12}$ clusters,\index{B$_{12}$ cluster} whereas covalent bonding prevails in the rigid boron cage~\cite{Slu_GrechnevBaranovskiy08}. Hence, all the RE dodecaborides discussed here are good metals with an exception of the intermediate-valence system YbB$_{12}$ (so-called Kondo insulator\index{Kondo insulator} with the Yb valence between $\sim$2.9 and 3)~\cite{Slu_IgaShimizu98, Slu_IgaTakakuwa84, Slu_KasuyaKasaya83, Slu_AlekseevNefeodova01}, which undergoes a MIT and becomes insulating at low temperatures.

\vspace{-2pt}\subsection{Metallic \textit{R}B$_\text{12}$}

During the last 40 years, stable B$_{12}$ nanoclusters\index{B$_{12}$ cluster} have been considered as basic structural elements of the fcc lattice of dodecaborides in band-structure calculations. The corresponding UB$_{12}$-type crystal structure is similar to the simple rock-salt lattice, where U atoms and B$_{12}$ cuboctahedra occupy the Na- and Cl-sites, respectively (see Chapters~\ref{Chapter:Shitsevalova} and \ref{Chapter:Bolotina} for details). Calculations of the energy band structure of LuB$_{12}$ by Harima~\textit{et al.} \cite{Slu_HarimaYanase85, Slu_HarimaKobayashi85} suggest that there are two conduction bands intersecting the Fermi level. The upper band corresponds to a simply connected Fermi surface located around the $X$ point. The other band shows certain similarities to the well-known noble metal ``monster'' Fermi surface \cite{Slu_Shoenberg84},\index{Fermi surface!monster@``monster''} with the difference that in LuB$_{12}$ the occupied electronic states lie near the boundary of the Brillouin zone, and the necks are wider~\cite{Slu_HeineckeWinzer95}. The calculations of Heinecke \textit{et al.} \cite{Slu_HeineckeWinzer95} confirmed the conduction band structure for LuB$_{12}$ and corrected the position of $4f$ levels from 4.9~eV \cite{Slu_HarimaYanase85, Slu_HarimaKobayashi85} to 6.4~eV below the Fermi level ($E_{\rm F}$). The difference $E_{\rm F}-E_{4f}$ obtained in Ref.~\cite{Slu_HeineckeWinzer95} is closer to the results of x-ray photoelectron spectroscopy (XPS)\index{photoelectron spectroscopy} measurements of TmB$_{12}$ and LuB$_{12}$ \cite{Slu_IgaTakakuwa84}, which show that the binding energy of the $4f$ level in these two dodecaborides is about 5 and 7.7~eV, respectively.

\begin{figure}[t!]
\centerline{\includegraphics[width=0.425\textwidth]{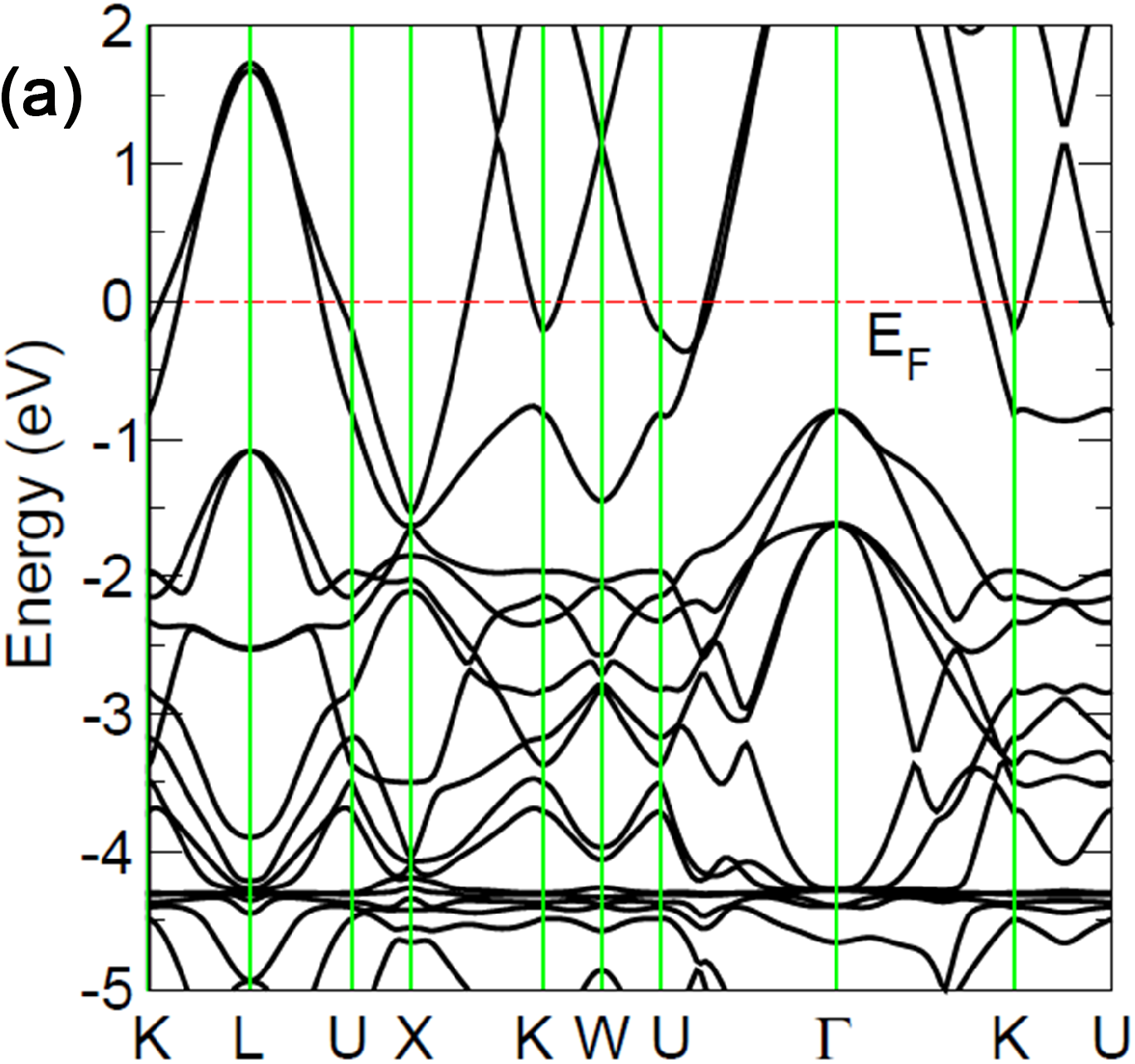}\hfill\raisebox{-3.5pt}{\includegraphics[width=0.565\textwidth]{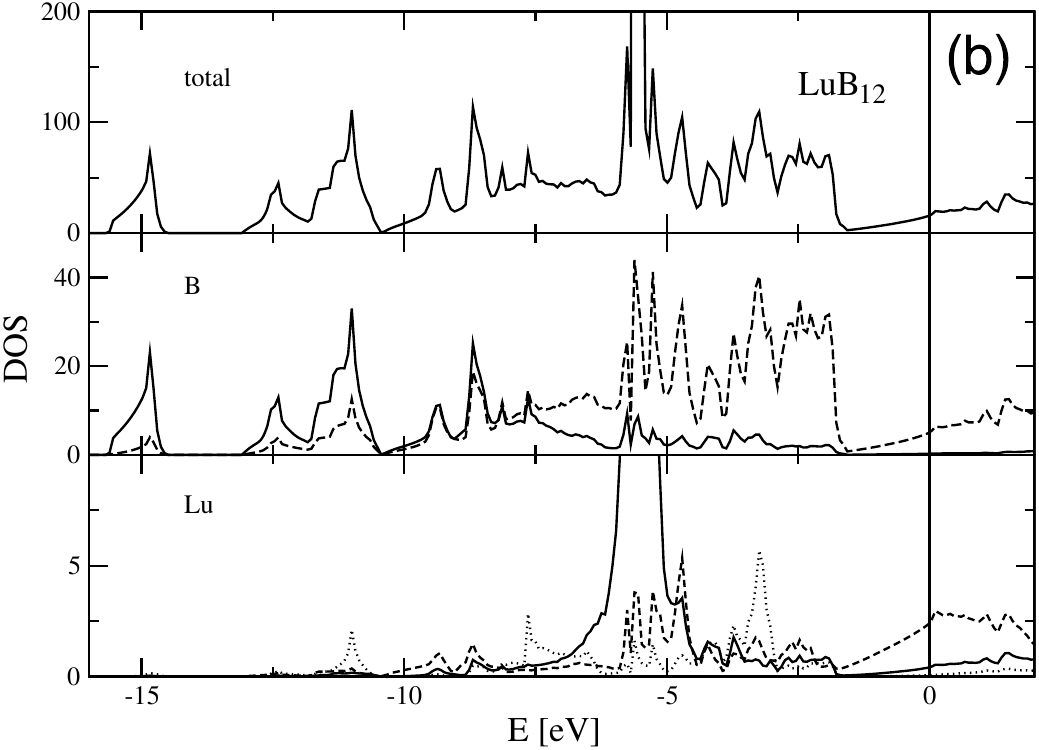}}}
\caption{(a)~Band structure for the nonmagnetic reference compound LuB$_{12}$. Courtesy of G.~Grechnev~\cite{Slu_Grechnev}. (b)~Total density of states (DOS) (top) and local partial DOS components for LuB$_{12}$ in units of states per Rydberg and per formula unit (center: B~$s$ states, solid line; B~$p$ states, dashed line; bottom: Lu~$e_g$ states, dotted line; Lu $t_{2g}$ states, dashed line; Lu~$f$ states, solid line). Reproduced from J\"{a}ger~\textit{et al.}~\cite{Slu_JaegerPaluch06}.}
\label{Slu:Fig1}\index{LuB$_{12}$!band structure}\index{LuB$_{12}$!density of states}
\end{figure}

Later on, \textit{ab initio} electronic structure calculations were carried out for the paramagnetic (PM), ferromagnetic (FM), and collinear AFM phases of $R$B$_{12}$ ($R$~=~Ho, Er, Tm) in Refs.~\cite{Slu_GrechnevBaranovskiy08, Slu_BaranovskiyGrechnev09} for a number of lattice parameters close to the experimental ones.
\begin{figure}[t!]
\centerline{\includegraphics[width=\textwidth]{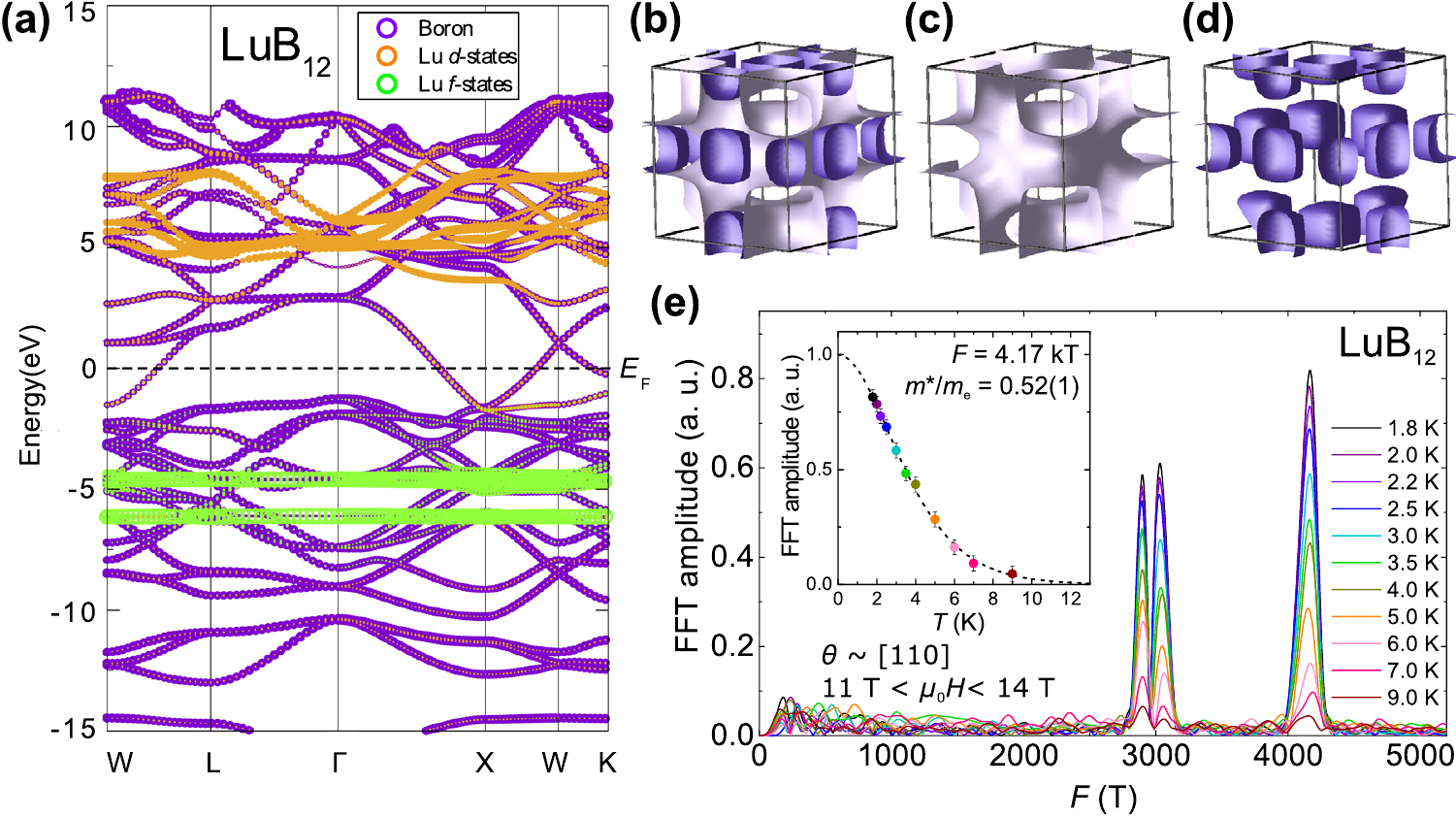}}
\caption{(a)~Calculated band structure of LuB$_{12}$. Several characters are projected out of the eigenvectors at each $\mathbf{k}$ point, and the resulting weight is indicated by a circle of proportional size. Green circles are Lu~$f$ states, orange circles are Lu~$d$-states, and violet circles are boron states. (b--d)~The Fermi surfaces of LuB$_{12}$ shown together in (b) and separately for clearer viewing in (c) and (d). (e)~Fast Fourier transform (FFT) of the magnetic field sweeps taken at different temperatures. The inset shows the FFT fit (dashed line) to the frequency $F=4.17$~kT. Reproduced from Liu \textit{et al.}~\cite{Slu_LiuHartstein18}.}
\label{Slu:Fig2}\index{LuB$_{12}$!band structure}\index{LuB$_{12}$!Fermi surface}\index{Fermi surface!of LuB$_{12}$}
\end{figure}
These calculations provided total ground-state energies and the corresponding equations of states $E(V)$ with sufficient accuracy. In this way, the magnetic stability of AFM ordering in $R$B$_{12}$ ($R$~=~Ho, Er, and Tm, see below) was confirmed by comprehensive total-energy calculations for PM, FM, and AFM phases. Also, more detailed calculations of the band structure, Fermi surface, total and partial densities of electronic states (DOS) were carried out for the reference compound LuB$_{12}$ \cite{Slu_JaegerPaluch06, Slu_GrechnevBaranovskiy08, Slu_BaranovskiyGrechnev09, Slu_LiuHartstein18}. They reveal principal features of the electronic spectra, which are common for the whole $R$B$_{12}$ series. The corresponding electronic structure and the total electron DOS of LuB$_{12}$ in the close vicinity of the Fermi level are presented in Fig.~\ref{Slu:Fig1}. It was found that in addition to the lower and upper conduction bands and the corresponding two Fermi surface sheets\,---\,the first multiply connected in the $(111)$ directions [$\Gamma L$ direction in the Brillouin zone, Fig.~\ref{Slu:Fig1}\,(a)] and the second simply connected which forms ``pancake''-like electron surfaces centered at $X$ symmetry points [see Figs.~\ref{Slu:Fig1} and \ref{Slu:Fig2}\,(a--b)], there is a third Fermi surface sheet that consists of small electronlike lenses centered at $K$ points of the Brillouin zone \cite{Slu_BaranovskiyGrechnev09}. As it follows from the results of calculations, the main features in the conduction band structure of LuB$_{12}$ are governed by hybridization of the RE $5d$ states (predominantly the Lu~$e_{\rm g}$ orbitals pointing toward the centers of the square faces of the B$_{12}$ cuboctahedron \cite{Slu_JaegerPaluch06}) with the $2p$ states of boron. At energies above about $-7$~eV and up to the Fermi level, mainly $\pi$ bonds between adjacent B$_{12}$ clusters\index{B$_{12}$ cluster} are found. These hybridized states are forming a conduction band of about 1.6~eV in width and exhibit a strong dispersion at the Fermi level, see Fig.~\ref{Slu:Fig1}.

The calculated effective masses of conduction electrons in LuB$_{12}$ appeared to be comparatively small with respect to the bare electron mass, \mbox{$m^\ast \leq m_0$}~\cite{Slu_BaranovskiyGrechnev09}. The $m^\ast$ estimations are confirmed in quantum-oscillation experiments \cite{Slu_HeineckeWinzer95, Slu_OkudaSuzuki00, Slu_LiuHartstein18},\index{quantum oscillations} where effective masses of $0.44\!-\!1.88\,m_0$ were deduced from the Fourier transform data analysis. It is worth noting that in addition to the fundamental peaks in the frequency spectrum of the de Haas\,--\,van Alphen (dHvA)\index{de Haas\,--\,van Alphen} oscillations (i.e. quantum oscillations in the magnetic susceptibility), distinctive higher harmonics appear in these spectra with an unusually large amplitude. As deduced by Pluzhnikov \textit{et al.}~\cite{Slu_PluzhnikovShitsevalova08}, the cyclotron masses for HoB$_{12}$, ErB$_{12}$, and TmB$_{12}$ are in the range of (0.4--1.45)\,$m_0$, in agreement with the dHvA results on LuB$_{12}$ \cite{Slu_OkudaSuzuki00, Slu_IkushimaKato00, Slu_LiuHartstein18}. Moreover, for ErB$_{12}$ and TmB$_{12}$, the same authors \cite{Slu_PluzhnikovShitsevalova08} found branches in the dHvA oscillation spectrum that originated from Fermi-surface fragments with spins along and opposite to the applied magnetic field. Thus, the authors have discussed the exchange splitting of the conduction band on spin-up and spin-down sub-bands. The splitting is proportional to the magnetic moment and the exchange integral, and the energy of the exchange interaction was estimated to be 8.2~meV in ErB$_{12}$ and 14.5, 28.5 and 18.3~meV for different branches in TmB$_{12}$. However, the splitting is not observed in HoB$_{12}$ with the largest value of the RE magnetic moment, but this fact was not explained in Ref.~\cite{Slu_PluzhnikovShitsevalova08}.\vspace{-3pt}

\subsection{Strongly correlated semiconductor YbB$_\text{12}$}\vspace{2pt}
\index{YbB$_{12}$!electronic structure}

The energy band structure of YbB$_{12}$ has been first calculated by Yanase and Harima~\cite{Slu_YanaseHarima92} using the self-consistent linearized augmented plane wave (LAPW) method in the local spin-density approximation (LSDA) including the spin-orbit interaction. YbB$_{12}$ was found to be a semimetal with a holelike Fermi surface around the $L$ points and an electronlike Fermi surface centered at the $W$ points. Later Antonov~\textit{et al.}~\cite{Slu_AntonovHarmon02} mentioned that LSDA calculations generally provide an inadequate description of the $4f$ electrons due to improper treatment of correlation effects. In particular, LSDA calculations cannot account for the splitting of filled and empty $f$ bands determined, for example, by photoelectron spectroscopy,\index{photoelectron spectroscopy} which is expected to be 7--8~eV in YbB$_{12}$ \cite{Slu_CampagnaWertheim76}. The authors \cite{Slu_YanaseHarima92} noted that they cannot definitely assign an energy-gap from LDA, which is only a ground-state theory.

\begin{figure}[b!]
\centerline{\includegraphics[width=\textwidth]{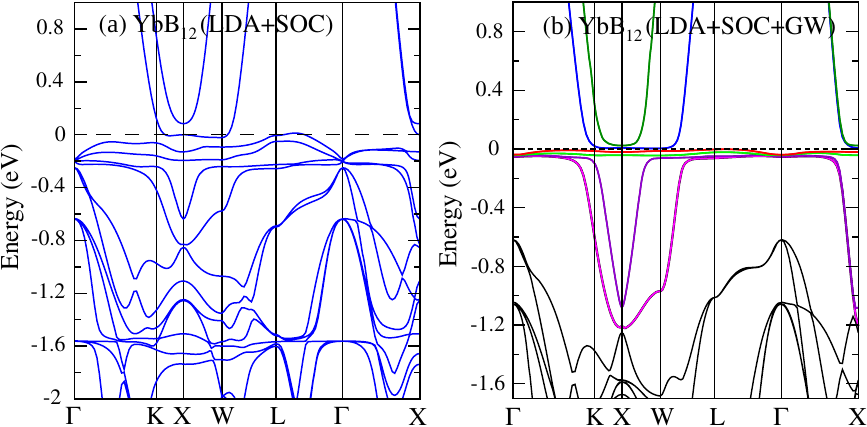}}
\caption{The band structure of YbB$_{12}$: (a)~obtained from LDA+SOC calculations and (b)~quasiparticle band structure calculated from LDA + SOC + Gutzwiller with $U = 6$~eV \cite{Slu_WengZhao14}.}
\label{Slu:Fig3}\index{YbB$_{12}$!band structure}
\end{figure}

In the detailed band-structure study of YbB$_{12}$~\cite{Slu_AntonovHarmon02}, three independent fully relativistic spin-polarized calculations were performed, considering 4$f$ electrons as (1)~itinerant electrons using the local spin-density approximation; (2)~fully localized putting them in the core and (3)~partly localized using the LSDA+$U$ approximation. The sharp Yb~$4f$ peaks in the DOS calculated within the LSDA cross the Yb~$5d$ states just below $E_{\rm F}$ and hybridize with them, resulting in a small direct gap of $\sim$65~meV at the Fermi level.\index{YbB$_{12}$!energy gap} It was found that there is a small hybridization gap at $E_\text{F}$ also in the case of LSDA+$U$ energy bands for divalent Yb ions, indicating a nonmagnetic semiconducting ground state in YbB$_{12}$ \cite{Slu_AntonovHarmon02}.

\begin{figure}[t!]\vspace{-1pt}
\centerline{\includegraphics[width=\textwidth]{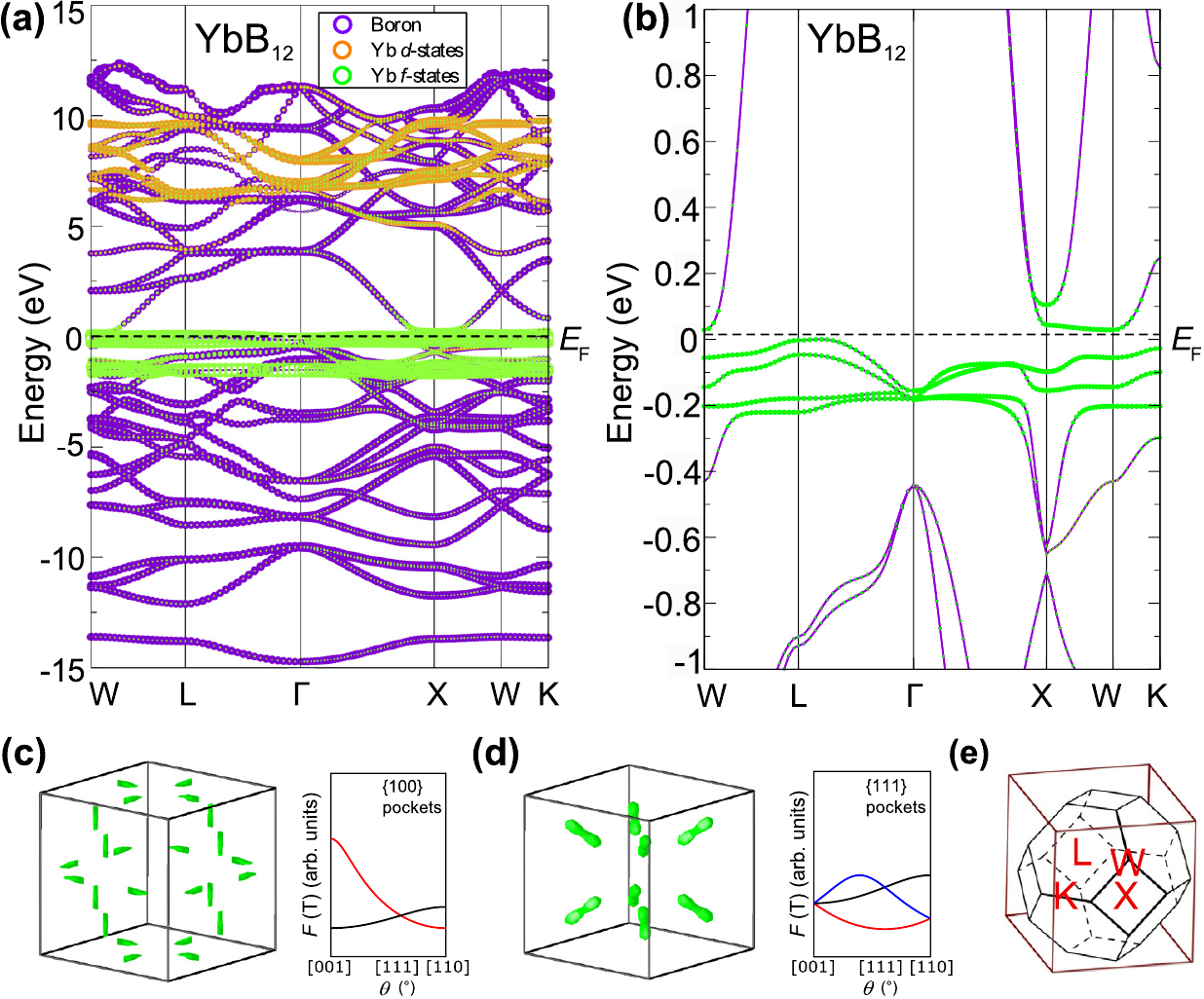}}\vspace{-1pt}
\caption{Calculated band structure of YbB$_{12}$ shown over a wide energy range in (a) with an expanded view around the Fermi energy $E_{\text F}$ in (b). Several characters are projected out of the eigenvectors at each $k$-point and the resulting weight is indicated by a circle of proportional size. Green circles are Yb~$f$-states, orange circles are Yb $d$-states, and violet circles are boron states. (c)~Small needle-shaped Fermi surfaces of YbB$_{12}$ with effective mass $m^{\ast}/m_0 \approx 6$ obtained using the modified Becke-Johnson potential for a small positive energy shift. Expected angular dependence of the quantum oscillation frequencies in the [001]--[111]--[110] rotation plane are shown by approximating the shown Fermi surfaces as prolate ellipsoids. (d) Small peanut-shaped Fermi surfaces of YbB$_{12}$ with effective mass $m^{\ast}/m_0 \approx 9$ obtained using the modified Becke-Johnson potential for a small negative energy shift. Expected angular dependence of the quantum oscillation frequencies in the [001]--[111]--[110] rotation plane are shown by approximating the shown Fermi surfaces as prolate ellipsoids. (e) A schematic of the conventional fcc Brillouin zone used for the band structures within the cubic Brillouin zone used for the Fermi surfaces \cite{Slu_LiuHartstein18}.}
\label{Slu:Fig4}\index{YbB$_{12}$!band structure}\index{YbB$_{12}$!Fermi surface}\index{Fermi surface!of YbB$_{12}$}
\end{figure}

For the trivalent Yb ion, thirteen $4f$ electron bands were found well below the Fermi energy and hybridized with the B~$2p$ states. The 14th unoccupied $4f$ hole level is about 1.5~eV above $E_{\text F}$ \cite{Slu_AntonovHarmon02}. Authors note that such an electronic structure is appropriate for the development of the Kondo scenario. The energy band structures of YbB$_{12}$ with $4f$ electrons in the core are characterized by partly occupied Yb~$5d$ bands which cross the Fermi level and overlap slightly with the B~$2p$ states. On the one hand, according to the conclusion of Alekseev \textit{et al.} \cite{Slu_AlekseevNefeodova01} about the trivalent state of Yb ion in YbB$_{12}$, the last two cases describing YbB$_{12}$ as a metallic strongly correlated electron system\index{strongly correlated electrons} with a $5d$-$2p$ conduction band of about 1~eV in width should be considered as preferable. On the other hand, the experimentally estimated energy gap in YbB$_{12}$ is less than 10~meV from the activation-energy measurements \cite{Slu_KasuyaKasaya83, Slu_IgaHiura99, Slu_SluchankoAzarevich12, Slu_SluchankoAzarevich19}, 25~meV from optical spectroscopy \cite{Slu_OkamuraKimura98, Slu_GorshunovHaas06}, and 200--300~meV from tunneling experiments \cite{Slu_EkinoUmeda99}, which stimulated a search for the mechanisms responsible for the MIT upon cooling below 70~K.

Recently, two investigations of the YbB$_{12}$ electronic structure were carried out.\index{YbB$_{12}$!electronic structure} In the first study, Weng \textit{et al.}~\cite{Slu_WengZhao14} applied the local density approximation with spin-orbit coupling (SOC) combined with the Gutzwiller (GW) density functional theory (DFT) to compute the ground state and the quasi-particle spectrum. The LDA-SOC-GW band structure, shown in Fig.~\ref{Slu:Fig3}, indicates (i)~two $5d$ bands strongly hybridized with the $2s$ and $2p$ states of boron, (ii)~the $4f$ occupation number $n_{\rm f} \approx 13.28$ and (iii)~the hybridization gap $E_\text{g}\approx 6$~meV between the $4f$ and itinerant $5d$ bands which is only slightly below the experimental values \cite{Slu_KasuyaKasaya83, Slu_IgaHiura99, Slu_SluchankoAzarevich12, Slu_SluchankoAzarevich19}. In the second study, Liu \textit{et al.}~\cite{Slu_LiuHartstein18} used DFT with the modified Becke-Johnson potential\index{Becke-Johnson potential} (a semi-local approximation to the exact exchange plus a screening term \cite{Slu_TranBlaha09}), resulting in a nonmagnetic ground state of YbB$_{12}$ with an indirect band gap of 21~meV and a direct gap of 80~meV (see Fig.~\ref{Slu:Fig4}). Integrating over all $\mathbf{k}$ points, it was found that out of the 14 states of the Yb~$f$ complex, 13.2 lie below the Fermi energy and 0.8 above (Fig.~\ref{Slu:Fig4}), which corresponds well with the experiment and band structure calculations \cite{Slu_WengZhao14}. Small needle-shaped Fermi surfaces of YbB$_{12}$ with an effective mass $m^\ast \approx 6 m_0$ were detected in the de Haas--van Alphen measurements and explained using the modified Becke-Johnson potential for a small positive energy shift \cite{Slu_LiuHartstein18}.

I would also like to mention recent observations of the Shubnikov--de Haas (SdH) oscillations\index{YbB$_{12}$!charge transport!Shubnikov--de Haas oscillations}\index{Shubnikov--de Haas effect} (i.e. quantum oscillations in the resistivity)\index{quantum oscillations} in YbB$_{12}$ \cite{Slu_XiangKasahara18, Slu_SatoXiang19} that have been discussed by the authors as a big surprise. Indeed, in spite of the large charge gap inferred from the insulating behavior of the resistivity, YbB$_{12}$, as well as the SmB$_6$ compound, apparently hosts a Fermi surface at high magnetic fields. The authors argue that in YbB$_{12}$, where the mean valence of the Yb ions is close to +3 ($4f^{13}$ state) \cite{Slu_AlekseevNefeodova01, Slu_YamaguchiSekiyama09}, the 3D nature of the SdH signal demonstrates that the quantum oscillations in the resistivity arise from the electrically insulating bulk. On the basis of the symmetry analysis and simulations, they deny the possibility of SdH oscillations arising from a minority portion of the sample or from metallic domains resulting from impurities or strain~\cite{Slu_XiangKasahara18, Slu_SatoXiang19}.

To summarize the results presented in this section, the main postulates of both the aforementioned band-structure calculations and the interpretation of quantum-oscillation experiments are (i)~homogeneous state in all $R$B$_{12}$ single crystals including the intermediate-valence YbB$_{12}$ compound and (ii)~the stable B$_{12}$ nanoclusters\index{B$_{12}$ cluster} which are considered as basic structural elements of the fcc lattice of dodecaborides. It will be shown below in this chapter that both these assertions are, strictly speaking, not valid for $R$B$_{12}$, and the dynamic JT instability\index{dynamic Jahn-Teller effect} of the boron cage is among the main factors responsible for phase-separation effects and inhomogeneity in these entangled strongly correlated electron systems.\index{strongly correlated electrons}

\vspace{-3pt}\section{The nonmagnetic reference compound LuB$_\text{12}$}\vspace{-2pt}
\label{Sec:Slu3}

\subsection{Charge transport}\index{LuB$_{12}$!charge transport|(}

\begin{figure}[t!]
\centerline{\includegraphics[width=0.9\textwidth]{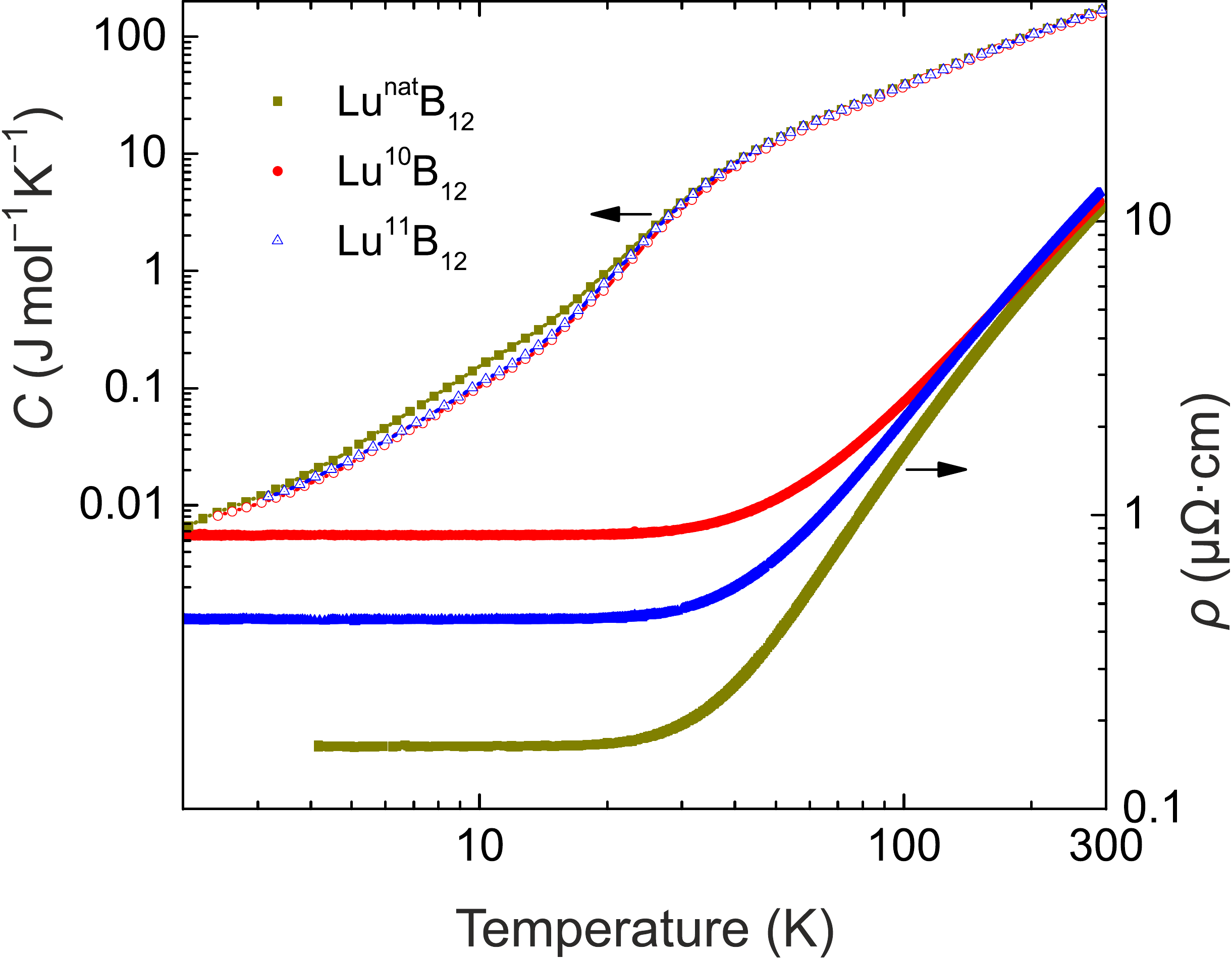}}
\caption{Temperature dependencies of the specific heat and resistivity of the isotopic Lu$^N$B$_{12}$ crystals with $N=10$, 11 and nat. Reproduced from Ref.~\cite{Slu_BolotinaDudka19}.\vspace{-3pt}}
\label{Slu:Fig5}\index{LuB$_{12}$!specific heat}\index{LuB$_{12}$!resistivity}\index{LuB$_{12}$!resistivity!isotope effect}\index{heat capacity!in LuB$_{12}$}\index{specific heat!LuB$_{12}$}
\end{figure}

\begin{figure}[t]
\centerline{\includegraphics[width=0.95\textwidth]{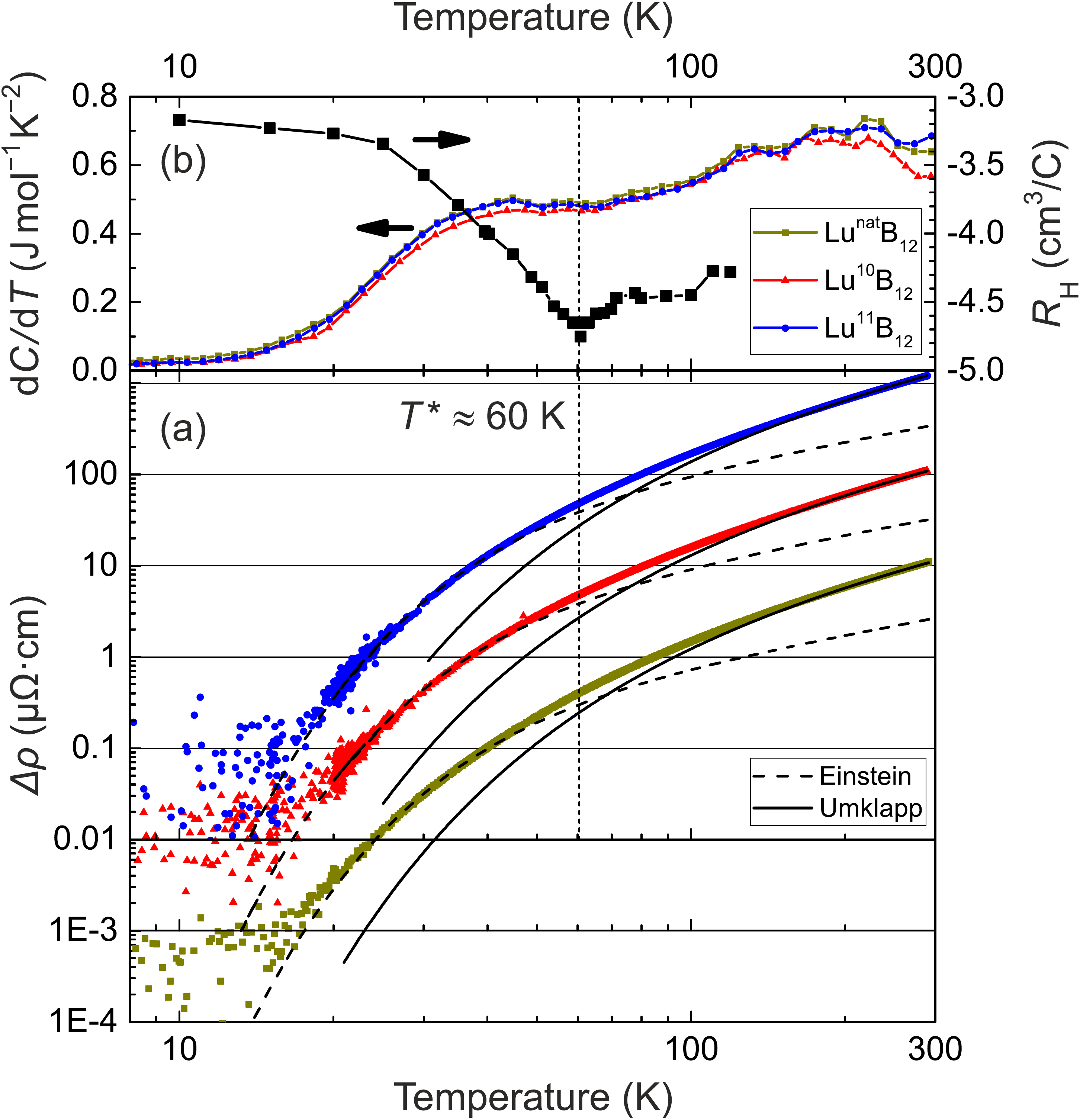}}
\caption{(a)~Temperature-dependent component of the resistivity curves $\Delta\rho(T)$ (curves are shifted for convenience). Fitting results from Eqs.~(\ref{Slu:Eq1}) and (\ref{Slu:Eq2}) are shown by dashed and solid lines, respectively. (b)~Temperature dependences of the derivative of specific heat ${\rm d}C/{\rm d}T$ and the Hall coefficient $R_\text{H}(T)$. The vertical dashed line is located at $T^{\ast} \approx 60$~K. Reproduced from Ref.~\cite{Slu_BolotinaDudka19}.}\index{LuB$_{12}$!specific heat}\index{LuB$_{12}$!resistivity}\index{specific heat!LuB$_{12}$}\index{heat capacity!in LuB$_{12}$}
\label{Slu:Fig6}
\end{figure}

Lutetium dodecaboride is a good metal with the measured residual resistivity ratio $\rho(300~\text{K})/\rho(10~\text{K}) \approx 70$ for the natural boron content (Lu$^\text{nat}$B$_{12}$, ``nat'' means the natural isotopic composition with 18.8\% $^{10}$B and 81.2\% $^{11}$B), as shown in Fig.~\ref{Slu:Fig5}. The high value confirms the high quality of the single crystals \cite{Slu_BatkoBatkova95, Slu_PadernoShitsevalova95, Slu_SluchankoAzarevich10}. For pure isotopic Lu$^{10}$B$_{12}$ and Lu$^{11}$B$_{12}$ crystals, the residual resistivity turns out to be considerably higher, with the maximum value of $\rho_0 \approx 0.83~\mu \Omega\cdot$cm observed in Lu$^{10}$B$_{12}$ \cite{Slu_SluchankoAzarevich10, Slu_BolotinaDudka19}. Bolotina \textit{et al.}~\cite{Slu_BolotinaDudka19} note that for all three boron-isotope compositions, the resistivity reaches its residual value below 20~K (Fig.~\ref{Slu:Fig5}), and the resistivity changes with temperature $\Delta\rho(T)$ can be analyzed in terms of the Einstein formula~\cite{Slu_Cooper74}:\index{Einstein formula}
\begin{equation}\label{Slu:Eq1}\vspace{-2pt}
\Delta\rho=\rho-\rho_0=\rho_{\rm E}=\frac{A}{T\left(e^{\Theta_{\rm E}/T}-1\right)\left(1-e^{-\Theta_{\rm E}/T}\right)},
\end{equation}
where $\Theta_{\rm E}$ is the Einstein temperature and $A$ is a proportionality constant. Eq.~(\ref{Slu:Eq1}) is valid in the range $T < T^{\ast}\approx 60$~K where the strong electron-phonon scattering on the quasi-local vibrations of the Lu$^{3+}$ ions dominates. Figure~\ref{Slu:Fig6} shows a fit of the resistivity data by Eq.~(\ref{Slu:Eq1}), which allows estimating the Einstein temperature $\Theta_{\rm E}=162$--170~K~\cite{Slu_BolotinaDudka19}. At higher temperatures, above $T^\ast$, Umklapp processes\index{Umklapp process} dominate in the charge-carrier scattering of LuB$_{12}$, and the relation\vspace{-2pt}
\begin{equation}\label{Slu:Eq2}
\Delta\rho=\rho_{\rm U}=B_0 T\exp(-T_0/T)
\end{equation}
may be considered a good approximation for resistivity in the temperature range 100--300~K. The authors \cite{Slu_BolotinaDudka19} note that the $T_0$ parameter in Eq.~(\ref{Slu:Eq2}) estimated from these fits is close to the Einstein temperatures $\Theta_{\rm E}$, and the crossover between these two regimes described by Eqs.~(\ref{Slu:Eq1}) and (\ref{Slu:Eq2}) coincides very well with the $T^\ast$ transition region (see Fig.~\ref{Slu:Fig6}). Taking into account that $T_0 \approx v_{\rm s} q/k_{\rm B}$, where $v_{\rm s}$ is the sound velocity, $q$ is the wave number of the Umklapp phonon, and $k_{\rm B}$ is the Boltzmann constant, the authors concluded that the quasi-local modes of Lu ions are the vibrations also prevailing in the Umklapp processes in LuB$_{12}$.\index{Umklapp process}

\begin{figure}[t]
\centerline{\includegraphics[width=1.05\textwidth]{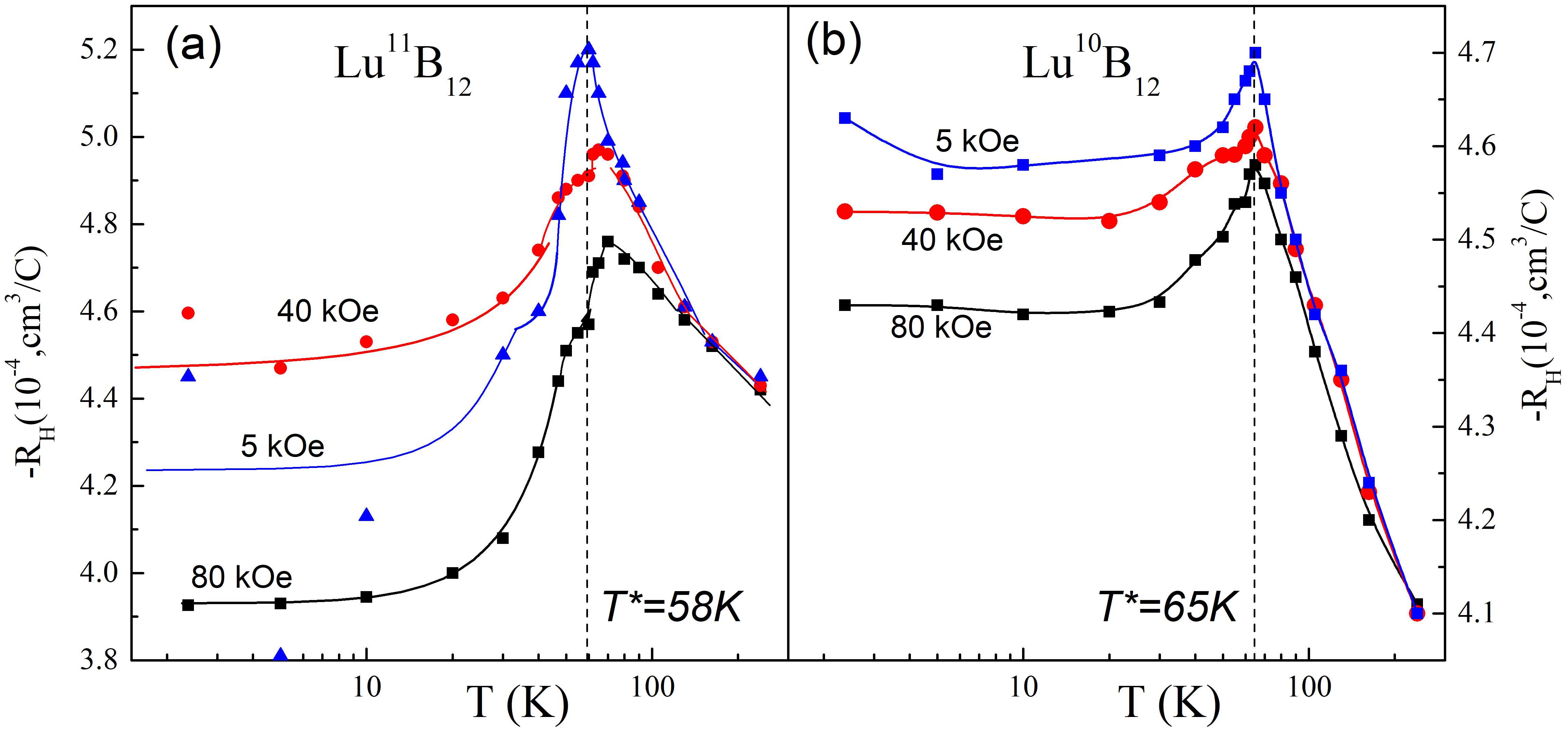}}\vspace{-1pt}
\caption{Temperature dependences of the Hall coefficient $R_{\text H}$ for (a)~Lu$^{11}$B$_{12}$ and (b)~Lu$^{10}$B$_{12}$ dodecaborides in magnetic fields $H=5$, 40, and 80~kOe \cite{Slu_SluchankoAzarevich10}.}
\label{Slu:Fig7}\index{LuB$_{12}$!Hall coefficient}
\end{figure}

The Hall effect in Lu$^N$B$_{12}$ has been studied in Refs.~\cite{Slu_SluchankoAzarevich10, Slu_SluchankoBogomolov06}. Figure~\ref{Slu:Fig7} shows the dependences of the Hall coefficient $R_{\text H}(T, H)$ obtained for the Lu$^{11}$B$_{12}$ and Lu$^{10}$B$_{12}$ crystals in fixed external magnetic fields. It follows from the data of Fig.~\ref{Slu:Fig7} and from the results of Ref.~\cite{Slu_SluchankoBogomolov06} for Lu$^{\text {nat}}$B$_{12}$ that the change in the Hall coefficient with temperature in the lutetium dodecaboride is essentially nonmonotonic with a strong anomaly near $T^{\ast}\approx60$~K, which is observed in the entire range of magnetic fields $H \leq 80$~kOe. Moreover, in strong magnetic fields (40 and 80~kOe curves in Fig.~\ref{Slu:Fig7}), the high-accuracy measurements allow discerning two closely spaced features near $T^{\ast}$ in the temperature dependences of $R_{\text H}$.

\begin{figure}[t]\vspace{-2pt}
\centerline{\includegraphics[width=0.8\textwidth]{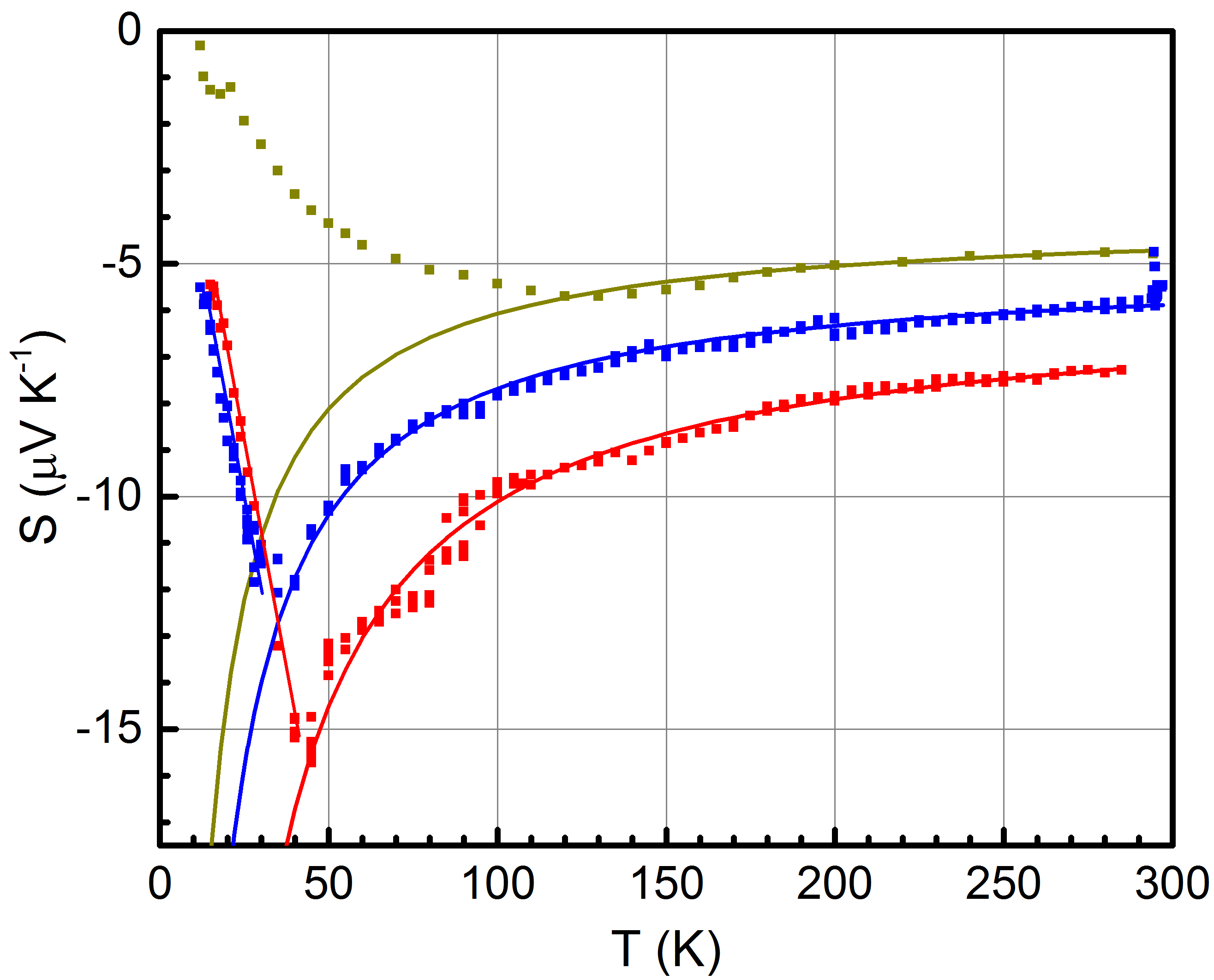}\vspace{-2pt}}
\caption{Temperature dependence of the Seebeck coefficient of the crystals Lu$^N$B$_{12}$ with $N=10$ (bottom), 11 (middle) and nat (top). Solid lines show the data approximation by the Mott dependence\index{Mott thermopower}\index{diffusive thermopower} ($T<40$~K) and phonon-drag\index{phonon drag} thermopower (at intermediate temperatures). Reproduced from Bolotina \textit{et al.}~\cite{Slu_BolotinaDudka19}.}
\label{Slu:Fig8}\index{LuB$_{12}$!Seebeck coefficient}\index{thermopower}
\end{figure}

The Seebeck coefficient $S(T)$ in Fig.~\ref{Slu:Fig8} demonstrates more or less pronounced negative minima at intermediate temperatures 30--150~K \cite{Slu_BolotinaDudka19}. The amplitude of the anomaly, which is observed for all the Lu$^{\text N}$B$_{12}$ crystals, is the largest in Lu$^{10}$B$_{12}$ and the smallest in Lu$^{\text {nat}}$B$_{12}$. At low temperatures, the Seebeck coefficient changes linearly, and the largest slope of this Mott (diffusive) thermopower\index{Mott thermopower}\index{diffusive thermopower}\index{thermopower} is detected in the isotopically pure lutetium dodecaborides. When discussing the $S(T)$ behavior presented in Fig.~\ref{Slu:Fig8}, the authors conclude that the negative minimum is a typical feature for metals with electron conduction and it appears as a crossover from phonon-drag\index{phonon drag} thermopower with the dependence $S_\text{g} \propto 1/T$ at higher temperatures to a linear diffusive low-temperature component $S=BT$. These two parts of thermopower are approximated by solid lines in Fig.~\ref{Slu:Fig8}, and the authors argue that electron-phonon scattering on quasi-local vibrations (Einstein modes)\index{Einstein mode} of Lu$^{3+}$ ions dominates in the intermediate temperature range, hence these Einstein modes are determinants for thermopower of the phonon dragging~\cite{Slu_BolotinaDudka19}.
\index{LuB$_{12}$!charge transport|)}

\vspace{-2pt}\subsection{Thermal properties}\index{LuB$_{12}$!thermal properties!specific heat}\index{LuB$_{12}$!thermal properties|(}\index{specific heat!LuB$_{12}$|(}\index{heat capacity!in LuB$_{12}$|(}

For Lu$^N$B$_{12}$ ($N=10$, 11, nat), temperature dependences of the specific heat $C(T)$ at a constant pressure and intermediate temperatures $T > 30$ K almost coincide with each other in the log-log plot (Fig.~\ref{Slu:Fig5}) whereas noticeable differences in $C(T)$ are observed at low temperatures~\cite{Slu_SluchankoAzarevich11, Slu_BolotinaDudka19, Slu_SluchankoAzarevich14}. The largest $C(T)$ values in the range of 2--20~K were detected for the Lu$^{\text {nat}}$B$_{12}$ assuming effects in the specific heat of this compound of random isotope substitution and boron vacancies. A steplike anomaly in the $C(T)$ dependence in the range 20--40~K for Lu$^{\text {nat}}$B$_{12}$ (Fig.~\ref{Slu:Fig5}) was observed and discussed previously~\cite{Slu_SluchankoAzarevich11, Slu_CzopnikShitsevalova04, Slu_CzopnikShitsevalova05} in terms of the Einstein type contribution to the specific heat from quasi-local vibrations of RE ions embedded in the large B$_{24}$ cavities in the rigid boron cage. According to the data in Fig.~\ref{Slu:Fig5}, boron isotope substitution affects only slightly the behavior of the Einstein component in the specific heat of Lu$^N$B$_{12}$ ($N=10$, 11, nat), which confirms the loosely bound state of the Lu$^{3+}$ ions.

\begin{figure}[t]
\centerline{\includegraphics[width=0.73\textwidth]{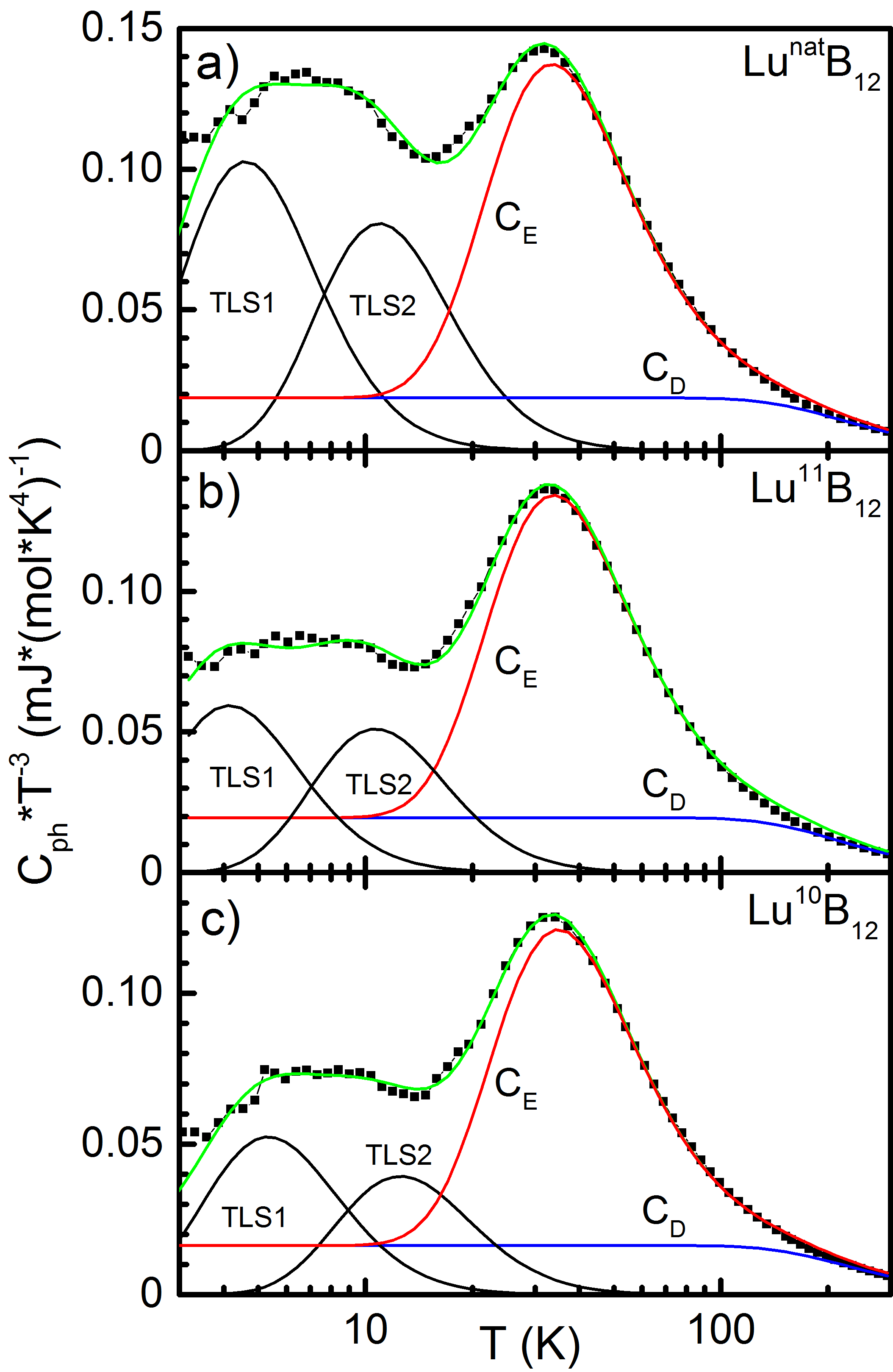}}
\caption{Separation of the low-temperature vibrational contribution $C_\text{ph}/T^3$ to the specific heat of Lu$^N$B$_{12}$ into Debye ($C_\text{D}$), Einstein ($C_\text{E}$) and two Schottky (TLS$_1$ and TLS$_2$) components. Reproduced from Ref.~\cite{Slu_BolotinaDudka19}.}
\label{Slu:Fig9}\index{LuB$_{12}$!thermal properties!specific heat!vibrational contribution}\index{Schottky specific heat}
\end{figure}

The contributions from B and Lu atoms in the vibration heat capacity have been considered in terms of the Debye and Einstein models\index{Debye model}\index{Einstein model} given by Eqs.~(\ref{Slu:Eq3}) and \ref{Slu:Eq4}), respectively \cite{Slu_SluchankoAzarevich11}:
\begin{equation}\label{Slu:Eq3}
\frac{C_\text{D}}{T^3} = \frac{9\,rR}{T_{\rm D}^3} \int_0^{\Theta_{\rm D}/T}\!\!\frac{x^4 e^{x}}{(e^{x}-1)^2}{\rm d}x
\end{equation}
\begin{equation}\label{Slu:Eq4}
\frac{C_\text{E}}{T^3} = \frac{3R}{\Theta_{\rm E}^3} \left(\frac{\Theta_{\rm E}}{T}\right)^5 \frac{e^{-\Theta_{\rm E}/T}}{(1-e^{-\Theta_{\rm E}/T})^2},
\end{equation}
where $R$ is the gas constant and $r=12$ is the number of boron atoms in the unit cell. The electronic specific heat $C_\text{el} = \gamma T$ with $\gamma \approx 3$~mJ/(mol$\cdot$K$^2$) and the Debye contribution with $\Theta_{\rm D} \approx 1100$~K detected from refined atomic displacement parameters (ADP)\index{atomic displacement parameters} of the structure model were applied in Ref.~\cite{Slu_BolotinaDudka19} to calculate the difference $C_\text{ph} = C - C_\text{el} - C_\text{D} = C_\text{E}(T) + \sum_{\,i=1,2}C_{\text{Sch}(i)}(T)$. The two Schottky terms $C_{\text {Sch}(i)}$\index{Schottky specific heat} were used additionally to approximate low-temperature anomalies in the specific heat by contributions from the two-level systems TLS$_1$ and TLS$_2$. It has been argued \cite{Slu_SluchankoAzarevich11, Slu_SluchankoAzarevich14} that these two Schottky components given by
\begin{equation}\label{Slu:Eq5}
\frac{C_{\text{Sch}(i)}}{T^3}=\frac{RN_i}{T^3} \left(\frac{\Delta E_i}{k_{\rm B}T}\right)^{\!2}\frac{e^{\Delta E_i/k_{\rm B}T}}{(e^{\Delta E_i/k_{\rm B}T}+1)^2}
\end{equation}
($N_i$ is the concentration of TLS$_i$) are necessary to describe the effect of boron vacancies (TLS$_2$) and divacancies (TLS$_1$) in the specific heat of $R$B$_{12}$. Indeed, in view of a weak coupling of the RE ions in the boron network in combination with a significant number of boron vacancies and other intrinsic defects in the UB$_{12}$ type structure \cite{Slu_FojudHerzig07}, the formation of various two-level systems arranged in double-well potentials (DWP)\index{double-well potential} should be expected at the displacements of Lu$^{3+}$ ions from their central positions in the B$_{24}$ cuboctahedra. All the above mentioned specific heat contributions are shown in Fig.~\ref{Slu:Fig9} in the $C(T)/T^3$ plot together with the experimental data~\cite{Slu_BolotinaDudka19}.

Based on the approach developed in Refs.~\cite{Slu_SluchankoAzarevich11, Slu_SluchankoAzarevich14}, the energy $\Delta E_2/k_{\rm B} = 54$--64~K in Eq.~(\ref{Slu:Eq5}) should be attributed to the barrier height in the DWP (or, in other words, the energy difference in the two-level system). The normalized concentration of TLS$_2$ of $N_2 = 0.047$--0.08 corresponds to the number of Lu ions displaced from the central positions in B$_{24}$ cells. This result approximately agrees with the concentration of 0.036 deduced from EXAFS measurements\index{EXAFS} of LuB$_{12}$ powders at low temperatures~\cite{Slu_MenushenkovYaroslavtsev13}. Taking into account that each boron vacancy\index{boron vacancy concentration} ensures the displacement of two neighboring RE ions from the center of the B$_{24}$ octahedron in the structure of $R$B$_{12}$, the number of boron vacancies was estimated as $n_{\text v} = N_2/2 = 2.4$--4\%, being dependent on the boron isotope composition~\cite{Slu_SluchankoAzarevich14}. According to Junod \textit{et~al.} \cite{Slu_JunodJarlborg83}, $(5/4)R\pi^4C_\text{ph}/T^3$ vs. $T$ gives an approximate picture of the one-dimensional phonon density of states (DOS) $\omega^{-2}F(\omega)$ for $\omega = 4.928\,T$, where $\omega$ is expressed in kelvins. On a logarithmic scale the response of $C_\text{ph}/T^3$ to a $\delta$-function (Einstein peak) is a bell-shaped peak, shown as the $C_\text{E}$ component in Fig.~\ref{Slu:Fig9}. In this way, the main contribution to $C_\text{ph}$ for LuB$_{12}$ at low temperatures \cite{Slu_SluchankoAzarevich11, Slu_BolotinaDudka19, Slu_CzopnikShitsevalova05}\nocite{Slu_Shitsevalova01} is related to the mode equal to 14.1~meV that is very close to the energy of the $\delta$-like peak (14.2~meV) from the neutron phonon spectrum\index{inelastic neutron scattering!on phonon modes} \cite{Slu_BouvetKasuya98} (see also Chapter~\ref{Chapter:Alekseev}) and from the point-contact spectrum (14.5~meV)~\cite{Slu_FlachbartSamuely02}.

When discussing the nature of the anomalies at $T^{\ast} \approx 60$~K (see Fig.~\ref{Slu:Fig6}), the authors \cite{Slu_SluchankoAzarevich11, Slu_BolotinaDudka19} concluded that it should be attributed to the order-disorder phase transition, and below $T^{\ast}$ the Lu$^{3+}$ ions are ``freezing'' at different positions of DWP\index{double-well potential} minima induced by the random distribution of boron vacancies, $^{10}$B and $^{11}$B atoms, and impurities. In this scenario the barrier height $\Delta E_2/k_{\rm B}$ in the DWP is close to the cage-glass\index{cage glass} transition temperature $T^\ast = 54$--65~K in crystals of Lu$^N$B$_{12}$ \cite{Slu_SluchankoAzarevich11, Slu_SluchankoAzarevich10}, and the disordered state below $T^{\ast}$ is a mixture of two components: the crystal (rigid boron covalent cage) and glass (clusters of Lu ions displaced from their central positions in the B$_{24}$ cuboctahedra).

\begin{figure}[t]
\centerline{\includegraphics[width=0.85\textwidth]{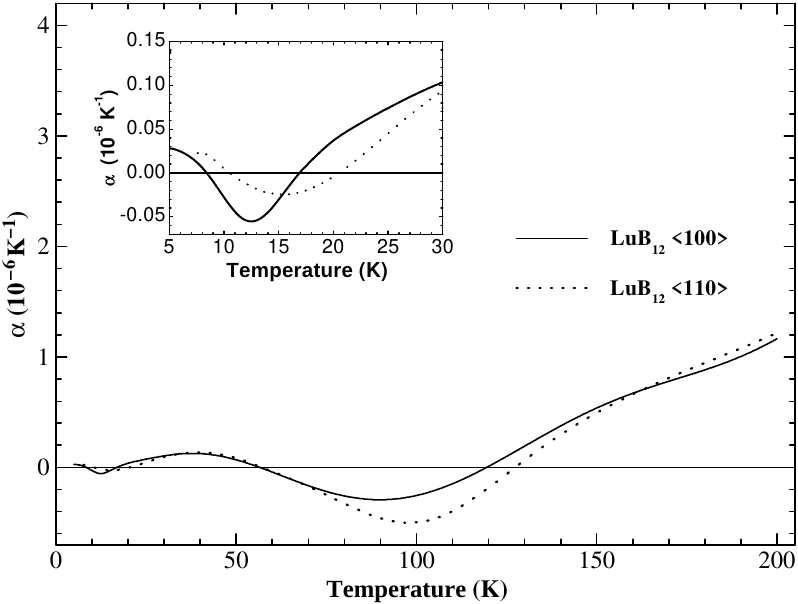}}
\caption{The temperature dependence of the thermal expansion coefficient $\alpha$ for the LuB$_{12}$ $\langle 100 \rangle$ and $\langle 110 \rangle$ single crystals. Inset: $\alpha$ versus $T$ in an expanded scale. Adapted from Shitsevalova~\cite{Slu_Shitsevalova01}.}
\label{Slu:Fig10}\index{LuB$_{12}$!thermal expansion}
\end{figure}

Thermal conductivity\index{thermal conductivity} measurements of Lu$^\text{nat}$B$_{12}$ display a large amplitude maximum of $\lambda(T)$ at $\sim$25~K~\cite{Slu_MisiorekMucha95}. The separation of electron and phonon contributions was calculated from the Wiedemann--Franz law, indicating strong thermal conductivity anomalies at 25~K both for $\lambda_{\rm e}$ and $\lambda_\text{ph}$ components and detecting at higher temperatures the $\lambda_{\text {ph}}\propto 1/T$ dependence which is typical for dielectrics \cite{Slu_MisiorekMucha95}. The ratio $\lambda_\text{ph}/\lambda_{\rm e}\approx 3$--4, that is also unusual for good metals, may be estimated from the thermal conductivity data in the range 4--300~K~\cite{Slu_MisiorekMucha95}.

Keeping in mind that vibrations in the DWP are anharmonic in principle,\index{double-well potential} it is important to investigate the thermal expansion originating from the anharmonic atom vibrations in LuB$_{12}$. The temperature dependences of the thermal expansion coefficient $\alpha$ of the Lu$^\text{nat}$B$_{12}$ single crystals are shown in Fig.~\ref{Slu:Fig10} \cite{Slu_CzopnikShitsevalova05}. Here $\alpha(T)$ reveals a negative thermal expansion (NTE)\index{negative thermal expansion} in two intervals: the ``low-temperature'' one with a minimum around 12--15~K and the ``high-temperature'' one with a minimum around 90--100~K. The authors \cite{Slu_CzopnikShitsevalova05} discuss a slight difference in the thermal expansion of the LuB$_{12}$ crystals with $\langle 100 \rangle$ and $\langle 110 \rangle$ orientations, which could be produced either by the anisotropy of the chemical bonds (the $\langle 100 \rangle$ axis corresponds to the Lu--B bond direction, and the $\langle 110 \rangle$ axis to the Lu--Lu bond), or the low-temperature distortion of the LuB$_{12}$ fcc crystal structure.

\begin{figure}[b!]
\centerline{\includegraphics[width=0.84\textwidth]{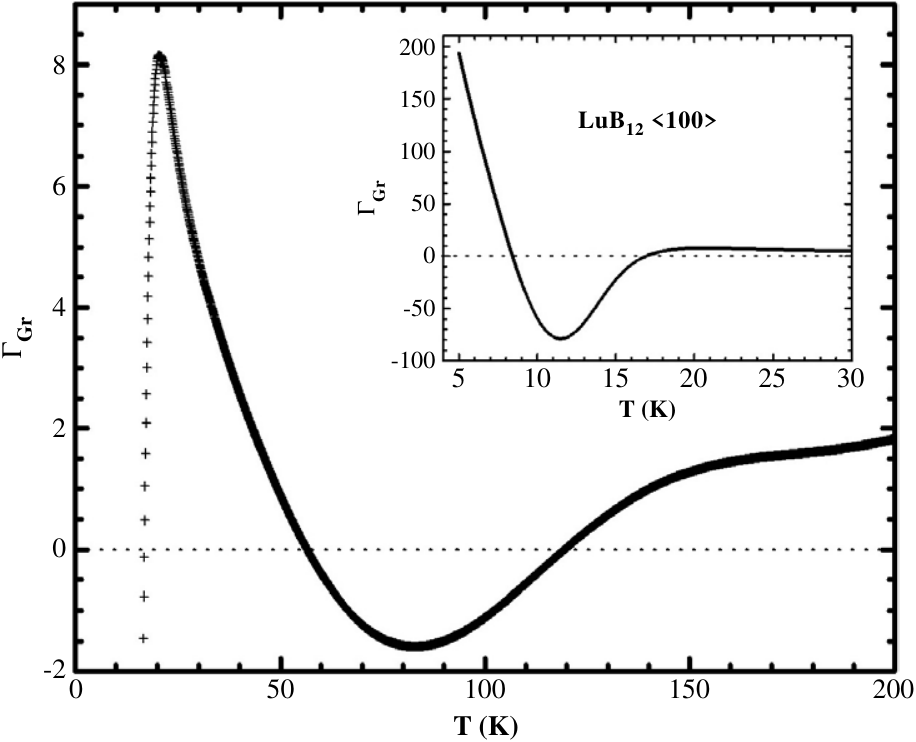}}
\caption{The LuB$_{12}$ generalized Gr\"{u}neisen parameter $\Gamma_\text{Gr}$. Inset: $\Gamma_\text{Gr}$ versus $T$ for LuB$_{12}$ in the expanded low-temperature scale 5--30~K~\cite{Slu_CzopnikShitsevalova05}.}
\label{Slu:Fig11}\index{LuB$_{12}$!Gr\"{u}neisen parameter}\index{Gr\"{u}neisen parameter}
\end{figure}

\begin{figure}[t]
\centerline{\includegraphics[width=\textwidth]{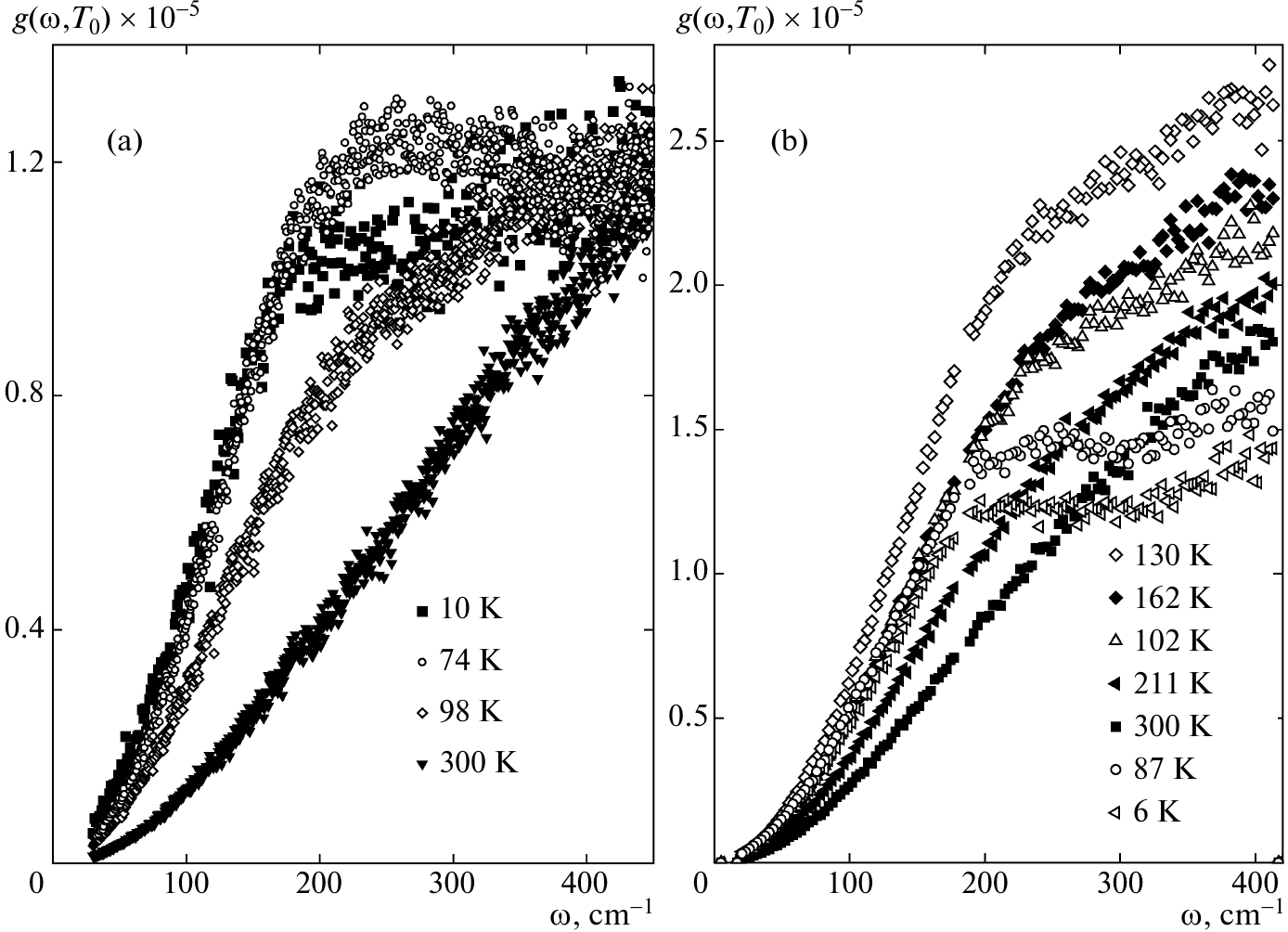}}
\caption{Temperature variation of the low-frequency vibrational density of states $g(\omega,T_0)$ for (a)~Lu$^{10}$B$_{12}$ and (b)~Lu$^{11}$B$_{12}$ dodecaborides. Reproduced from Ref.~\cite{Slu_SluchankoAzarevich11}.}
\label{Slu:Fig12}\index{LuB$_{12}$!vibrational density of states}
\end{figure}

Based on the experimental heat capacity, thermal expansion and bulk modulus ($c_\text{b}$) data, Czopnik \textit{et~al.} \cite{Slu_CzopnikShitsevalova05} evaluated the temperature dependence of the generalized Gr\"{u}neisen parameter\index{Gr\"{u}neisen parameter} $\Gamma_\text{Gr} = 3{\kern.5pt}\alpha{\kern.5pt}c_\text{b}V{\kern-1pt}/C_\text{v}$ (where $V$ is the molar volume) in LuB$_{12}$. In Fig.~\ref{Slu:Fig11}, $\Gamma_\text{Gr}(T)$ is presented for LuB$_{12}$ in the direction $\langle 100 \rangle$. In the 8--17~K and 60--130~K ranges, $\Gamma_\text{Gr}$ for LuB$_{12}$ is negative in accordance with its two ranges of NTE, whereas $C_\text{p}$ from 2~K to 300~K is a monotonically increasing function of temperature (Fig.~\ref{Slu:Fig5}), and $c_\text{b}$ is a smooth monotonically decreasing function of temperature that changes from 229.5 down to 224.9~GPa \cite{Slu_CzopnikShitsevalova05}. As a result, the generalized Gr\"{u}neisen parameter $\Gamma_\text{Gr}$, but not the heat capacity, mainly determines the $\alpha(T)$ behavior for LuB$_{12}$ in both $\langle 100 \rangle$ and $\langle 110 \rangle$ crystal orientations. The authors \cite{Slu_CzopnikShitsevalova05} suggested that a defect-induced soft mode is responsible both for the NTE and the Schottky anomalies\index{Schottky specific heat} of heat capacity at temperatures below 20~K, having the same origin\,---\,a formation of the two-level tunneling systems based on the metal ions and defects \cite{Slu_TakegaharaKasuya85}. As well, according to \cite{Slu_CzopnikShitsevalova05}, the existence of the negative $\Gamma_\text{Gr}$ at intermediate temperatures 60--130~K should be attributed to the presence in the phonon spectra of LuB$_{12}$ of a flat transverse acoustic mode characterized by $\Theta_{\text E} \approx 164$~$\text{K} \approx 14.1$~meV (see also Chapter~\ref{Chapter:Alekseev}), as it was predicted for the first time by Dayal~\cite{Slu_Dayal44} and Barron~\cite{Slu_Barron55} for tetrahedral semiconductors Si and Ge. Dove and Hong Fang pointed out recently~\cite{Slu_DoveFang16} that in many cases NTE is associated with a combination of transverse acoustic and optical modes in compounds where the lattice instability is developed.

Then, it is worth noting here two characteristic temperatures in the $\alpha(T)$ and $\Gamma_\text{Gr}(T)$ dependences of LuB$_{12}$ (Figs.~\ref{Slu:Fig10} and \ref{Slu:Fig11}). The sign inversion at 60~K on these two curves matches precisely both with the cage-glass\index{cage glass} transition $T^{\ast}$ and the barrier height in DWP $\Delta E_2/k_{\rm B}\approx 60$~K.\index{double-well potential} Additionally, the knee-type anomaly observed on these curves near 130--150~K correlates very well with the renormalization of the low-frequency vibration spectra (see, for example, Fig.~\ref{Slu:Fig12}) and to the appearance of the boson peak in the Raman spectra\index{LuB$_{12}$!Raman spectra}\index{Raman scattering} attributed to the disordering in the positions of the RE ions in the LuB$_{12}$ matrix~\cite{Slu_SluchankoAzarevich11}. Applying the approach proposed in Refs.~\cite{Slu_DuvalBoukenter86, Slu_MalinovskyNovikov88}, the position of the boson peak in the Raman spectra of disordered systems was used in Ref.~\cite{Slu_SluchankoAzarevich11} to evaluate quantitatively the spatial size $D_N$ of regions with low-frequency quasi-local vibrations (vibrational clusters)\index{vibrational cluster} in the $R$B$_{12}$ structure. Using the relation for boson peak frequency $\omega_\text{max}$~$\approx$~(0.7--0.85)~$v_{\rm s}/cD_N$ \cite{Slu_DuvalBoukenter86, Slu_MalinovskyNovikov88, Slu_BuchenauPrager86} and the sound velocity $v_{\text s} \approx 962$~m/s found for LuB$_{12}$ at $T = 78$~K \cite{Slu_GrechnevBaranovskiy08}, authors \cite{Slu_SluchankoAzarevich11} obtain $D_{10, 11} = 12$--15~\r{A} for Lu$^{10}$B$_{12}$ and Lu$^{11}$B$_{12}$, and $D_\text{nat} = 18$--22~\r{A} for Lu$^\text{nat}$B$_{12}$, arguing that in the presence of additional substitutional $^{10}$B\hspace{0.5pt}-$^{11}$B disorder, the correlation length in the system of interacting harmonic oscillators increases by a factor of about 1.5 and reaches $D_\text{nat} \approx 3a_0$ ($a_0\approx7.5$~\r{A} is the lattice constant).

It was noted in Ref.~\cite{Slu_SluchankoAzarevich11} that the development of the lattice instability with a decrease in temperature leads to a sharp increase in the vibrational DOS $g(\omega,T_0)$ that is plotted in Fig.~\ref{Slu:Fig12}. Around the temperature $T \approx T_{\text E} = 5T_{C_\text{max}}\approx 150$~K (here $T_{C_\text{max}}$ is the temperature of the heat-capacity maximum), the mean free path of phonons reaches the Ioffe-Regel limit\index{Ioffe-Regel limit} and becomes comparable with their wavelength~\cite{Slu_Parshin94}. Near $T_\text{E}$, in addition to the maximum in the vibrational DOS $g(\omega,T)$ (see Fig.~\ref{Slu:Fig12}), the sharp maximum of the relaxation rate is observed in muon-spin relaxation ($\mu$SR) experiments with $R$B$_{12}$ dodecaborides ($R$~=~Er, Yb, Lu) and Lu$_{\text {1-x}}$Yb$_x$B$_{12}$ solid solutions~\cite{Slu_KalviusNoakes02, Slu_KalviusNoakes03}. The authors suggested that the large amplitude dynamic features arise from atomic motions within the B$_{12}$ clusters.
\index{LuB$_{12}$!thermal properties|)}\index{specific heat!LuB$_{12}$|)}\index{heat capacity!in LuB$_{12}$|)}

\subsection{Optical properties}\index{LuB$_{12}$!optical properties|(}

\nocite{Slu_DresselGruener02, Slu_OkamuraMatsunami00}
\begin{figure}[b!]
\centerline{\includegraphics[width=0.85\textwidth]{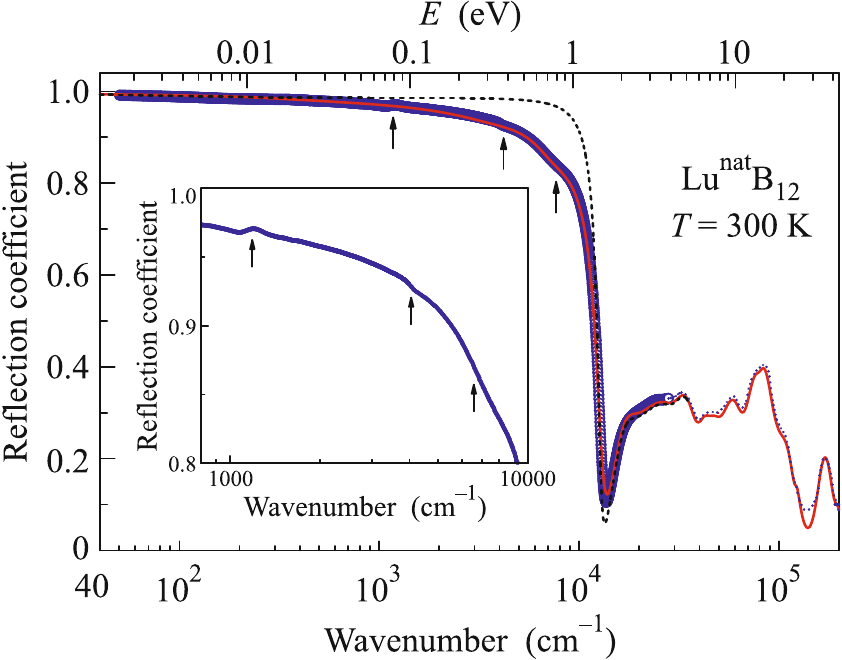}}
\caption{Room-temperature reflection coefficient spectrum of the Lu$^{\text {nat}}$B$_{12}$ crystal. Dots show experimental data obtained using the Fourier-transform spectrometer and ellipsometer. The dotted line corresponds to high-frequency reflectivity data from \cite{Slu_KimuraOkamura99}. The solid red line shows the results of fitting the spectrum using the Drude term for the free charge carrier response and Lorentzians responsible for absorption resonances. The dashed line shows best fit of the spectrum that can be obtained using the Drude conductivity term alone with $\sigma_\text{DC}=95\,000$~$\Omega^{-1}$cm$^{-1}$ and $\gamma_\text{D}=290$~cm$^{-1}$. Kink-like features observed in the spectrum are indicated by arrows and presented in more detail in the inset. Reproduced from Gorshunov \textit{et al.}~\cite{Slu_GorshunovZhukova18}.}
\label{Slu:Fig13}\index{LuB$_{12}$!optical properties!reflectivity}
\end{figure}

The optical reflectivity $R(\omega)$ experiments on single crystals of LuB$_{12}$ have been conducted for the first time by Okamura \textit{et~al.} \cite{Slu_OkamuraKimura98}, and the optical conductivity $\sigma(\omega)$ spectra were obtained from $R(\omega)$ using the Kramers-Kronig relations~\cite{Slu_DresselGruener02}. It was found in Refs.~\cite{Slu_OkamuraKimura98, Slu_OkamuraMatsunami00} that $R(\omega)$ spectra have a clear plasma cutoff ($\omega_\text{p}$) near 1.6~eV and sharp structures above 2~eV due to interband transitions~\cite{Slu_KimuraOkamura99}. Below $\omega_\text{p}$, the optical conductivity demonstrates a sharp rise due to a metallic response of free carriers.

\begin{figure}[b!]\vspace{-2pt}
\centerline{\includegraphics[width=0.85\textwidth]{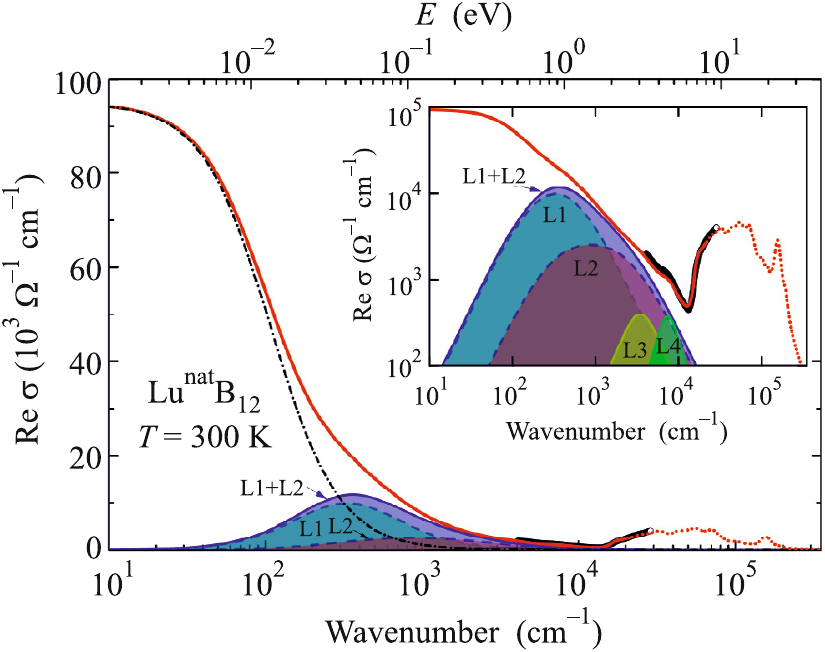}}
\caption{Room-temperature spectrum of the real part of the optical conductivity of the Lu$^\text{nat}$B$_{12}$ single crystal (solid red line). The dashed spectrum above 20\,000~cm$^{-1}$ corresponds to the result obtained by fitting the reflectivity data from Ref.~\cite{Slu_KimuraOkamura99} (dashed line in Fig.~\ref{Slu:Fig13}). Dots at 4000--27\,000~cm$^{-1}$ represent conductivity spectrum from ellipsometry measurements. The spectrum is obtained by least-square fitting of the reflection coefficient spectrum shown in Fig.~\ref{Slu:Fig13} using Drude conductivity term and four Lorentzian terms. The Drude term is shown separately by a dash-dotted line and the Lorentzian terms by dashed lines. The L1\,+\,L2 line corresponds to the sum of L1 and L2 contributions. To visualize separate contributions, same spectra are shown in a double logarithmic plot in the inset. Reproduced from Gorshunov \textit{et al.}~\cite{Slu_GorshunovZhukova18}.\vspace{-1pt}}
\label{Slu:Fig14}\index{LuB$_{12}$!optical properties!optical conductivity}
\end{figure}

Recently, more detailed optical studies of the high-quality Lu$^N$B$_{12}$ single crystals with different boron isotopes ($N = 10$, 11, nat) were performed at room temperature by Gorshunov \textit{et~al.} \cite{Slu_GorshunovZhukova18}. Figure~\ref{Slu:Fig13} shows a broad-band spectrum of the reflection coefficient of Lu$^{\text {nat}}$B$_{12}$ (dots). At low frequencies, the spectrum looks typically metallic \cite{Slu_DresselGruener02} with the plasma edge at $\sim$14\,000~cm$^{-1}$ ($\sim$1.75~eV). It was found in Ref.~\cite{Slu_GorshunovZhukova18} that it is impossible to model the measured spectrum within the Drude conductivity model \cite{Slu_DresselGruener02} alone (see Fig.~\ref{Slu:Fig13}), keeping the measured dc conductivity value fixed to $\sigma_\text{dc} = 95\,000$~$\Omega^{-1}$cm$^{-1}$ and varying only the scattering rate $\gamma_{\rm D}$. This result is a strong indication that the infrared optical response of LuB$_{12}$ is not determined by just free charge carriers which provide its metallic conductivity, and that there are additional IR excitations and specific mechanisms governing the electronic properties of the compound. Additional evidence for such excitations is seen in the reflectivity spectrum as kink-like features at $\sim$1100, 4500, and 7000~cm$^{-1}$ (indicated by arrows in Fig.~\ref{Slu:Fig13} and enlarged in the inset), which are not completely screened by the charge carriers. The authors introduce, together with the free-carrier Drude term, the \emph{minimal} set of excitations (Lorentzians) that provides a suitable model of the measured reflectivity spectrum (red solid line in Fig.~\ref{Slu:Fig13}). Figure~\ref{Slu:Fig14} presents the obtained broad-band conductivity spectrum of LuB$_{12}$ including five additive contributions detected in Ref.~\cite{Slu_GorshunovZhukova18}. The excitations L1 and L2 (Fig.~\ref{Slu:Fig14}) are the dominant and rather unusual in having unexpectedly large dielectric contributions, especially the L1 peak ($\Delta{\varepsilon} = 8000 \pm 4000$), and being strongly overdamped ($\gamma/v_0>2.5$).\enlargethispage{1pt}

\begin{figure}[t]
\centerline{\includegraphics[width=0.9\textwidth]{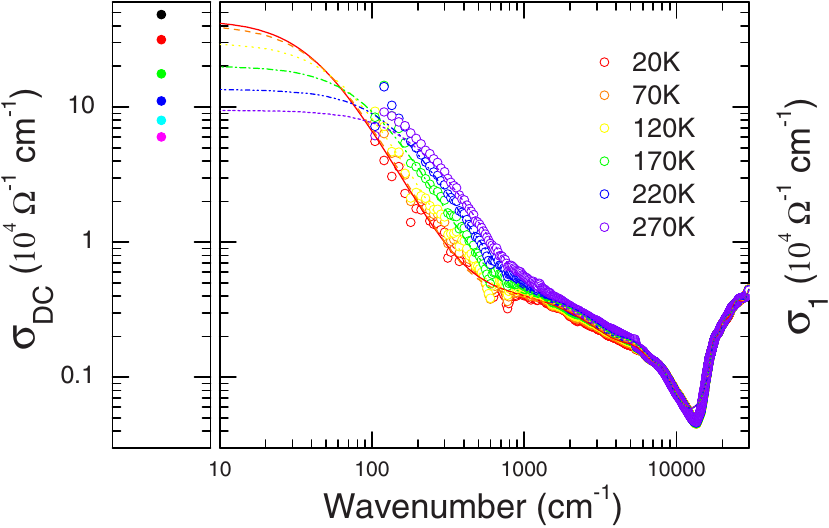}}\vspace{-1pt}
\caption{Temperature dependence of the real part of the wide-frequency-range optical conductivity spectra of LuB$_{12}$. Direct current conductivity values obtained from resistivity measurements are shown in the left panel. Reproduced from Teyssier \textit{et al.}~\cite{Slu_TeyssierLortz08}.}
\label{Slu:Fig15}\index{LuB$_{12}$!optical properties!optical conductivity}
\end{figure}

Gorshunov \textit{et al.}~\cite{Slu_GorshunovZhukova18} estimate the concentration of conduction electrons involved in the free carrier conductivity and in the formation of the collective excitations L1 and L2 using an expression $v_\text{pl}^2 = ne^2/(\pi m^\ast)=f$ for the plasma frequency and oscillator strength $f$ (here $n$ is the concentration of free electrons, $e$\,--\,elementary charge). With the Drude unscreened plasma frequency \mbox{$v_\text{pl} = 21\,700\pm3200$~cm$^{-1}$} (\mbox{$\varepsilon_\infty \approx 2.5$}) and \mbox{$m^{\ast} = 0.5 m_0$}~\cite{Slu_HeineckeWinzer95, Slu_LiuHartstein18, Slu_PluzhnikovShitsevalova08}, they obtained \mbox{$n_\text{D}=2.6 \cdot 10^{21}$~cm$^{-3}\pm30$\%}. From the combined oscillator strength of the L1 and L2 Lorentzian terms, $f_\text{L1+L2} = 1.3 \cdot 10^9$~cm$^{-2}$ assuming \mbox{$m^\ast$ = 0.5$m_0$}, they get the concentration of the charge carriers\index{LuB$_{12}$!charge-carrier concentration}\index{charge-carrier concentration} participating in the formation of the collective excitation \mbox{$n_\text{L1+L2} = 7.2 \cdot 10^{21}$~cm$^{-3}\pm30\%$} and the total concentration of charges in the conduction band $n_\text{tot} = n_\text{D} + n_\text{L1+L2} = 9.8 \cdot 10^{21}$~cm$^{-3}\pm30\%$. The obtained value $n_\text{tot}$ coincides very well with the concentration of Lu-ions $n(\text{Lu}) = 9.6 \cdot 10^{21}$~cm$^{-3}$ in LuB$_{12}$ assuming that every Lu ion delivers one electron in the conduction band. According to the above estimates, about 70\% of charges in the conduction band are involved in the formation of the collective excitation. Similar estimates were done in Ref.~\cite{Slu_GorshunovZhukova18} for Lu$^{10}$B$_{12}$ and Lu$^{11}$B$_{12}$ crystals with the parameters giving the percentage of about 70\% and 80\%, respectively. Thus, it was concluded that nonequilibrium electrons involved in the collective modes dominate in the LuB$_{12}$ charge transport. The temperature dependence of the broad-band dynamic conductivity spectra has been studied in Lu$^{\text {nat}}$B$_{12}$ by Teissier \textit{et~al.}~\cite{Slu_TeyssierLortz08} where the narrowing of the Drude term frequency range with the temperature lowering down to 20~K (Fig.~\ref{Slu:Fig15}) was observed, providing better conditions for the separation of Drude and collective-mode contributions. Our crude estimations allow concluding that during cooling both the oscillator strength (intensity) of the Lorentzian describing the broad bump seen between 1000 and 10\,000~cm$^{-1}$ (see Fig.~\ref{Slu:Fig15}) and the squared plasma frequency of the Drude contribution slightly increase. So, the collective mode (overdamped oscillator), which is present even at 20~K, may be considered as a fingerprint of a quantum motion (zero-temperature vibrations) in the matrix of LuB$_{12}$.\enlargethispage{1pt}

It is worth noting also that the positions of L1 ($\sim$250~cm$^{-1}\approx 31$~meV) and L2 ($\sim$1000~cm$^{-1}=125$~meV) collective modes in the IR spectra of LuB$_{12}$ (Fig.~\ref{Slu:Fig14}) correlate very well with those reported by Bouvet \textit{et~al.}~\cite{Slu_BouvetKasuya98}, Nemkovski \textit{et~al.}~\cite{Slu_NemkovskiAlekseev04}, and Rybina \textit{et~al.}~\cite{Slu_RybinaAlekseev07} in the inelastic neutron scattering (INS)\index{inelastic neutron scattering!on phonon modes} studies of the boron phonon modes in Lu$^{11}$B$_{12}$ (see Chapter \ref{Chapter:Alekseev}) and by Werheit \textit{et~al.}~\cite{Slu_WerheitFilipov11} in the Raman spectra of Lu$^N$B$_{12}$ with $N=10$, 11, nat (see Chapter~\ref{Chapter:Ponosov}).
\index{LuB$_{12}$!optical properties|)}

\subsection{Magnetoresistance anisotropy and dynamic charge stripes}\index{LuB$_{12}$!magnetoresistance anisotropy|(}\index{LuB$_{12}$!dynamic charge stripes}\index{dynamic charge stripes}\index{charge stripes}\index{anisotropic magnetoresistance!in LuB$_{12}$}

The transverse magnetoresistance (MR)\index{transverse magnetoresistance}\index{LuB$_{12}$!transverse magnetoresistance} of LuB$_{12}$ was measured for current directions $\mathbf{I}\parallel\langle110\rangle$ and $\langle100\rangle$ by Heinecke \textit{et~al.}~\cite{Slu_HeineckeWinzer95}. The strongest magnetic anisotropy was observed at $\mu_0 H = 12$~T and $T = 0.5$~K, reaching the values between 4 for $\mathbf{H}\parallel\langle111\rangle$ and 18 for $\mathbf{H}\parallel\langle001\rangle$. As the authors found a saturation of MR with $\mathbf{I}\parallel\langle110\rangle$ and $\mathbf{H}\parallel\langle111\rangle$, they concluded that LuB$_{12}$ is an uncompensated metal with more than one direction of open orbits on the Fermi surface. The local MR maxima were suggested to be caused by open orbits in the current direction. The drift and Hall mobilities of the charge carriers were estimated to be $\sim$2000~\mbox{cm$^2$\hspace{0.8pt}V$^{-1}$\hspace{0.5pt}s$^{-1}$}, and it was concluded that the impurity scattering dominates in LuB$_{12}$ below 20~K~\cite{Slu_SluchankoBogach09a}.

\begin{figure}[t!]
\centerline{\includegraphics[width=0.9\textwidth]{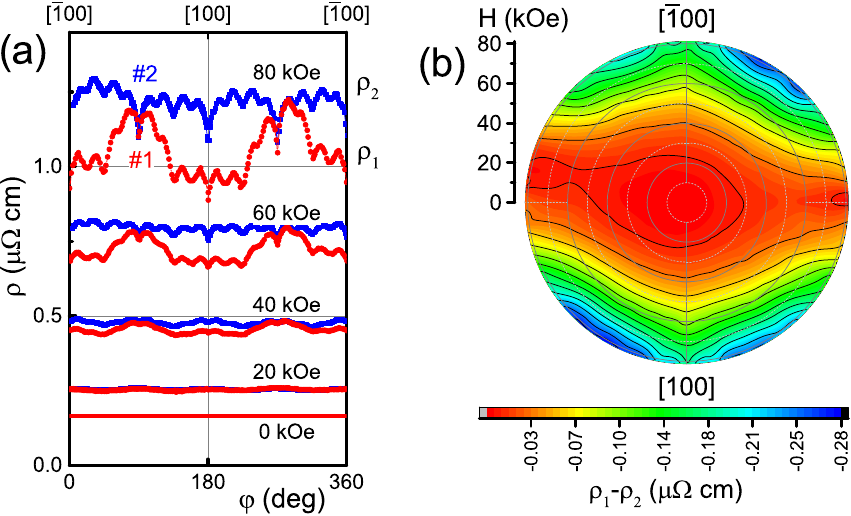}}
\caption{(a)~Field-angular dependences of resistivity $\rho_1(\phi)$ and $\rho_2(\phi)$ obtained by rotating the crystals \#1 and \#2 around their current axes $\mathbf{I}_1\parallel[010]$ and $\mathbf{I}_2\parallel[001]$, respectively, in various magnetic fields up to 80~kOe at temperatures of 2.0--4.2~K. (b)~The anisotropy of magnetoresistance $\rho_1(\phi)-\rho_2(\phi) = f(\phi,H)$ presented in polar coordinates. Reproduced from Ref.~\cite{Slu_SluchankoBogach18}.}
\label{Slu:Fig16}\index{LuB$_{12}$!resistivity!field-angular dependence}\index{LuB$_{12}$!magnetoresistance anisotropy}\index{anisotropic magnetoresistance!in LuB$_{12}$}\index{field-angular anisotropy!in magnetoresistance}
\end{figure}

An alternative interpretation was proposed in recent transverse MR studies of LuB$_{12}$ \cite{Slu_SluchankoBogach18, Slu_BolotinaDudka18}. The transverse MR anisotropy has been measured for two [010]-\;and [001]-elongated rectangular mono\-domain single crystals cut from one ingot of Lu$^\text{nat}$B$_{12}$ with equally oriented [100], [010], and [001] faces \cite{Slu_SluchankoBogach18}. A significant anisotropy (up to 20\%) was observed below $T^{\ast} \approx 60$~K at $H = 80$~kOe for one of these crystals, in spite of the fact that the field directions [001] and [100] are symmetry equivalent in the cubic lattice. In addition, a minimum was detected below $T^{\ast}$ in the temperature dependences of the resistivity, $\rho$($T$, $H$\,=\,80~kOe), with the growth of resistivity at low temperatures. To shed more light on the nature of the unusual anisotropy, angular dependences $\rho(\phi)$ were measured at 2.0--4.2~K in the magnetic fields up to 80~kOe by rotating these two crystals around their current axes \cite{Slu_SluchankoBogach18}. Families of curves $\rho_{1,2}(\phi)$  for the samples are shown in Fig.~\ref{Slu:Fig16}\,(a) demonstrating the great difference for $\mathbf{H}\parallel[100]$ ($\phi = 0^{\circ}$) and $\mathbf{H}\parallel[001]$ ($\phi = 90^{\circ}$), although these directions are equivalent in cubic crystals. It is clearly discerned from the polar plot in Fig.~\ref{Slu:Fig16}\,(b) that conduction channels\index{conduction channels} appear in the LuB$_{12}$ matrix \cite{Slu_SluchankoBogach18}. The orientation of the conduction channels was detected certainly in Ref.~\cite{Slu_BolotinaDudka18}, where the detailed x-ray diffraction\index{x-ray diffraction} study was carried out in combination with the angular MR measurements of Lu$^\text{nat}$B$_{12}$ crystals. A new approach was developed that consists in the difference Fourier synthesis\index{difference Fourier synthesis} of the residual electron density (ED) as well as in the reconstruction of the ED distribution using maximal entropy method\index{maximal entropy method} (see Chapter~\ref{Chapter:Bolotina}). It was found~\cite{Slu_BolotinaDudka18} that the ED peaks become stronger with the temperature lowering and form a filamentary structure of conduction channels\,---\,unbroken charge stripes oriented almost along the $[110]$ axis (see Figs.~\ref{Fig:LuB12-MEM-maps} and \ref{Fig:LuB12-Magnetoresistance} in Chapter~\ref{Chapter:Bolotina}). These observations are in accordance with the cubic symmetry distortions of the LuB$_{12}$ crystals \cite{Slu_BolotinaDudka18} (see Fig.~\ref{Fig:Bolotina13-14} in Chapter~\ref{Chapter:Bolotina}). It has been shown also that the asymmetric ED distribution correlates very accurately with the anisotropy of transverse MR. Thus, the same conduction channels along the direction $[110]$ observed from both the x-ray and charge-transport data should be considered as the reason for the strong increase of MR in the direction $\mathbf{H}\parallel[001]$ which is transverse to the dynamic charge stripes.\index{dynamic charge stripes}\index{charge stripes} Below we are discussing the nature of both the lattice instability and inhomogeneity of electron density distribution in the nonmagnetic reference compound LuB$_{12}$.
\index{LuB$_{12}$!magnetoresistance anisotropy|)}

\subsection{The origin of electron and lattice instability and the energy scales in LuB$_\text{12}$}\index{LuB$_{12}$!lattice instability}\index{LuB$_{12}$!electron instability}

\begin{figure}[b!]
\centerline{\includegraphics[width=\textwidth]{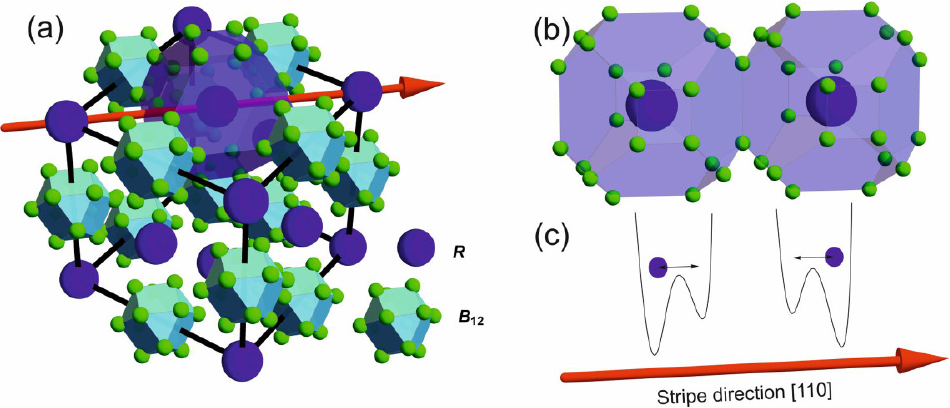}}
\caption{(a)~Crystal structure of $R$B$_{12}$. (b)~B$_{24}$ truncated cuboctahedra surrounding two adjacent $R^{3+}$ ions. (c)~Schematic view of $R$ ions vibrations in the double-well potentials. The red arrow shows the direction $[110]$ of dynamic charge stripes in $R$B$_{12}$. Reproduced from Ref.~\cite{Slu_KhoroshilovKrasnorussky19}.}
\label{Slu:Fig17}\index{LuB$_{12}$!crystal structure}\index{double-well potential}
\end{figure}

The mechanism responsible for the development of electron and lattice instability in $R$B$_{12}$ was discussed in Refs.~\cite{Slu_BolotinaDudka19, Slu_SluchankoBogach18} in terms of the dynamic cooperative JT effect on B$_{12}$ clusters.\index{Jahn-Teller effect!in B$_{12}$ clusters}\index{B$_{12}$ cluster!Jahn-Teller effect} More specifically:\nocite{Slu_KhoroshilovKrasnorussky19}
\begin{enumerate}
\item It has been shown in Ref.~\cite{Slu_SluchankoBogach18} that because of triple orbital degeneracy of the electronic ground state, the B$_{12}$ molecules are JT active and thus their structure is labile due to JT distortions (see Chapter~\ref{Chapter:Bolotina} for details). The quantum-chemistry calculations and geometry optimizations for a charged [B$_{12}$]$^{2-}$ cluster, whose doubly negative charge state is regarded as the most relevant in $R$B$_{12}$ compounds, allow the conclusion in favor of strong trigonal and tetragonal distortions and a mixture via the electron-vibronic nonadiabatic coupling of electronic states on each B$_{12}$ cluster~\cite{Slu_Bersuker06}.
\item In the dodecaboride matrix, the collective JT effect on the lattice of these B$_{12}$ complexes is at the origin of both the collective dynamics of boron clusters and large amplitude vibrations of the RE ions embedded in cavities of the boron cage as it is shown in Fig.~\ref{Slu:Fig17}.
\item Strong coupling of these Lu rattling modes\index{rattling phonon mode} that are located at 110~cm$^{-1}$~\cite{Slu_RybinaNemkovski10} and the JT vibrations is the reason both for the lattice instability with a dramatic increase in the vibrational DOS near $T_\text{E}\approx150$~K (see Fig.~\ref{Slu:Fig12}) and for the emergence of the collective excitation in the optical spectra (Fig.~\ref{Slu:Fig14}).
\item The Lu rattling modes at 110~cm$^{-1}$ \cite{Slu_SluchankoAzarevich11, Slu_RybinaNemkovski10, Slu_Alekseev15} are responsible also for the damping in the Drude term (solid line in Fig.~\ref{Slu:Fig13}, $\gamma_{\rm D} = 90 \pm 30$~cm$^{-1}$~\cite{Slu_GorshunovZhukova18}). These large-amplitude Einstein oscillators in the DWP,\index{double-well potential} schematically illustrated in Fig.~\ref{Slu:Fig17}\,(c), necessarily initiate strong changes in the $5d$-$2p$ hybridization between the RE and the electronic states of boron. Because the states in the conduction band are composed by the $2p$ orbitals of the B$_{12}$ cluster and the $5d$ orbitals of the Lu atoms \cite{Slu_GrechnevBaranovskiy08, Slu_HarimaYanase85, Slu_HarimaKobayashi85, Slu_YanaseHarima92, Slu_HeineckeWinzer95, Slu_BaranovskiyGrechnev09}, the variation in the $5d$-$2p$ hybridization will lead to the modulation of conduction bandwidth (see Chapter~\ref{Chapter:Bolotina}) and consequently generate numerous (up to $\sim$70\% of the total number of conduction band electrons \cite{Slu_GorshunovZhukova18}) nonequilibrium (hot) charge carriers manifested in the collective mode in the optical conductivity spectra at room temperature (Fig.~\ref{Slu:Fig14}).
\item In the cage-glass\index{cage glass} state of $R$B$_{12}$ at $T<T^{\ast} \approx 60$~K \cite{Slu_SluchankoAzarevich11}, two additional factors appear: (i)~the positional disorder in the arrangement of Lu$^{3+}$ ions in the B$_{24}$ truncated cuboctahedra (static displacements of $R^{3+}$ ions in the DWP), which is accompanied by the formation of vibrationally coupled nanometer-size clusters in the $R$B$_{12}$ matrix, and (ii)~the emergence of dynamic charge stripes\index{dynamic charge stripes|(} (ac-current with a frequency $\sim$200~GHz \cite{Slu_SluchankoAzarevich19}) directed along the single $[110]$ axis in LuB$_{12}$~\cite{Slu_BolotinaDudka18} that accumulate a considerable part of nonequilibrium conduction electrons in the filamentary structure of fluctuating charges.
\end{enumerate}

Based on the precise investigations of the crystal structure, heat capacity, and charge transport, Bolotina \textit{et al.}~\cite{Slu_BolotinaDudka19} argue that in the family of Lu$^N$B$_{12}$ crystals ($N$~=~10, 11, nat), Lu$^\text{nat}$B$_{12}$ has the strongest local atomic disorder in combination with long-range JT trigonal distortions, providing optimal conditions for the formation of the dynamic charge stripes below $T^{\ast}\approx 60$~K. As a result, the resistivity and Seebeck coefficient of the Lu$^\text{nat}$B$_{12}$ heterogeneous media decrease strongly in comparison with the characteristics detected for pure boron-isotope enriched crystals (see Figs.~\ref{Slu:Fig5} and \ref{Slu:Fig8}). Thus, the authors concluded that defects are supposedly used as the centers of pinning facilitating the formation of additional ac conductive channels\,---\,the dynamic charge stripes in the metallic matrix of $R$B$_{12}$.

To summarize, three energy scales that determine the properties of RE dodecaborides at intermediate and low temperatures can be highlighted. The JT splitting of the triply degenerate electronic ground state of the B$_{12}$ clusters is estimated to be $\sim$100--200~meV \cite{Slu_SluchankoBogach18} (see also Chapter~\ref{Chapter:Bolotina}), so the cooperative JT dynamics\index{Jahn-Teller effect!cooperative} is expected to be observed in the range of 100--1500~cm$^{-1}$. The energy of the rattling mode\index{rattling phonon mode} \mbox{$\Theta_{\text E}\approx 110$~cm$^{-1} \approx 15$}~meV is very close to the characteristic temperature $T_\text{E}\approx 150$~K of the fcc lattice instability, which develops in $R$B$_{12}$ and leads to a strong increase in the vibrational DOS in approach to the Ioffe-Regel regime\index{Ioffe-Regel limit} of the lattice dynamics. The order-disorder cage-glass\index{cage glass} transition at $T^{\ast}\approx 60$~K is regulated by the barrier height in the DWP\index{double-well potential} for RE ions, and below this temperature, the static displacements of $R^{3+}$ ions in the fcc lattice are accompanied by an emergence of dynamic charge stripes\index{dynamic charge stripes|)} in the dodecaboride matrix. Taking into account the loosely bound state of the $R$ ions in the rigid boron covalent sublattice, it is natural to expect that all these energy scales along with nonequilibrium and many-body effects produced by the JT effect in the B$_{12}$ clusters will be realized also in magnetic and semiconducting RE dodecaborides, as will be discussed in the following sections.

\vspace{-3pt}\section{Magnetic dodecaborides \textit{R}B$_\text{12}$ (\textit{R}\,=\,Tb, Dy, Ho, Er, Tm) and the solid solutions \textit{R}$_x$Lu$_{\text{1}-x}$B$_\text{12}$}\vspace{-2pt}
\label{Sec:Slu4}

\subsection{Magnetic properties}\index{rare-earth dodecaborides!magnetic properties|(}

The temperature dependence of the magnetic susceptibilities $\chi(T)$ of $R$B$_{12}$ ($R$~=~Tb\,--\,Yb) was investigated over the temperature range 90--1200~K by Moiseenko and Odintsov~\cite{Slu_MoiseenkoOdintsov79}. The linear Curie-Weiss type behavior of $\chi^{-1}(T)$ was found for all compounds, and the authors concluded that the paramagnetic properties of these dodecaborides are caused mainly by the $4f$ electrons with AFM exchange interaction between the localized magnetic moments of the RE ions. The calculated effective magnetic moments $\mu_\text{eff} = 4.3$--10.6~$\mu_\text{B}$ were close to the values of the corresponding 3+ charged free RE ions, and the RKKY indirect exchange\index{RKKY interaction} is suggested to be responsible for the interaction between them~\cite{Slu_MoiseenkoOdintsov79}. The low-temperature magnetization measurements of $R$B$_{12}$ ($R$~=~Tb, Dy, Ho, Er, Tm) \cite{Slu_GabaniBatko98, Slu_Shitsevalova01} indicated the AFM phase transitions with Ne\'{e}l temperatures of 22, 16.4, 7.4, 6.7, and 3.3~K, respectively. It was mentioned by Kohout \textit{et al.}~\cite{Slu_KohoutBatko04} that in the AFM state toward low temperature, the magnetization does not approach zero; almost independent of the crystalline orientation, the magnetization extrapolates to approximately 70\% of its maximum at $T_\text{N}$. Thus, the magnetization data for single crystals resemble the response from a powder sample with equal mixing of the transverse and longitudinal components indicating a significant disorder in the orientation of the $4f$ magnetic moments. The high-field magnetization measurements have been carried out for HoB$_{12}$ and TmB$_{12}$ by Siemensmeyer \textit{et al.}~\cite{Slu_SiemensmeyerHabicht07}. The saturation is not fully reached even in 14~T field due to very strong interactions. The comparison of the experimental results with calculation allows concluding in favor of a triplet $\Gamma_5^1$ ground state of Ho and Tm ions which is isotropic in low magnetic field. At high fields, the induced magnetization and thus the single-ion energy in a magnetic field depend strongly on the field orientation and the $\langle111\rangle$ direction is preferred over $\langle110\rangle$ and especially the $\langle001\rangle$ direction~\cite{Slu_SiemensmeyerHabicht07}.

The substitution of Ho atoms by nonmagnetic Lu in Ho$_x$Lu$_{1-x}$B$_{12}$ solid solutions has been studied by Gab\'{a}ni \textit{et~al.}~\cite{Slu_GabaniBatko13}. They showed that the $T_\text{N}(x)$ dependence demonstrates a linear increase from a quantum critical point (QCP)\index{quantum criticality} with $T_\text{N} = 0$ detected at $x_{\rm c}\approx 0.1$. At temperatures above $3T_{\text N}$, the measured $\chi(T)$ dependences obey the Curie-Weiss law and the effective magnetic moments per formula unit $\mu_\text{eff}$ were well fitted by the relation $\mu_\text{eff}$(Ho$_x$Lu$_{1-x}$B$_{12}$)~=~$\mu_{\text {eff}}$(HoB$_{12}$)$\sqrt{1-x}$, indicating the suppression of indirect exchange interaction between Ho$^{3+}$ moments with an increase in the concentration of nonmagnetic Lu ions.
\index{rare-earth dodecaborides!magnetic properties|)}

\begin{figure}[b!]
\centerline{\includegraphics[width=\textwidth]{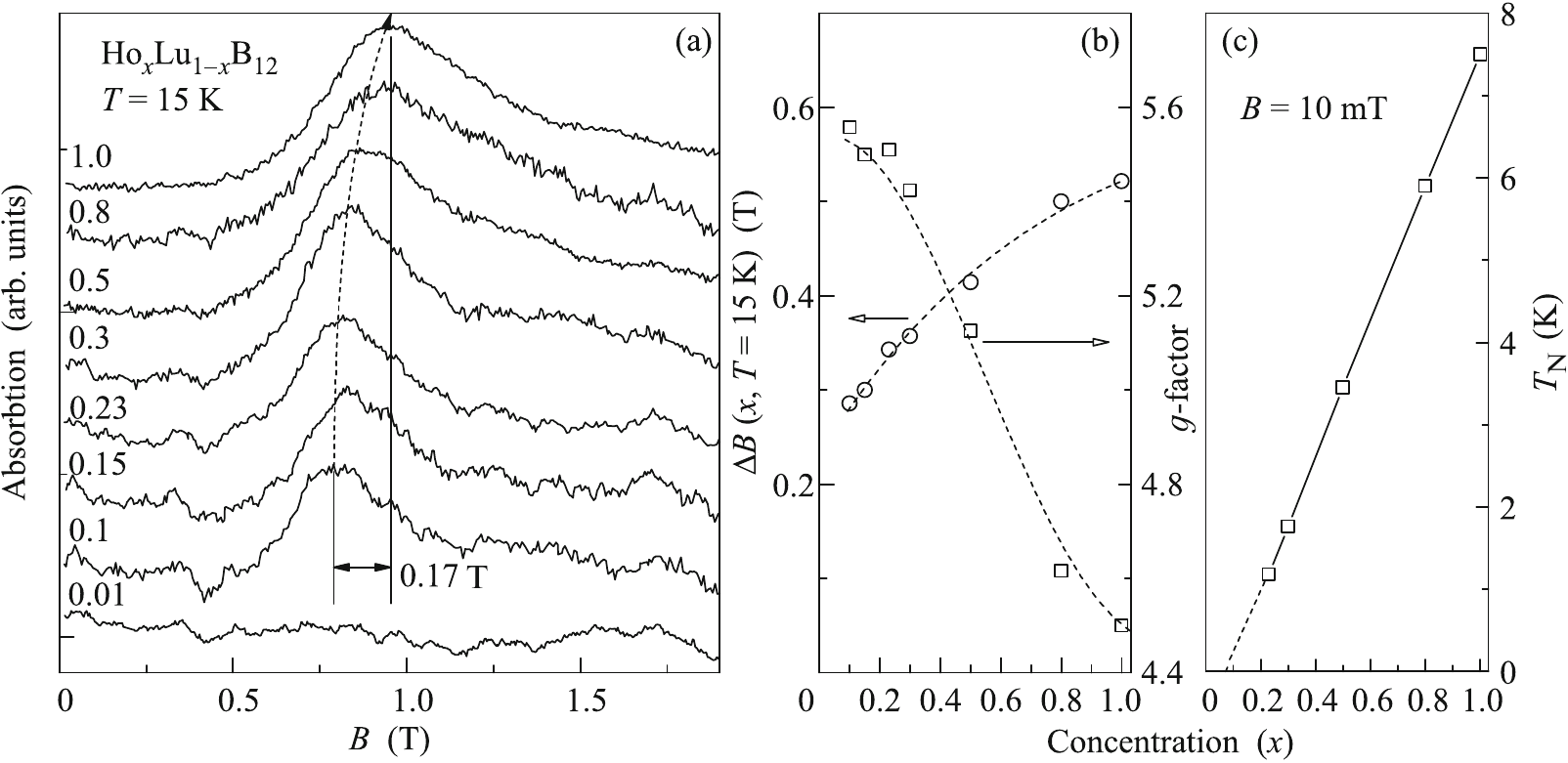}}
\caption{(a)~Evolution of the EPR spectra with the concentration $x$ in Ho$_x$Lu$_{1-x}$B$_{12}$ at $T=15$~K~\cite{Slu_GilmanovDemishev19}. (b)~Concentration dependence of the EPR linewidth (left vertical axis) and $g$-factor (right vertical axis). (c)~Concentration dependence of the Ne\'{e}l temperature~\cite{Slu_SluchankoKhoroshilov18}.}
\label{Slu:Fig18}\index{Ho$_x$Lu$_{1-x}$B$_{12}$!electron paramagnetic resonance}
\end{figure}

\subsection{Electron paramagnetic resonance}\index{electron paramagnetic resonance}

Recently, measurements of the high-frequency (60~GHz) electron paramagnetic resonance (EPR) were carried out in Ho$_x$Lu$_{1-x}$B$_{12}$ solid solutions at intermediate and low temperatures in a wide range of the holmium content \mbox{$0.01 \leq x \leq 1$}~\cite{Slu_GilmanovDemishev19}. For the samples with \mbox{$x \geq 0.1$}, it was demonstrated (see Fig.~\ref{Slu:Fig18}) that the electron paramagnetic resonance in the form of a single broad line with a $g$-factor of $\sim$5 appears in the cage-glass\index{cage glass} phase at $T < T^{\ast} \approx 60$~K because of the decrease in the relaxation rate for magnetic moments of Ho$^{3+}$ ions. For the compounds with $x > 0.3$, the pronounced broadening of the resonance line and a steep decrease in the $g$-factor related to AFM correlations (short-range order effects) are observed on cooling below 30~K. The performed simulation of the EPR spectra has indicated \cite{Slu_GilmanovDemishev19} that the exchange interaction and positional disorder in a location of magnetic ions play a decisive role in the specific features of spin dynamics in Ho$_x$Lu$_{1-x}$B$_{12}$.

\subsection{Charge transport}\index{rare-earth dodecaborides!charge transport|(}

\paragraph{Resistivity}\index{rare-earth dodecaborides!charge transport!resistivity}
In the magnetic dodecaborides HoB$_{12}$, ErB$_{12}$, and TmB$_{12}$, the temperature dependence of the resistivity $\rho(T)$ is metallic-like with a power-law behavior $\rho(T) \propto T^{\alpha}$ at intermediate temperatures with the exponent varying in the range from $\alpha \approx 1.2$ for HoB$_{12}$ to $\alpha \approx 0.8$ for TmB$_{12}$. In the range from 10 to 20~K, the $\rho(T)$ curves exhibit a minimum, and the resistivity demonstrates a moderate increase with the temperature lowering to $T_\text{N}$. The transition to the AFM state at 7.4, 6.7, and 3.2~K for HoB$_{12}$, ErB$_{12}$, and TmB$_{12}$, respectively, is accompanied by a steplike upturn in $\rho(T)$, and in the case of ErB$_{12}$, a second magnetic phase transition appears at 5.8~K~\cite{Slu_SluchankoBogach09a}. At liquid helium (LHe) temperatures, the crossover to the coherent regime of charge transport in the magnetically ordered phase of $R$B$_{12}$ is accompanied by a monotonic decrease in resistivity.

In an external magnetic field, three characteristic regimes of MR behavior have been revealed in the AFM metals: (i)~the positive MR similar to the case of LuB$_{12}$ is observed in the paramagnetic phase at $T>25$~K; (ii)~in the range $T_\text{N} \leq T \leq 15$~K, the MR becomes negative and depends quadratically on the external magnetic field; and, finally, (iii)~upon the transition to the AFM phase ($T < T_\text{N}$), the positive MR is again observed and its amplitude reaches 150\% for HoB$_{12}$~\cite{Slu_SluchankoBogach09a}. The low-temperature resistivity increase in the range $T_\text{N} \leq T \leq 15$~K in combination with a strong negative MR has been explained in terms of charge carrier scattering on nanosize magnetic domains (short-range AFM order of the RE-ion moments) and spin-polarization effects~\cite{Slu_SluchankoBogach09a, Slu_SluchankoKhoroshilov15}, which are suppressed dramatically in external magnetic field. It was found \cite{Slu_SluchankoBogach09a, Slu_SluchankoKhoroshilov15, Slu_KhoroshilovAzarevich16} that the negative MR component in magnetic $R$B$_{12}$ is proportional to the square of the local magnetization $-\Delta\rho/\rho \propto M_\text{loc}^2$, and in the vicinity of the Ne\'{e}l temperature it demonstrates the critical behavior $\Delta\rho/\rho \propto (1-T/T_{\text N})^{2\beta}$ with the exponent $\beta = 0.36$--0.44 which is about equal to that one predicted for magnetization and observed in model antiferromagnets~\cite{Slu_Ma76}.

\begin{figure}[t!]
\centerline{\includegraphics[width=0.85\textwidth]{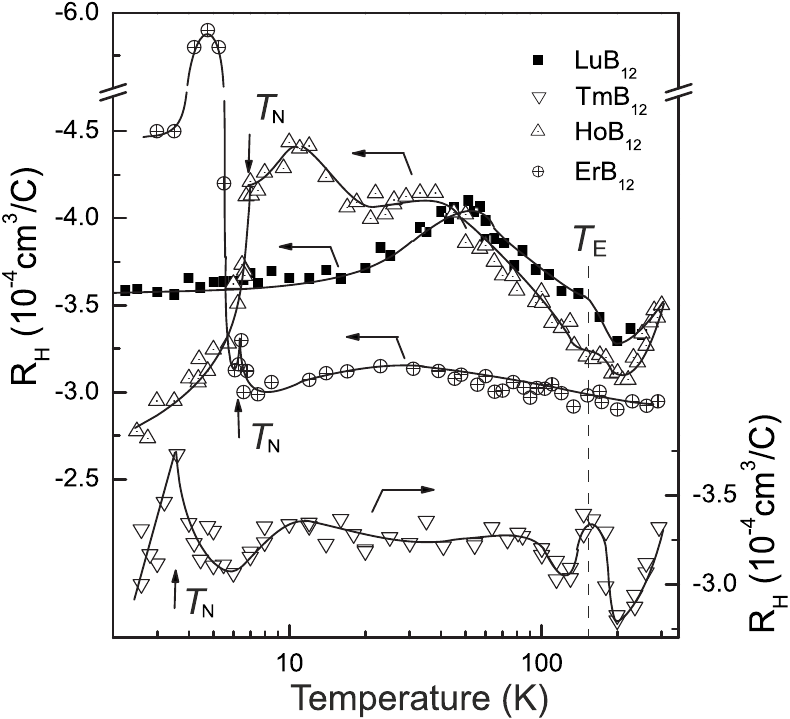}}
\caption{Temperature dependences of Hall coefficient $R_\text{H}(T)$ for $R$B$_{12}$ compounds ($R$~=~Ho, Er, Tm, Lu). Reproduced from Ref.~\cite{Slu_SluchankoBogomolov06}.}
\label{Slu:Fig19}\index{rare-earth dodecaborides!charge transport!Hall coefficient}
\end{figure}

\paragraph{Hall effect}\index{rare-earth dodecaborides!charge transport!Hall effect}
The temperature-independent behavior of the Hall coefficient, which was observed in the nonmagnetic metal LuB$_{12}$ at low temperatures $T<20$~K and small magnetic fields (Fig.~\ref{Slu:Fig19}), is not typical for the magnetic dodecaborides HoB$_{12}$, ErB$_{12}$ and TmB$_{12}$ \cite{Slu_SluchankoBogomolov06, Slu_SluchankoSluchanko07}. Indeed, distinct maxima of $R_{\text H}(T)$ are observed at $T_\text{max}\approx 10$--12~K in HoB$_{12}$ and TmB$_{12}$ (Fig.~\ref{Slu:Fig19}). Additionally, the pronounced anomalies of the Hall coefficient associated with AFM transitions can be clearly seen in the temperature dependences of $R_\text{H}(T)$ for all the magnetic compounds $R$B$_{12}$ in the vicinity of their N\'{e}el temperatures, $T_{\text N}$ (Fig.~\ref{Slu:Fig19}). The estimated values of $T_{\text N}$ agree very well with the results of magnetic and thermal measurements of $R$B$_{12}$ \cite{Slu_CzopnikShitsevalova04, Slu_KohoutBatko04, Slu_GabaniBatko99}. However, the observed discrepancy between the increase of $R_{\text H}$ in ErB$_{12}$ and the decrease of $R_{\text H}$ in HoB$_{12}$ and TmB$_{12}$ below $T_{\text N}$ (Fig.~\ref{Slu:Fig19}) was not explained in Ref.~\cite{Slu_SluchankoBogomolov06}. Additional anomalies were observed in the range of 130--300~K where several extrema were detected in the temperature dependence of the Hall coefficient \cite{Slu_SluchankoBogomolov06, Slu_SluchankoSluchanko07}. High-precision measurements of the Hall effect have been carried out for HoB$_{12}$ at LHe and intermediate temperatures for different field directions in magnetic fields up to 80~kOe~\cite{Slu_SluchankoSluchanko07}. It has been shown that, along with the normal component of $R_\text{H}(H,T)$, the strong (up to 30\%) anomalous magnetic contribution to the Hall effect, which is suppressed in high magnetic field, is observed in both the paramagnetic and N\'{e}el phases of HoB$_{12}$. The ferromagnetic component of the Hall signal was revealed above 20~kOe in the AF1 phase of HoB$_{12}$~\cite{Slu_SiemensmeyerHabicht07}. The singularities in the Hall coefficient $R_\text{H}(T,H)$ were compared with the $H$-$T$ magnetic phase diagram of HoB$_{12}$ and explained by the considerable spin-polaron effects that are typical for strongly correlated electron systems~\cite{Slu_SluchankoSluchanko07}.\index{strongly correlated electrons} A similar suppression of the temperature-dependent contribution to the Hall effect was found in the strong-magnetic-field studies of Ho$_x$Lu$_{1-x}$B$_{12}$ solid solutions \cite{Slu_GabaniBatko13} where it was shown that below $T^{\ast} \approx 60$~K, the $R_\text{H}(T)$ becomes temperature-independent at high magnetic fields $\sim$80~kOe. These changes in the charge transport in the cage-glass\index{cage glass} state of Ho$_x$Lu$_{1-x}$B$_{12}$ were explained in terms of spin-polaron effects and reconstruction of the filamentary structure of the conduction channels\index{conduction channels} (dynamic charge stripes) in strong magnetic field.\index{dynamic charge stripes}\index{charge stripes}

\begin{figure}[t!]
\centerline{\includegraphics[width=\textwidth]{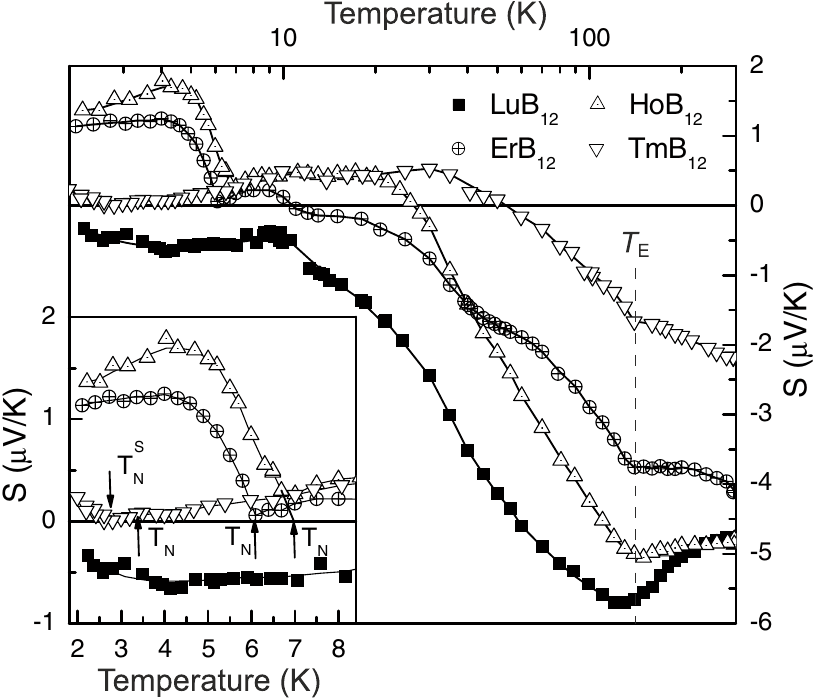}}
\caption{Temperature dependences of Seebeck coefficient S($T$) for $R$B$_{12}$ compounds ($R$ = Ho, Er, Tm, Lu). The vicinity of magnetic transitions is presented in the inset on larger scale. Reproduced from Ref.~\cite{Slu_SluchankoBogomolov06}.}
\label{Slu:Fig20}\index{rare-earth dodecaborides!Seebeck coefficient}
\end{figure}

\paragraph{Seebeck coefficient}\index{rare-earth dodecaborides!charge transport!Seebeck coefficient}\index{Seebeck coefficient}
The temperature dependences of the Seebeck coefficient $S(T)$ for magnetic compounds $R$B$_{12}$ ($R$~=~Ho, Er, Tm) demonstrate a singularity at $T_\text{E} \approx 130$--150~K (see Fig.~\ref{Slu:Fig20}), which is connected both with the Hall-effect minimum (Fig.~\ref{Slu:Fig19}) and the dramatic changes in the $\mu$SR scattering rate~\cite{Slu_KalviusNoakes03}. The amplitude of the negative minimum in $S(T)$ near $T_\text{E}$ is the largest for the nonmagnetic LuB$_{12}$ and decreases along the HoB$_{12}$\,--\,TmB$_{12}$ sequence, as can be seen from Fig.~\ref{Slu:Fig20}. When lowering the temperature, the thermopower\index{thermopower} of all magnetic dodecaborides changes sign. The authors \cite{Slu_SluchankoBogomolov06, Slu_GlushkovDemishev08} noted that the different signs of thermopower and Hall effect observed at low temperatures for magnetic dodecaborides can be ascribed to the peculiarities of strong quasiparticle interactions similar to the canonical heavy-fermion system CeB$_6$~\cite{Slu_SluchankoBogach07}. The AFM transition in $R$B$_{12}$ is accompanied by the appearance of a large positive contribution to the Seebeck coefficient, preceded by the positive spin-polaron component (Fig.~\ref{Slu:Fig20}) resulting from short-range order effects.

It was noted in Refs.~\cite{Slu_SluchankoBogomolov06, Slu_GlushkovDemishev08} that the change in the $4f$-shell population in the sequence HoB$_{12}$\,--\,LuB$_{12}$ starting from HoB$_{12}$ ($4f^n$, $n=10$) induces a significant decrease in the charge-carrier mobility\index{charge-carrier mobility} and in the amplitude of the Seebeck coefficient at \mbox{$T>T^{\ast}\approx60$}~K, as both parameters depend monotonically on the filling number $n$ in the range \mbox{$10 < n < 13$}~\cite{Slu_SluchankoGlushkov08}. This behavior of the transport characteristics evidently contradicts the drastic reduction of the de-Gennes factor $(g - 1)^2 J(J + 1)$, where $J$ is the total angular momentum of the $4f$ shell, which is generally used to characterize the magnetic scattering of charge carriers. This unusual tendency seems to be connected with the enhancement of spin fluctuations in the magnetic $R$B$_{12}$ compounds that will be discussed in detail in the next section.
\index{rare-earth dodecaborides!charge transport|)}

\subsection{Thermal conductivity}\index{rare-earth dodecaborides!thermal conductivity}\index{thermal conductivity}

\begin{figure}[b!]
\centerline{\includegraphics[width=0.85\textwidth]{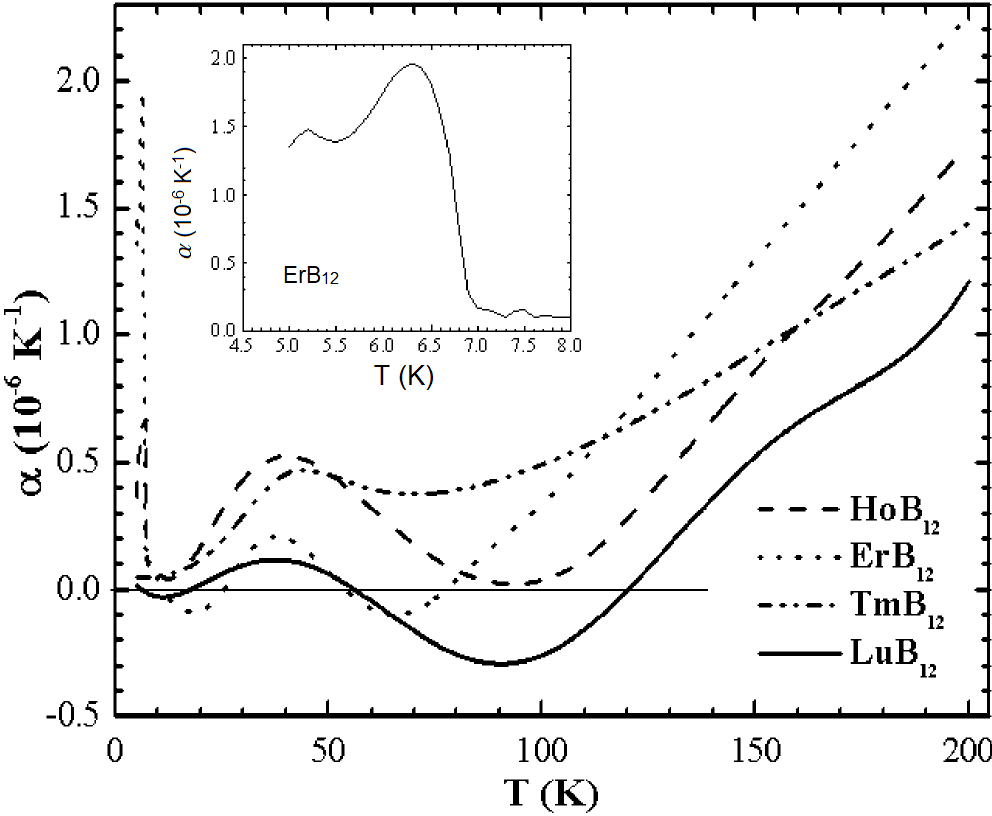}}
\caption{Temperature dependences of the thermal expansion coefficient for $R$B$_{12}$. Inset: low-temperature part of $\alpha(T)$ for ErB$_{12}$~\cite{Slu_CzopnikShitsevalova04, Slu_Shitsevalova01}.}
\label{Slu:Fig21}\index{rare-earth dodecaborides!thermal expansion}
\end{figure}

The thermal conductivity values for magnetic dodecaborides $R$B$_{12}$ ($R$~=~Dy, Ho, Er, Tm) are quite close to $\lambda$(LuB$_{12}$) in the intermediate temperature range (150--300~K) \cite{Slu_PadernoShitsevalova95, Slu_MisiorekMucha95}, in accordance with the development of the fcc lattice instability that leads to a strong increase of the vibrational DOS in approach to the Ioffe-Regel regime\index{Ioffe-Regel limit} of the lattice dynamics near $T_\text{E} \approx 150$~K \cite{Slu_SluchankoAzarevich11}. On the contrary, a very strong ($\sim$10 times) suppression of $\lambda(T)$ is observed in these magnetic compounds at low temperatures~\cite{Slu_PadernoShitsevalova95}. Besides, the authors note that the main feature of all the magnetic dodecaborides is a sharp decrease in the thermal conductivity within the AFM state in the vicinity of the antiferromagnet-paramagnet (AF-P) phase boundary as the magnon heat conductivity near the magnetic phase transition falls to a negligibly small value \cite{Slu_PadernoShitsevalova95, Slu_MisiorekMucha95}. They concluded also that the electronic thermal conductivity $\lambda_\text{e}$ dominates over the phonon component $\lambda_\text{ph}$ with the ratio $\lambda_\text{e}/\lambda_\text{ph} \approx 2$--3 within the temperature range of 4--300~K~\cite{Slu_MisiorekMucha95}.

\subsection{Thermal expansion and heat capacity}
\index{rare-earth dodecaborides!thermal expansion}\index{rare-earth dodecaborides!specific heat}\index{thermal expansion}\index{specific heat}\index{heat capacity}

Temperature dependences of the linear thermal expansion coefficient $\alpha$ of HoB$_{12}$, ErB$_{12}$, and TmB$_{12}$ are shown in Fig.~\ref{Slu:Fig21}~\cite{Slu_Shitsevalova01, Slu_CzopnikShitsevalova04}. The behavior of $\alpha(T)$ with two minima at low and intermediate temperatures is quite similar to that one observed for LuB$_{12}$. It is worth noting also that one of the $\alpha(T)$ inflection points in $R$B$_{12}$ is detected near the cage-glass\index{cage glass} transition temperature $T^{\ast} \sim 60$~K. In the ordered state of HoB$_{12}$ and ErB$_{12}$, an additional anomaly in the thermal expansion occurs at the N\'{e}el temperature (see Fig.~\ref{Slu:Fig21}, inset).

\begin{figure}[b!]
\centerline{\includegraphics[width=0.97\textwidth]{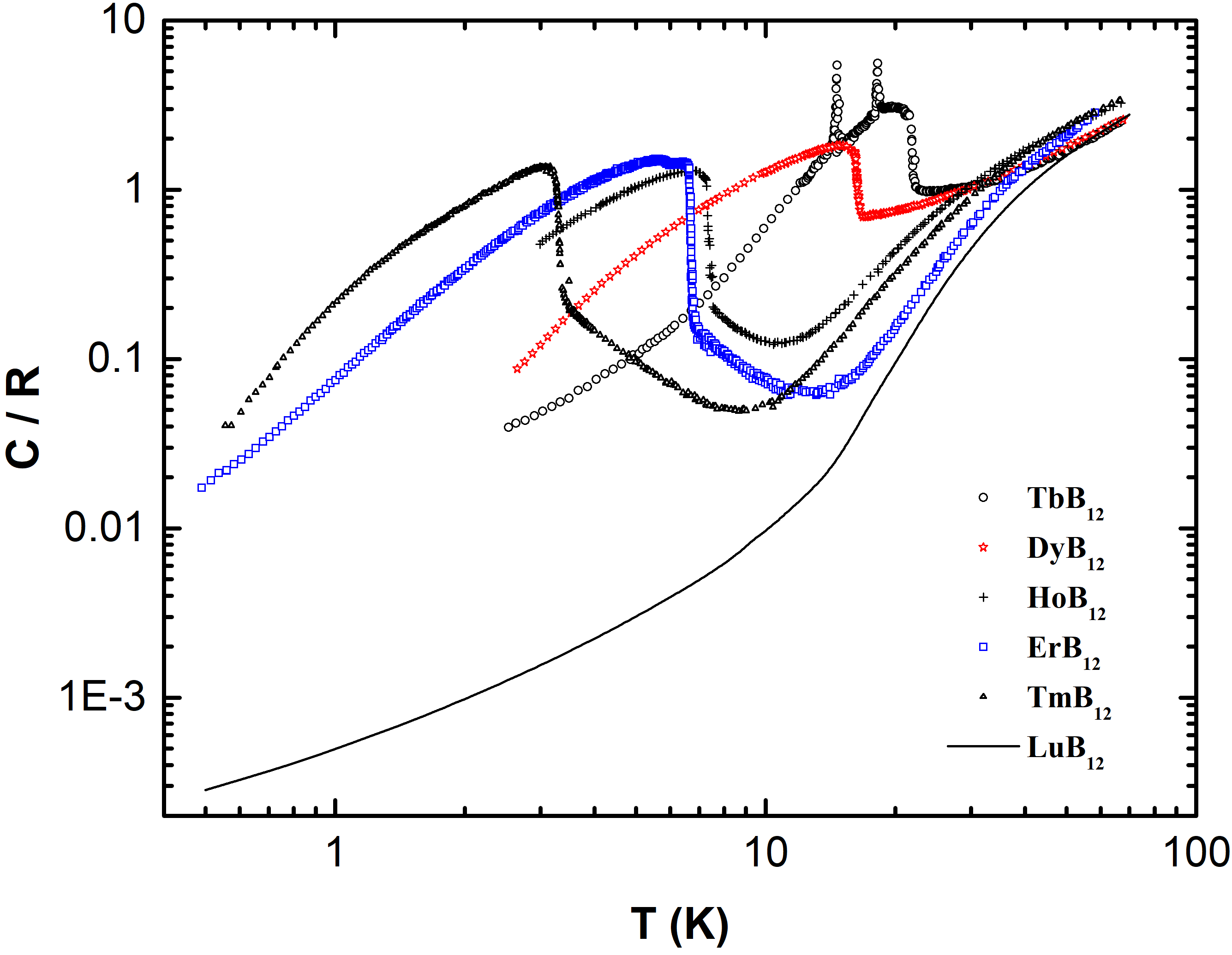}}
\caption{Temperature dependence of the heat capacity of the magnetic RE dodecaborides~\cite{Slu_CzopnikShitsevalova04, Slu_Shitsevalova01}.}
\label{Slu:Fig22}\index{rare-earth dodecaborides!specific heat}
\end{figure}

Temperature dependences of the heat capacity of magnetic dodecaborides have been studied in Refs.~\cite{Slu_Shitsevalova01, Slu_CzopnikShitsevalova04} (see Fig.~\ref{Slu:Fig22}). The most important features on the $C(T)$ curves are a discontinuity $\Delta C$ at the AFM transition and the heat-capacity maximum below $T_\text{N}$. In the ordered state, TbB$_{12}$ also reveals two first-order metamagnetic transitions~\cite{Slu_MurasikCzopnik02}, whereas ErB$_{12}$ shows a small anomaly of unclear origin below the N\'{e}el temperature~\cite{Slu_CzopnikShitsevalova04, Slu_Shitsevalova01}. In the paramagnetic region of HoB$_{12}$\,--\,TmB$_{12}$, a significant contribution from short-range correlations is observed in the wide vicinity of $T_\text{N}$. A Schottky contribution\index{Schottky specific heat} to the specific heat above 40~K~\cite{Slu_CzopnikShitsevalova04, Slu_Shitsevalova01} has been analyzed in terms of the CEF\index{crystal electric field} splitting of the magnetic ground state.

The observed behavior of the heat capacity in the critical region of all AFM dodecaborides is intrinsic for amplitude-modulated (AM)\index{amplitude-modulated magnetic structure|(} magnetic systems. Indeed, in the framework of the periodic field model (PFM)~\cite{Slu_BlancoGignoux90}, $\Delta C_\text{AM}$ of a system ordered with an amplitude-modulated magnetic structure is equal to 2/3 of the corresponding $\Delta C_\text{EM}$ expected for the structure with equal magnetic moments. In the absence of the crystal-field contribution, $\Delta C_\text{AM}$ is described by \cite{Slu_BlancoGignoux91}
\begin{equation}\label{Slu:Eq6}
\frac{\Delta C_\text{AM}}{R}=\frac{10}{3}\frac{J(J+1)}{2J^2+2J+1},
\end{equation}
where $J$ is the total angular momentum. The first excited CEF level in HoB$_{12}$, ErB$_{12}$ and TmB$_{12}$ is located too high in energy, so that practically only the ground state ($\Gamma_5^1$ triplet for HoB$_{12}$ and TmB$_{12}$ \cite{Slu_CzopnikShitsevalova04, Slu_Shitsevalova01, Slu_CzopnikMurasik00, Slu_AlekseevNemkovski14}, and $\Gamma_8^3$ quartet for ErB$_{12}$ \cite{Slu_CzopnikShitsevalova04, Slu_Shitsevalova01, Slu_AlekseevMignot04}) is thermally populated at temperatures near $T_\text{N}$. In this case, HoB$_{12}$ and TmB$_{12}$ may be characterized by an effective spin $S_\text{eff} = 1$ and $\Delta C_\text{AM} = 11.09$~J/(mol$\cdot$K), ErB$_{12}$ by $S_\text{eff} = 3/2$ and $\Delta C_\text{AM} = 12.23$~J/(mol$\cdot$K), while the experimental $\Delta C_\text{AM}$ values are equal to 10.75, 11.36, and 12.14~J/(mol$\cdot$K), respectively, in good agreement with the $C(T)$ measurements and the PFM estimation.

The low-temperature specific heat of Ho$_x$Lu$_{1-x}$B$_{12}$ solid solutions with $0 < x \leq 1$ has been studied in the cage-glass\index{cage glass} state~\cite{Slu_SluchankoKhoroshilov18}. For the Ho$_{0.01}$Lu$_{0.99}$B$_{12}$ composition in the regime of isolated magnetic impurities, the authors have found the specific heat anisotropy becoming as high as 15\% in an applied magnetic field up to 9~T. The increase in specific heat for the magnetic field orientation $\mathbf{H}\parallel[001]$ has been attributed to the existence of dynamic charge stripes (i.e., high-frequency, $\sim$200~GHz, \cite{Slu_SluchankoAzarevich19} oscillations of the inhomogeneous charge carrier density) along the direction $[110]$ in the fcc lattice of dodecaborides.\index{dynamic charge stripes}\index{charge stripes}

\subsection{Magnetic structure}
\index{magnetic structure}\index{rare-earth dodecaborides!magnetic structure|(}

The amplitude-modulated magnetic structures in TbB$_{12}$ \cite{Slu_MurasikCzopnik02}, HoB$_{12}$ \cite{Slu_KohoutBatko04, Slu_SiemensmeyerHabicht07, Slu_SiemensmeyerFlachbart06, Slu_FlachbartBauer07}, ErB$_{12}$ \cite{Slu_SiemensmeyerFlachbart06}, and TmB$_{12}$ \cite{Slu_CzopnikMurasik00, Slu_SiemensmeyerFlachbart06} have been confirmed by elastic neutron scattering.\index{neutron diffraction} The structure can be described as a modulation of magnetic moments that have parallel orientation within the $\langle 111 \rangle$ sheets and antiparallel orientation between the neighboring sheets, which propagates along the three crystallographic directions.

In the case of the terbium compound, the modulation persists well below the lower first-order transition. In this temperature range, the two magnetic phases, described by the same type of the propagation vector\index{magnetic propagation vector}\index{TbB$_{12}$!magnetic propagation vector}\index{TbB$_{12}$!magnetic structure} \mbox{$\mathbf{q}=(\frac{1}{2}\!\pm\!\delta,~\frac{1}{2}\!\pm\!\delta,~\frac{1}{2}\!\pm\!\delta)$} with different incommensurabilities \mbox{$\delta_1 = 0.022$} and \mbox{$\delta_2 = 0.059$}, coexist.\index{magnetic incommensurability} The phase transition at 14.6~K is due to the disappearance of an additional magnetic phase and the simultaneous rearrangement of the $\delta$-components of the $\mathbf{q}$ vector describing the sine-wave modulation of magnetic moments. Even the most careful examination of the neutron data revealed no difference in the temperature range where the second ($T_2 = 18.2$~K) first-order transition was observed in specific heat. The authors \cite{Slu_CzopnikShitsevalova04, Slu_Shitsevalova01} concluded that, most likely, TbB$_{12}$ preserves the sine-wave ordering down to 0~K. Therefore, the CEF\index{crystal electric field} ground state has to be a nonmagnetic singlet. In fact, the $\Gamma_2$ singlet ground state has been proposed also from an analysis of the Schottky heat capacity~\cite{Slu_CzopnikShitsevalova04, Slu_Shitsevalova01}.\index{Schottky specific heat}

On the contrary, in HoB$_{12}$, ErB$_{12}$, and TmB$_{12}$, the CEF ground state is the magnetic one~\cite{Slu_CzopnikShitsevalova04, Slu_Shitsevalova01, Slu_CzopnikMurasik00, Slu_AlekseevNemkovski14, Slu_AlekseevMignot04}. In this case, due to entropy effects associated with the modulation of the magnetic-moment amplitude, the magnetic system should either (i)~suddenly jump through a first-order transition from the amplitude-modulated to an equal-magnetic-moment structure, or (ii)~evolve to an antiphase structure at $T=0$ through a progressive squaring up of the modulation. This latter process should be accompanied by the growth of higher-order harmonics of the propagation vector.

Powder neutron diffraction\index{neutron diffraction}\index{HoB$_{12}$!magnetic propagation vector}\index{HoB$_{12}$!magnetic structure} on HoB$_{12}$ in zero magnetic field revealed an incommensurate amplitude-modulated magnetic structure below $T_\text{N}$ with the basic reflections that can be indexed by \mbox{$\mathbf{q}=(\frac{1}{2}\!\pm\!\delta,~\frac{1}{2}\!\pm\!\delta,~\frac{1}{2}\!\pm\!\delta)$} with \mbox{$\delta = 0.035$}~\cite{Slu_KohoutBatko04}. The data presented in Ref.~\cite{Slu_KohoutBatko04} show three phases in an applied field. The authors~\cite{Slu_KohoutBatko04} showed that the two AF1 and AF2 phases can be characterized by the same incommensurate ordering vector, and the difference between them is in the ferromagnetic moment that is present along the field direction in the AF2 phase, whereas in the zero-field AF1 phase a significant amount of magnetic moments remain disordered. The latter result supports an incommensurate amplitude-modulated structure and seems to be in qualitative agreement with the observation of Kalvius \textit{et~al.}~\cite{Slu_KalviusNoakes03} using $\mu$SR. Indeed, according to the $\mu$SR experiments, the mean local magnetic field and its distribution width are approximately of the same magnitude, indicating a significant spin disorder on the short-range scale ($\leq 5$ lattice constants) in ErB$_{12}$. Assuming that the spin disorder is predominantly directional, the $\mu$SR-data point toward a complex amplitude-modulated spin structure with a disordered component~\cite{Slu_KalviusNoakes03}.

The magnetic structure of Ho$^{11}$B$_{12}$, Er$^{11}$B$_{12}$ and Tm$^{11}$B$_{12}$ has been investigated by neutron diffraction on isotopically pure single crystals~\cite{Slu_SiemensmeyerFlachbart06}. Results in zero field as well as in magnetic fields up to 5~T reveal modulated incommensurate magnetic structures in these compounds with the basic reflections indexed with $\mathbf{q} = (\frac{1}{2}\!\pm\!\delta,~\frac{1}{2}\!\pm\!\delta,~\frac{1}{2}\!\pm\!\delta)$, where $\delta = 0.035$ both for HoB$_{12}$ and TmB$_{12}$, and with $\mathbf{q} = (\frac{3}{2}\!\pm\!\delta,~\frac{1}{2}\!\pm\!\delta,~\frac{1}{2}\!\pm\!\delta)$, where $\delta = 0.035$ for ErB$_{12}$. The authors suggested that the formation of these structures can be understood taking into account the interplay between the direct RKKY interaction,\index{RKKY interaction} which favors AFM ordering with $\mathbf{q}_\text{AFM} = (\frac{1}{2}\frac{1}{2}\frac{1}{2})$, and the dipole-dipole interaction that splits this basic propagation vector \cite{Slu_SiemensmeyerFlachbart06}. As long as the amplitude-modulated magnetic structure\index{amplitude-modulated magnetic structure|)} is not stable, a tendency towards squaring upon decreasing temperature is expected in these $R$B$_{12}$ compounds with a magnetic CEF\index{crystal electric field} ground state, so that the different moment values (magnitudes) of the magnetic structure get compensated. In this case, harmonics of basic reflections should appear, and the $3^\text{rd}$ harmonic satellites were observed in HoB$_{12}$ indicating that there is no ``simple'' amplitude modulation, and that likely a multi-$\mathbf{q}$ structure of the type $\pm(\frac{1}{2}\!-\!\delta,~\frac{1}{2}\!-\!\delta,~\frac{1}{2}\!-\!\delta)$ has to be considered~\cite{Slu_SiemensmeyerFlachbart06}. With increasing magnetic field up to 5~T, the basic propagation vector $(\frac{1}{2}\!-\!\delta,~\frac{1}{2}\!-\!\delta,~\frac{1}{2}\!-\!\delta)$ remains in the higher-magnetic-field phase AF2 with the ferromagnetic component~\cite{Slu_SiemensmeyerFlachbart06}. In ErB$_{12}$ only AFM reflections of the type $q = (\frac{3}{2}\!\pm\!\delta,~\frac{1}{2}\!\pm\!\delta,~\frac{1}{2}\!\pm\!\delta)$ have been observed and explained in terms of the possible changes in the crystal structure (symmetry breaking from fcc to \textit{tetragonal}) well below $T_\text{N}$~\cite{Slu_SiemensmeyerFlachbart06}.

A more detailed investigation of the magnetic structure was undertaken for HoB$_{12}$~\cite{Slu_SiemensmeyerHabicht07} below and above the N\'{e}el temperature. By studying the harmonics to the basic $\mathbf{q}_\text{AFM}$ vector along the $(111)$ direction, the authors concluded against the multi-domain single-$\mathbf{q}$ structure and argued in favor of the quadruple-$\mathbf{q}$ magnetic ordering at least in zero field. It was shown in Ref.~\cite{Slu_SiemensmeyerHabicht07} that in the field-induced AFM phase the quadruple-$\mathbf{q}$ structure changes to a double-$\mathbf{q}$ structure with a ferromagnetic component, which coexists with the quadruple-$\mathbf{q}$ order.

\begin{figure}[t]
\centerline{\includegraphics[width=\textwidth]{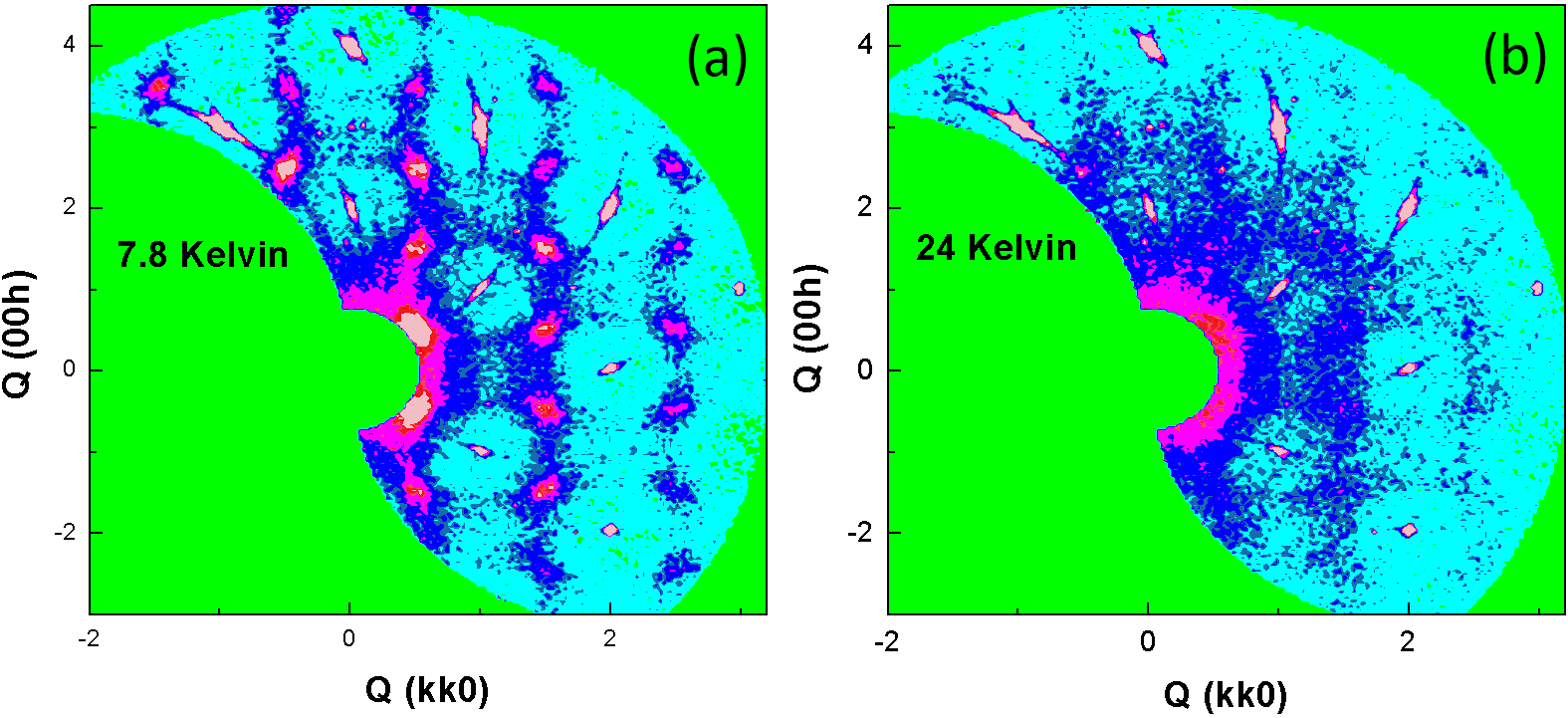}}
\caption{Diffuse neutron-scattering patterns measured on HoB$_{12}$ at 7.8 and 24~K above $T_\text{N} = 7.4$~K \cite{Slu_SiemensmeyerHabicht07, Slu_FlachbartBauer07, Slu_SiemensmeyerPrivate}.}
\label{Slu:Fig23}\index{HoB$_{12}$!diffuse neutron scattering}
\end{figure}

Figure~\ref{Slu:Fig23} shows the diffuse neutron-scattering patterns observed in HoB$_{12}$ above $T_\text{N}$~\cite{Slu_SiemensmeyerHabicht07, Slu_SiemensmeyerFlachbart06}. As follows from these patterns, a strong modulation of diffuse scattering appears at former magnetic reflections, e.g. at $(\frac{3}{2}\frac{3}{2}\frac{3}{2})$, but not at the typical crystal reflections. This fact points to strong correlations between the magnetic moments of Ho$^{3+}$ ions. The scattering patterns observed in Refs.~\cite{Slu_SiemensmeyerHabicht07, Slu_SiemensmeyerFlachbart06} can be explained by the appearance of correlated one-dimensional spin chains (short chains of Ho$^{3+}$-ion moments placed on space diagonals of the elementary unit), similar to those in low-dimensional magnets~\cite{Slu_TennantNagler97}. They can be resolved both well above $T_\text{N}$ (up to 70~K) and below the N\'{e}el temperature, where the 1D chains seem to condense into the ordered AFM structure~\cite{Slu_SiemensmeyerHabicht07, Slu_SiemensmeyerFlachbart06, Slu_FlachbartBauer07}.

\begin{figure}[t]
\centerline{\includegraphics[width=0.825\textwidth]{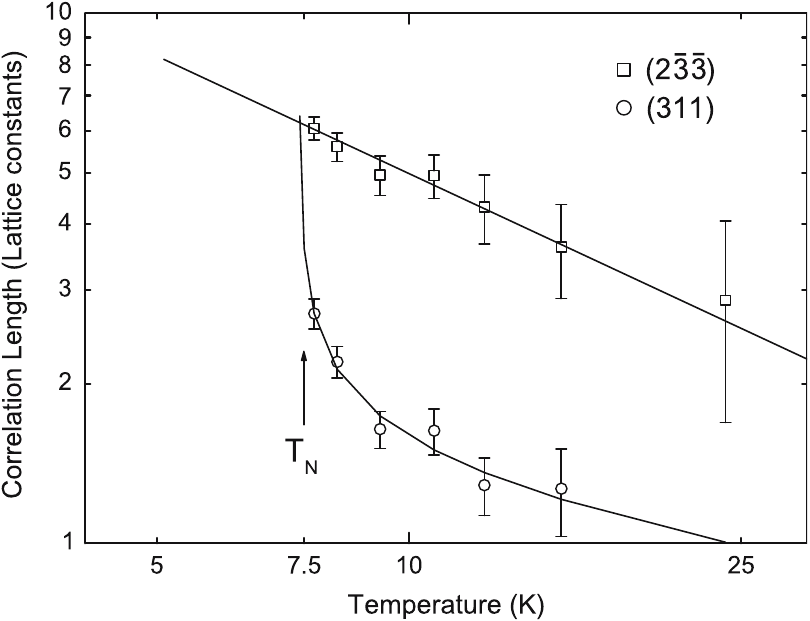}}
\caption{Correlation length versus temperature above $T_\text{N} = 7.4$~K along with two mutually perpendicular crystal directions: $[2\bar{3}\bar{3}]$ and $[311]$. The data are derived by numerical integration of the diffuse scattering along the directions indicated. The straight line indicates a $T^{-1}$ behavior. For the $[311]$ direction, a fit to $(T-T_\text{N})^{-1/4}$ was used to construct the curved line. Adapted from Siemensmeyer \textit{et al.}~\cite{Slu_SiemensmeyerHabicht07}.}
\label{Slu:Fig24}\index{HoB$_{12}$!magnetic correlation length}
\end{figure}

Thus, the direct interpretation of the results of diffuse scattering in HoB$_{12}$ is based on the fact that the AFM correlation length along the $\langle011\rangle$ and $\langle100\rangle$ directions is much shorter than along $\langle111\rangle$, and it demonstrates a quite different temperature dependence along $\langle111\rangle$ and transverse to this direction (see Fig.~\ref{Slu:Fig24} and Ref.~\cite{Slu_SiemensmeyerHabicht07} for more details). There is no diffuse signal at reflections of the crystal structure above $T_\text{N}$, thus there are no ferromagnetic correlations. The authors \cite{Slu_SiemensmeyerHabicht07} discussed the following scenario for the occurrence of long-range order in HoB$_{12}$: far away from $T_\text{N}$, strong interactions lead to correlations along $\langle111\rangle$, they are essentially one-dimensional and would not lead to long-range order at finite temperature. As $T_\text{N}$ is approached, the 1D-correlated regions grow in the perpendicular directions (Fig.~\ref{Slu:Fig24}), possibly due to other interactions. One may imagine \textit{cigar-shaped AFM-correlated regions} that become more spherical when $T_\text{N}$ is approached. Within this picture, the ordering temperature is determined by the point where spherical symmetry is reached. Only then 3D behavior sets in, and the system exhibits long-range AFM order.
\index{rare-earth dodecaborides!magnetic structure|)}

\vspace{-2pt}\subsection{Magnetic $H$-$T$-$\phi$ phase diagrams}\index{rare-earth dodecaborides!magnetic phase diagrams|(}\index{rare-earth dodecaborides!field-angular anisotropy|(}\index{rare-earth dodecaborides!$H$-$T$-$\phi$ phase diagrams|(}

\begin{figure}[t!]\vspace{-3pt}
\centerline{\includegraphics[width=0.74\textwidth]{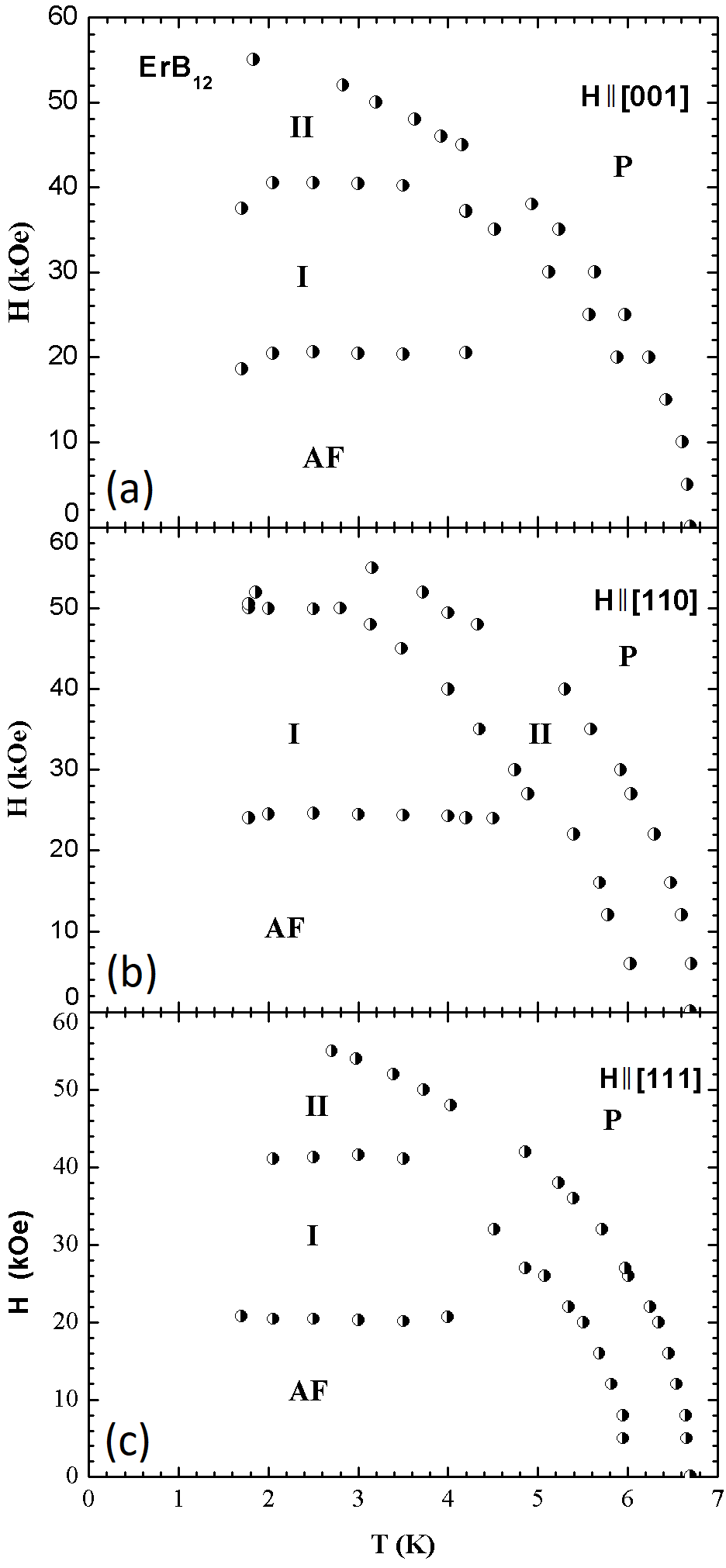}}\vspace{-2pt}
\caption{Magnetic $H$-$T$ phase diagrams of ErB$_{12}$ for the field applied along three principal directions in the fcc lattice. Reproduced from Refs.~\cite{Slu_Shitsevalova01, Slu_PluzhnikovShitsevalova08}.\vspace{-3em}}
\label{Slu:Fig25}\index{ErB$_{12}$!magnetic phase diagram}
\end{figure}

Results of the thermodynamic and charge-transport experiments were used to construct the $H$-$T$ phase diagrams of $R$B$_{12}$ ($R$~=~Ho, Er, and Tm) \cite{Slu_CzopnikShitsevalova04, Slu_PluzhnikovShitsevalova08, Slu_SluchankoBogach09a, Slu_Shitsevalova01, Slu_KohoutBatko04, Slu_SiemensmeyerHabicht07, Slu_SluchankoKhoroshilov15, Slu_BogachBogomolov08, Slu_BogachDemishev09, Slu_BogachDemishev09a} with three AFM phases in an external magnetic field applied along the three principal directions in the fcc lattice, \mbox{$\mathbf{H}\parallel\langle001\rangle$}, \mbox{$\mathbf{H}\parallel\langle110\rangle$}, and \mbox{$\mathbf{H}\parallel\langle111\rangle$} (see, for example, Fig.~\ref{Slu:Fig25} for ErB$_{12}$~\cite{Slu_Shitsevalova01, Slu_PluzhnikovShitsevalova08}). Scans at various temperatures in magnetic fields between 0 and 5~T were carried out in the neutron diffraction study of HoB$_{12}$~\cite{Slu_SiemensmeyerHabicht07} in order to test whether there are indications for phase boundaries between the AFM phases. It was concluded~\cite{Slu_SiemensmeyerHabicht07} that the principal reflections remain; i.e., no change in the AFM propagation vectors in a low magnetic field was observed. Also, it was deduced~\cite{Slu_SiemensmeyerHabicht07} that the changes in the neutron-scattering data of HoB$_{12}$, both as a function of temperature and field, correlate well with the location of the phase boundaries found in the specific-heat and magnetization studies, thus confirming the phase boundaries inferred from these experiments.

\begin{figure}[b!]\vspace{-2pt}
\centerline{\hspace{-3pt}\includegraphics[width=1.05\textwidth]{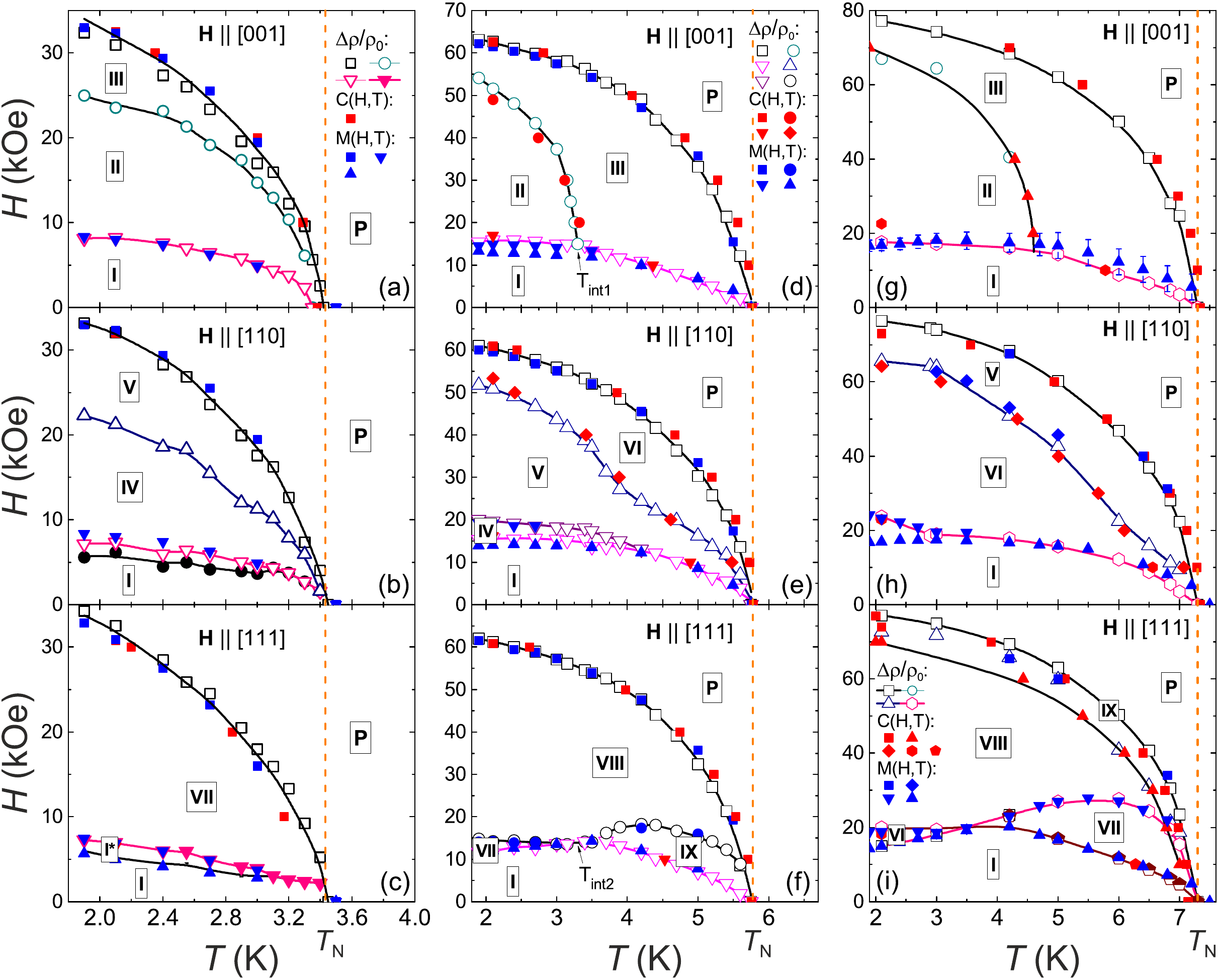}\vspace{-2pt}}
\caption{Magnetic $H$-$T$ phase diagrams of Ho$_{0.5}$Lu$_{0.5}$B$_{12}$ (left), Ho$_{0.8}$Lu$_{0.2}$B$_{12}$ (middle), and HoB$_{12}$ (right) for three principal field directions in the fcc lattice \cite{Slu_KhoroshilovKrasnorussky19, Slu_SluchankoKhoroshilov19, Slu_Khoroshilov19}.\vspace{-1pt}}
\label{Slu:Fig26}\index{HoB$_{12}$!magnetic phase diagram}\index{Ho$_x$Lu$_{1-x}$B$_{12}$!magnetic phase diagram}
\end{figure}

\begin{figure}[t!]
\centerline{\includegraphics[width=\textwidth]{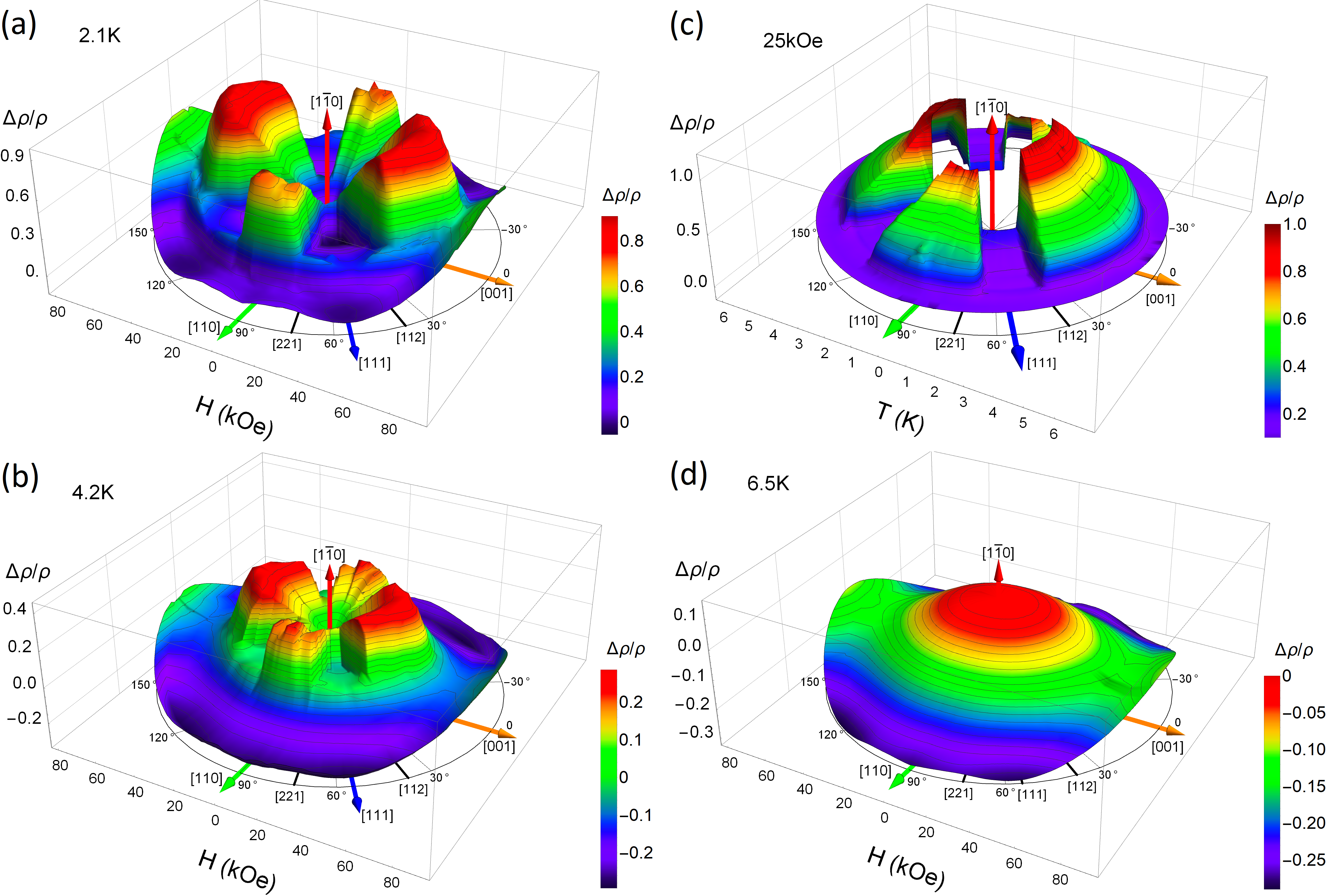}}
\caption{3D view of magnetoresistance dependences of Ho$_{0.8}$Lu$_{0.2}$B$_{12}$ on magnetic field (a)~$T=2.1$~K, (b)~$T=4.2$~K and (d)~$T=6.5$~K, and on temperature (c) at $H = 25$~kOe. The rotation axis is $\mathbf{I} \parallel [1\overline{1}0]$, the three principal cubic directions in the $[1\overline{1}0]$ plane are shown by arrows. Reproduced from Ref.~\cite{Slu_KhoroshilovKrasnorussky19}.}
\label{Slu:Fig27}\index{Ho$_x$Lu$_{1-x}$B$_{12}$!magnetoresistance}\index{anisotropic magnetoresistance!in Ho$_x$Lu$_{1-x}$B$_{12}$}\index{field-angular anisotropy!in magnetoresistance}
\end{figure}

\begin{figure}[t]
\includegraphics[width=\textwidth]{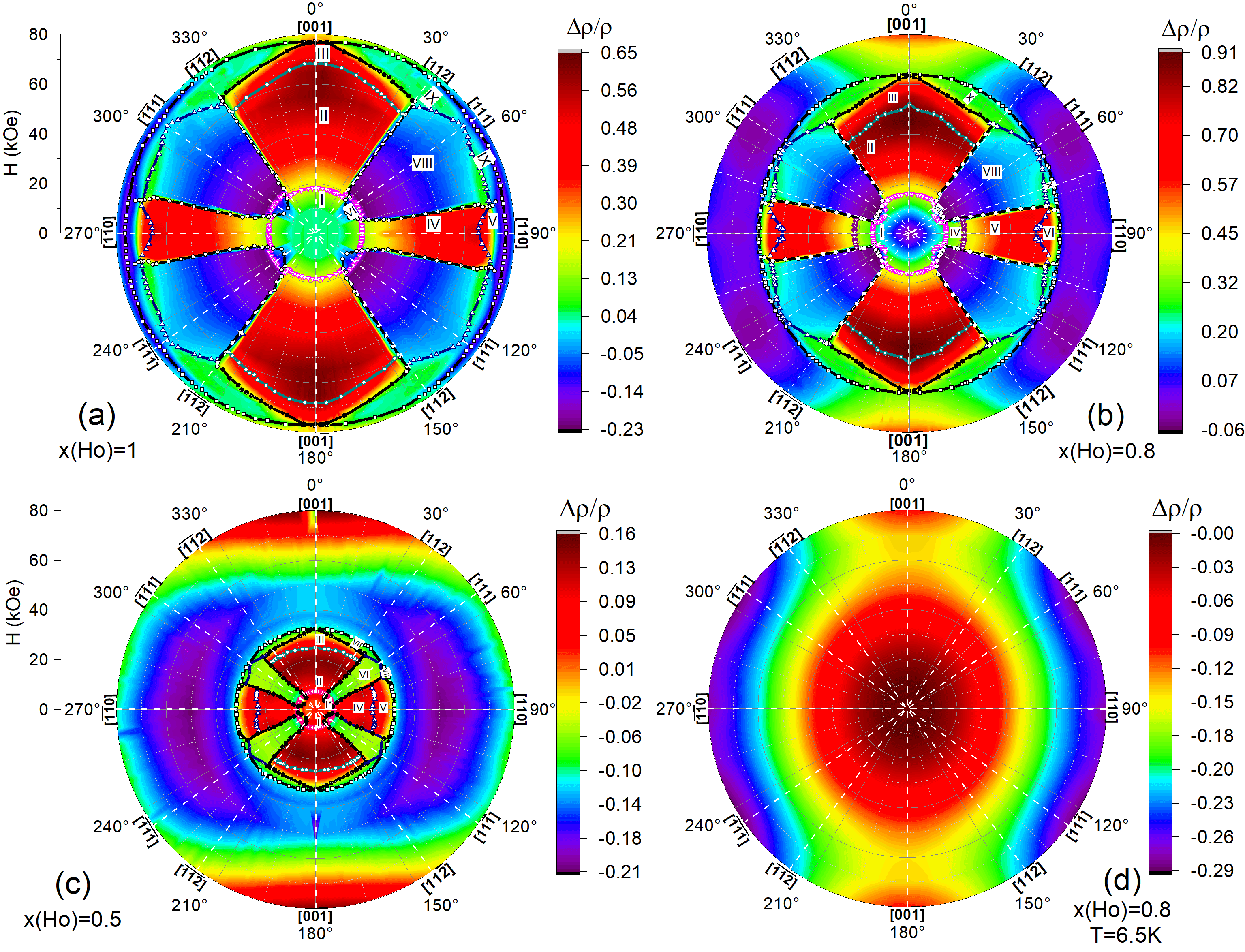}
\caption{Color plots of magnetoresistance vs. magnitude ($H$) and direction ($\phi$) of magnetic field as it is rotated in the $[1\overline{1}0]$ plane for (a)~HoB$_{12}$, (b)~Ho$_{0.8}$Lu$_{0.2}$B$_{12}$, and (c)~Ho$_{0.5}$Lu$_{0.5}$B$_{12}$ at a temperature of 2.1~K. For comparison, panel (d) demonstrates the anisotropy of magnetoresistance in the paramagnetic state of Ho$_{0.8}$Lu$_{0.2}$B$_{12}$ at $T=6.5$~K. The phase boundaries (solid lines) are derived from the magnetoresistance data (see symbols in Fig.~\ref{Slu:Fig26}). Roman numerals are the same as in Fig.~\ref{Slu:Fig26} and denote various AFM phases \cite{Slu_KhoroshilovKrasnorussky19, Slu_SluchankoKhoroshilov19, Slu_Khoroshilov19}.}
\label{Slu:Fig28}\index{Ho$_x$Lu$_{1-x}$B$_{12}$!magnetoresistance}\index{Ho$_x$Lu$_{1-x}$B$_{12}$!magnetic phase diagram}\index{field-angular anisotropy!in magnetoresistance}\index{anisotropic magnetoresistance!in Ho$_x$Lu$_{1-x}$B$_{12}$}
\end{figure}

Much more detailed and complex magnetic phase diagrams resulted from the higher-precision studies of specific heat, magnetization, and MR of the mono-domain single crystals of HoB$_{12}$~\cite{Slu_SluchankoKhoroshilov15}, Ho$_{0.8}$Lu$_{0.2}$B$_{12}$~\cite{Slu_KhoroshilovKrasnorussky19} and Ho$_{0.5}$Lu$_{0.5}$B$_{12}$~\cite{Slu_KhoroshilovAzarevich16, Slu_SluchankoKhoroshilov19, Slu_Khoroshilov19}. Among the number of magnetic phases in the AFM state of Ho$_x$Lu$_{1-x}$B$_{12}$, only the low-field one (phase~I in Fig.~\ref{Slu:Fig26}) is common for all three principal directions of magnetic field: \mbox{$\mathbf{H}\parallel\langle001\rangle$}, \mbox{$\mathbf{H}\parallel\langle110\rangle$}, and \mbox{$\mathbf{H}\parallel\langle111\rangle$}, but each of these directions has its own individual set of magnetically ordered states in higher magnetic fields~\cite{Slu_KhoroshilovKrasnorussky19, Slu_SluchankoKhoroshilov19, Slu_Khoroshilov19}. In these measurements, summarized for Ho$_{0.8}$Lu$_{0.2}$B$_{12}$ in Fig.~\ref{Slu:Fig27}, the authors detected sharp radial phase boundaries separating three main sectors $\Delta\phi_{001}$, $\Delta\phi_{110},$ and $\Delta\phi_{111}$ in the angular phase diagrams (see Fig.~\ref{Slu:Fig28}). It was shown~\cite{Slu_KhoroshilovKrasnorussky19, Slu_SluchankoKhoroshilov19, Slu_Khoroshilov19} that the boundaries form a Maltese cross\index{Maltese cross anisotropy} in the (110) plane (see Fig.~\ref{Slu:Fig28}), implying that the magnetic phases corresponding to these sectors differ significantly from each other in their magnetic structure. Such a complicated and low-symmetry magnetic $H$-$T$-$\phi$ diagrams are unusual and unexpected for a RE dodecaboride with the fcc crystal structure. When discussing the nature of the observed dramatic symmetry lowering, the authors \cite{Slu_KhoroshilovKrasnorussky19, Slu_SluchankoKhoroshilov19, Slu_Khoroshilov19} noted that significant MR changes are observed already in the paramagnetic phase, as one can see, for example, in Fig.~\ref{Slu:Fig27}\,(d) for Ho$_{0.8}$Lu$_{0.2}$B$_{12}$, $T_0 = 6.5$~K. They argue that numerous magnetic phases observed below $T_\text{N}$ and a lot of phase transitions in the $H$-$\phi$-$T$ phase diagram in the AFM state are the result of two factors: (i)~formation of the dynamic charge stripes\index{dynamic charge stripes}\index{charge stripes} along the $\langle110\rangle$ axis in the matrix of dodecaborides which leads to uniaxial scattering anisotropy in the paramagnetic phase and (ii)~the magnetic ordering due to indirect AFM exchange interaction mediated by the conduction electrons according to the RKKY-mechanism,\index{RKKY interaction} see Fig.~\ref{Slu:Fig29}. The conclusion is also confirmed by the results of more detailed studies~\cite{Slu_SluchankoKhoroshilov18a}, where it was shown that the MR in the paramagnetic phase of Ho$_{0.8}$Lu$_{0.2}$B$_{12}$ can be represented as a sum of (i)~isotropic negative component associated with charge-carrier scattering on magnetic clusters of Ho$^{3+}$ ions (nanometer-size domains with AFM short-range order) and (ii)~anisotropic positive contribution which is due to the cooperative dynamic JT effect\index{dynamic Jahn-Teller effect} in the boron sublattice. As one can see from Fig.~\ref{Slu:Fig28}, the Maltese-cross anisotropy was detected for charge-carrier scattering, and two more MR mechanisms dominant in the AFM state were separated and analyzed quantitatively~\cite{Slu_KhoroshilovKrasnorussky19}. It was suggested~\cite{Slu_SluchankoKhoroshilov15, Slu_SluchankoKhoroshilov18a, Slu_KhoroshilovKrasnorussky19} that the main positive linear in magnetic field contribution may be attributed to charge-carrier scattering on a spin density wave (SDW), i.e. the periodic modulation in the $5d$-component of the magnetization density that contributes to the magnetic structure in addition to the more localized $4f$ moments, whereas the negative quadratic term $\Delta\rho/\rho \propto H^2$ is due to the scattering on local $4f$-$5d$ spin fluctuations of the Ho$^{3+}$ ions. These two (SDW and localized moments of magnetic ions) components in the AFM ordering in combination with dynamic charge stripes\index{dynamic charge stripes}\index{charge stripes} are the main factors responsible for the complexity of the magnetic phase diagrams with numerous phases and phase transitions \cite{Slu_KhoroshilovKrasnorussky19, Slu_Khoroshilov19}.
\index{rare-earth dodecaborides!magnetic phase diagrams|)}\index{rare-earth dodecaborides!field-angular anisotropy|)}\index{rare-earth dodecaborides!$H$-$T$-$\phi$ phase diagrams|)}

\subsection{The root of the complexity of magnetic phase diagrams of \textit{R}B$_\text{12}$}\index{rare-earth dodecaborides!magnetic phase diagrams}

\begin{figure}[b!]
\centerline{\includegraphics[width=1.02\textwidth]{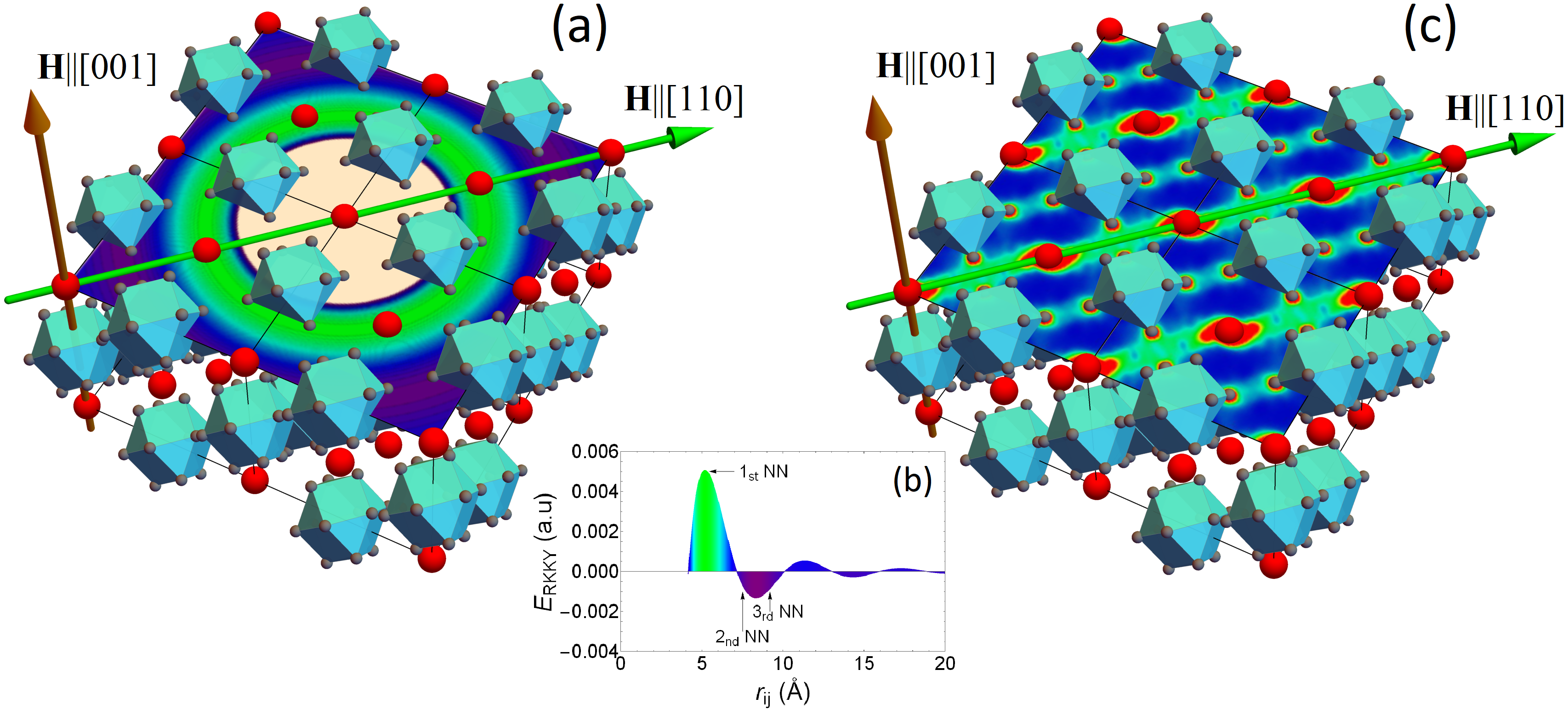}}
\caption{(a)~Crystal structure of $R$B$_{12}$. The color shows spin-density oscillations of conduction electrons that mediate the RKKY interactions, as shown in panel (b). (c)~Dynamic charge stripes along the $\langle110\rangle$ direction destroy the indirect RKKY exchange between the nearest-neighbour ($1^\text{st}$ NN) localized magnetic moments of the $R^{3+}$ ions. Courtesy of K.~Krasikov.}
\label{Slu:Fig29}\index{rare-earth dodecaborides!dynamic charge stripes}\index{rare-earth dodecaborides!RKKY interaction}\index{RKKY interaction}
\end{figure}

To summarize in more detail, in the cage-glass\index{cage glass} state of the RE magnetic dodecaborides at $T < T^\ast \approx 60$~K \cite{Slu_SluchankoAzarevich11}, two additional factors appear: (i)~the positional disorder in the arrangement of magnetic $R^{3+}$ ions in the B$_{24}$ truncated cuboctahedra (static displacements of $R^{3+}$ ions in the DWP,\index{double-well potential} see Fig.~\ref{Slu:Fig17}), which is accompanied by the formation of vibrationally coupled magnetic nanometer-size clusters in the $R$B$_{12}$ matrix and (ii)~the emergence of dynamic charge stripes\index{dynamic charge stripes}\index{charge stripes} (ac-current with a frequency $\sim$200~GHz \cite{Slu_SluchankoAzarevich19}) directed along the single $[110]$ axis in $R$B$_{12}$, which accumulate a considerable part of nonequilibrium conduction electrons in the filamentary structure of fluctuating charges~\cite{Slu_GorshunovZhukova18, Slu_BolotinaDudka18}. Thus, due to the cooperative dynamic JT effect,\index{dynamic Jahn-Teller effect}\index{Jahn-Teller effect!cooperative} when the B$_{12}$ polyhedra are consistently distorted (ferrodistortive effect, see Fig.~\ref{Fig:Bolotina_RB12-JTdistortions} in Chapter~\ref{Chapter:Bolotina} for details),\index{ferrodistortive effect} both static displacements of magnetic RE ions and $^{10}$B-$^{11}$B substitutional disorder provide centers of pinning that facilitate the formation of additional ac conductive channels\,---\,the dynamic charge stripes.\index{dynamic charge stripes}\index{charge stripes} The positional disorder in the arrangement of magnetic $R^{3+}$ ions in B$_{24}$ truncated cuboctahedra in the cage-glass\index{cage glass} state ($T < T^{\ast} \approx 60$~K) leads to a significant dispersion of exchange constants (through indirect exchange, RKKY mechanism)\index{RKKY interaction} and formation of both nanometer-size domains of magnetic ions in the $R$B$_{12}$ matrix (short-range AFM order effect above $T_\text{N}$) and creation of strong local $4f$-$5d$ spin fluctuations responsible for the polarization of $5d$ conduction band states (the spin-polaron effect). The last one produces spin polarized sub-nanometer-size ferromagnetic domains\index{ferromagnetic nanodomains} (``ferrons'',\index{ferrons} according to the terminology used in Refs.~\cite{Slu_Nagaev68, Slu_KaganKugel01}) resulting in the stabilization of SDW antinodes in the $R$B$_{12}$ matrix. The spin-polarized $5d$-component of the magnetic structure (ferrons) is on the one hand very sensitive to the external magnetic field, and, on the other hand, the applied field suppresses $4f$-$5d$ spin fluctuations by destroying the spin-flip scattering process. Moreover, along the direction of the dynamic charge stripes $[110]$, the huge charge-carrier scattering destroys the indirect exchange between the nearest-neighbor localized magnetic moments of $R^{3+}$ ions (see Fig.~\ref{Slu:Fig29}) and renormalizes the RKKY interaction,\index{RKKY interaction} accumulating a noticeable part of charge carriers in the filamentary electronic structure. Thus, the complex $H$-$T$-$\phi$ phase diagrams of $R$B$_{12}$ antiferromagnets may be explained in terms of the formation of a composite magnetically ordered state of localized $4f$ moments of $R^{3+}$-ions in combination with spin-polarized local areas of the $5d$ states\,--\,ferrons involved in the formation of SDW in the presence of a filamentary structure of dynamic charge~stripes.\index{dynamic charge stripes}\index{charge stripes}

\subsection{Quantum critical behavior in HoB$_\text{12}$}
\index{quantum criticality}\index{HoB$_{12}$!quantum criticality}

When long-range order (in the case of HoB$_{12}$, magnetic order below $T_\text{N}$) is suppressed to zero temperature by tuning an external variable, such as pressure or magnetic field, the system is said to cross a QCP~\cite{Slu_Stewart06, Slu_LoehneysenRosch07}. The MR measurements were carried out at 0.5~K and in the magnetic field up to 12~T, revealing the QCP for HoB$_{12}$ at $B_\text{c} \approx 8.2$~T~\cite{Slu_FlachbartBauer07}. The analysis of obtained electrical resistivity dependencies (using the $\rho(T) \propto T^x$ relation) at temperatures below 4.2~K shows that in magnetic fields close to $B_\text{c}$ the electrical resistivity exhibits an unusual $T^3$-dependence, where theoretical models expect a power law with $x = 3/2$~\cite{Slu_Stewart06, Slu_LoehneysenRosch07}. For $B \geq 10$~T a $\rho(T) \propto T^2$ dependence was observed~\cite{Slu_FlachbartBauer07}, which points to the fact that in higher fields, the Fermi-liquid behavior is recovered.

\vspace{-2pt}
\section{Metal-insulator transition in YbB$_\text{12}$ and solid solutions Yb$_x$\textit{R}$_{\text{1}-x}$B$_\text{12}$ (\textit{R}~=~Lu, Tm)}
\label{Sec:Slu5}
\index{metal-insulator transition|(}

\subsection{Metal-insulator transition in YbB$_\text{12}$}
\index{YbB$_{12}$!metal-insulator transition|(}

\paragraph{Yb valence}\index{YbB$_{12}$!fluctuating valence}\index{intermediate valence}\index{fluctuating valence}
YbB$_{12}$ is an archetypal strongly correlated electron system\index{strongly correlated electrons} that has been actively investigated since the 1980s as a fluctuating-valence semiconductor~\cite{Slu_IgaShimizu98, Slu_IgaTakakuwa84, Slu_KasuyaKasaya83}. It was mentioned that the studies of fluctuating-valence Yb compounds are especially attractive because the single $4f$ hole of the Yb$^{3+}$ ion is in the $J=7/2$ ground state, separated from the $J=5/2$ excited state by an energy splitting of 15\,000~K, while the Yb$^{2+}$ ions are in the $^1$S$_0$ nonmagnetic state, leading to a simple energy-level scheme similar to that of the Ce-based compounds~\cite{Slu_KasayaIga85}. It was found that at low temperatures YbB$_{12}$ shows semiconducting-like and nonmagnetic (paramagnetic) properties similar to the fluctuating-valence compound SmB$_6$ \cite{Slu_IgaTakakuwa84, Slu_KasayaIga85}. At the same time, existing data on M\"{o}ssbauer effect of YbB$_{12}$ suggest that the Yb ion is nearly trivalent at 4.2~K~\cite{Slu_BonvilleImbert78}. Moreover, as a bulk-sensitive technique, the Yb~L$_\text{III}$ ($2p_{3/2}$) edge x-ray absorption spectroscopy (XAS)\index{x-ray absorption spectroscopy} has been performed, and the Yb valence is estimated to be very close to 3.0 (larger than 2.95) at 20~K~\cite{Slu_AlekseevNefeodova01}. The Yb valence $v(\text{Yb}) = 2.92$--2.93 has been estimated in YbB$_{12}$ at room temperature both by a bulk-sensitive photoelectron spectroscopy (PES)~\cite{Slu_HagiwaraTakeno17}\index{photoelectron spectroscopy} and hard x-ray PES (HARPES)~\cite{Slu_YamaguchiSekiyama09, Slu_RousuliSato17}. The detected $v$(Yb) values show a moderate decrease to 2.9 on cooling down to 20~K~\cite{Slu_YamaguchiSekiyama09, Slu_HagiwaraTakeno17, Slu_RousuliSato17}. Thus, YbB$_{12}$ in which the Yb ion is nearly in a trivalent state and in the so-called Kondo regime, can be considered to be a fluctuating-valence material characterized by the formation of a gap at the Fermi energy~\cite{Slu_KasayaIga85}.

\paragraph{Charge transport}\index{YbB$_{12}$!charge transport|(}
As the temperature decreases from 300~K down to 15~K, resistivity $\rho$ increases exponentially with the activation energy $E_\text{g}/2k_\text{B}\approx 62$~K. With further cooling, $\rho$ increases with a smaller activation energy $E_\text{p}/k_\text{B} \approx 28$~K and becomes nearly temperature independent upon subsequent temperature lowering~\cite{Slu_KasayaIga85}. Similar values $E_\text{g}/2k_\text{B} \approx 68$~K and $E_\text{p}/k_\text{B} \approx 25$~K were estimated in studies of high-quality YbB$_{12}$ single crystals, where $\rho(T)$ increased by five orders of magnitude with the temperature lowering from 300 to 1.3~K~\cite{Slu_IgaShimizu98}. The authors also found two intervals 15~K~<~$T$~<~40~K and 7~K~<~$T$~<~15~K of the Arrhenius-type behavior in the Hall coefficient measurements with the activation energies $E_\text{g}/2k_\text{B} \approx 90$~K and $E_\text{p}/k_\text{B} \approx 28$~K, respectively, which have been attributed to the opening of indirect charge gap ($E_\text{g}$) and to the charge transport through the intra-gap states ($E_\text{p}$). The mobility $\mu_\text{H}$ of the single-band conduction electrons was estimated to be as small as 3 and 10~cm$^2$/(V$\cdot$s) at 300~and 1.7~K, respectively, and the $\mu_\text{H}(T)$ dependence demonstrates the maximum values of $\sim$50~cm$^2$/(V$\cdot$s) near the temperature $T \approx 15$~K, separating these two activation intervals~\cite{Slu_KasayaIga85}.

The temperature and field dependencies of longitudinal MR of YbB$_{12}$ have been studied in Ref.~\cite{Slu_IgaLeBihan99} for three principal directions in the cubic lattice. At 1.5~K, MR in all these directions shows a large negative value up to about 70\% with an increase of magnetic field up to 15~T. With increasing temperature, the negative MR decreases gradually and becomes zero at 70~K for $\langle110\rangle$ and $\langle111\rangle$. Only for $\langle100\rangle$, the positive MR appears most remarkably in the range 5--15~K, where the smaller energy gap appears in the resistivity and the strong MR anisotropy persists at least up to 70~K~\cite{Slu_IgaLeBihan99}. A similar anisotropic contribution in positive MR of YbB$_{12}$ at 88~mK was reported by Kawasaki \textit{et~al.}~\cite{Slu_KawasakiTakamoto00}. Taking into account that the $4f$-state of Yb ion is assumed to be the $\Gamma_8$ quartet, and the wave function of these $4f$-electrons is elongated in the $\langle100\rangle$ direction, Iga \textit{et al.}~\cite{Slu_IgaSuga10} have attributed the MR anisotropy to the anisotropic distribution of the $4f$ electron density. At the same time, it is worth noting the similarity with the largest positive MR in the direction $\mathbf{H}\parallel\langle100\rangle$ discussed previously in this Chapter both for LuB$_{12}$ and the magnetic $R$B$_{12}$ RE compounds. The analogy allows us to seek a general explanation of the MR anisotropy in terms of the formation of dynamic charge stripes in the RE dodecaborides.\index{dynamic charge stripes}\index{charge stripes}

The Seebeck coefficient $S(T)$\index{Seebeck coefficient!in YbB$_{12}$} exhibits a giant negative peak of $\sim$140~$\mu$V/K at 10~K in addition to the second peak at 35~K and a shoulder at the temperature slightly above 100~K. Between 40 and 80~K, $S(T)$ demonstrates semiconducting behavior, being proportional to $1/T$~\cite{Slu_IgaSuemitsu01}. According to an estimation in the framework of the single-impurity Anderson model~\cite{Slu_ZlaticCosti93}, it was concluded~\cite{Slu_IgaSuemitsu01} that the maximum temperature $T_{S_\text{max}}\approx 35$~K can be related to the Kondo (or spin fluctuation) temperature $T_\text{sf} = 3T_{S_\text{max}}\approx 100$~K.
\index{YbB$_{12}$!charge transport|)}

\paragraph{Magnetic properties}\index{YbB$_{12}$!magnetic properties}
The high-temperature susceptibility $\chi(T)$ obeys the Curie-Weiss law in the range 150--1200~K~\cite{Slu_IgaShimizu98, Slu_MoiseenkoOdintsov79, Slu_KasayaIga85}, and the effective Bohr magneton number $\mu_{\text {eff}}$ was evaluated to be 4.3--4.4~$\mu_\text{B}$, which corresponds to 2.9--2.94 for the Yb valence (the calculated value is 4.54~$\mu_\text{B}$ for Yb$^{3+}$). The large negative values of the paramagnetic Curie temperature $\Theta_\text{p}$ between $-112$ and $-135$~K~\cite{Slu_IgaShimizu98, Slu_MoiseenkoOdintsov79} indicate the AFM exchange interaction between Yb ions, and it is ascribed to the spin-fluctuation effects ($T_\text{sf} \approx \Theta_\text{p}$) because the RKKY interaction\index{RKKY interaction} in YbB$_{12}$ is only of the order of 1~K~\cite{Slu_MoiseenkoOdintsov79}. As the temperature decreases, $\chi(T)$ demonstrates a pronounced peak at 75~K and then decreases rather rapidly corresponding to the magnetic gap opening in YbB$_{12}$~\cite{Slu_KasayaIga85}. Below 20~K a slight upturn is observed in the $\chi(T)$ curves of single crystals, which is usually discussed in terms of a small amount of isolated Yb$^{3+}$ impurity ions in the matrix of YbB$_{12}$.

\paragraph{Nuclear magnetic resonance}\index{YbB$_{12}$!nuclear magnetic resonance}\index{nuclear magnetic resonance!in YbB$_{12}$}
To separate the impurity component in the susceptibility, it is important to investigate the \textit{intrinsic} $\chi(T)$ at low temperatures. For this purpose, nuclear magnetic resonance (NMR) Knight-shift measurements on $^{11}$B~\cite{Slu_KasayaIga85} and $^{171}$Yb \cite{Slu_IkushimaKato00} have been performed. The local symmetry of boron sites in $R$B$_{12}$ is orthorhombic \textit{mm2}. However, the line shape observed in the $^{11}$B NMR was found to have an essentially axial symmetry \cite{Slu_KasayaIga85}. The authors analyzed the Knight shift in terms of the isotropic $K_\text{iso}$ and axial (anisotropic) $K_\text{ax}$ parts. It was found that the proportionality between $\chi(T)$ and $K_\text{ax}(T)$ exists in the temperature range 30--300~K; the $^{11}$B hyperfine coupling constant $A \approx 650$~Oe/$\mu_\text{B}$ has been estimated and attributed to the dipolar field at $^{11}$B sites which comes from the induced dipole moment on Yb ions~\cite{Slu_KasayaIga85}. The authors noted that the constant value of K$_\text{ax}$ below 20~K is the characteristic of the nonmagnetic ground state of the fluctuating-valence compound. The large isotropic Knight shift of about 66\% was observed below 10~K in studies of the $^{171}$Yb NMR on an YbB$_{12}$ single crystal~\cite{Slu_IkushimaKato00}. The detected hyperfine coupling constant of 1150~kOe/$\mu_\text{B}$ agrees very well with the calculated value for the $J = 7/2$ state of free Yb$^{3+}$ ions, indicating that the magnetic susceptibility in the low-temperature limit is dominated by the Van Vleck contribution within the $J = 7/2$ multiplet.

The presence of an activation spin gap in YbB$_{12}$ was clearly established from the $^{11}$B NMR measurements of the $^{10}$B and $^{11}$B nuclear spin-lattice relaxation rate, $^{11}(1/T_1)$ \cite{Slu_KasayaIga85, Slu_ShishiuchiKato02}. It was shown in \cite{Slu_KasayaIga85, Slu_ShishiuchiKato02} that $^{11}(1/T_1)$ drastically decreases below 80~K, which was explained by an excitation gap in the DOS at the Fermi level of about 80--100~K, depending on the external magnetic field. The $^{11}(1/T_1)$, however, has a minimum value near 15~K which becomes deeper and moves to lower temperature when the field increases up to 16~T~\cite{Slu_IkushimaKato00, Slu_ShishiuchiKato02}. The ratio of $(1/T_1)$ for $^{11}$B and $^{10}$B nuclei indicates that the anomaly below 15~K is caused by dilute paramagnetic impurities assisted by nuclear spin diffusion. It was argued that the minimum in $(1/T_1)$ suggests that the impurity moments are created only after most of the bulk magnetic excitations die away to make a stable singlet spin-gap state. This suggests that the origin of the moments is likely to be associated with defects of magnetic states or nonstoichiometry of conduction electrons as opposed to the extrinsic magnetic ions such as Gd~\cite{Slu_ShishiuchiKato02}.

No anomaly below 15~K was observed at the Yb sites. The absence of such process for Yb nuclei can be explained by the strong hyperfine coupling of Yb nuclei to the $4f$ electrons~\cite{Slu_IkushimaKato00}, since the resonance frequency of Yb nuclei near impurities should be largely shifted, and the mutual spin-flip process will be prohibited~\cite{Slu_ShishiuchiKato02}. Moreover, the nuclear spin-lattice relaxation rate at the Yb sites shows an activated temperature dependence below 15~K with the activation energy of 87~K, which is completely different from the behavior at the B sites~\cite{Slu_IkushimaKato00, Slu_ShishiuchiKato02}. The authors noted that the recovery of the nuclear magnetization is not strictly exponential but shows a fast relaxing component with small amplitude, indicating inhomogeneous distribution of $^{171}(1/T_1)$. The appearance of the full gap $E_\text{g}$ activation in the spin-lattice relaxation on Yb sites in the temperature interval $T < 15$~K, where the charge transport through the intra-gap states is observed, cannot be explained in terms of the hybridization gap scenario. It will be shown below that, on the contrary, the formation of vibrationally coupled Yb-Yb pairs should be considered as a natural interpretation for this kind of behavior.

\paragraph{Electron paramagnetic resonance}\index{electron paramagnetic resonance}\index{YbB$_{12}$!electron paramagnetic resonance}
The EPR has been studied on Gd ions as spin markers inserted in YbB$_{12}$. In the framework of the exciton dielectric model, Altshuler \textit{et al.}~\cite{Slu_AltshulerBresler02} determined the temperature dependence of the spin gap $E_\text{g} \approx 12$~meV, which is nearly temperature independent below 40~K and then decreases and disappears at $\sim$115~K. Additional singularity in the temperature behavior of the EPR line width was found at 13--14~K, pointing to the existence of intra-gap states with a finite electron density~\cite{Slu_AltshulerBresler02}. It was suggested that the intra-gap states may be due to many-body excitations~\cite{Slu_AltshulerBresler02}. In the follow-up work~\cite{Slu_AltshulerGoryunov03}, an EPR signal from Yb ions with an integer valence of 3+ was detected in the single crystals of YbB$_{12}$ intermediate-valence compound at LHe temperatures. The authors~\cite{Slu_AltshulerGoryunov03} observed two main lines symmetrically displaced from a $g$ value of 2.55 and exhibiting a modulation with 5\% anisotropy when the magnetic field was rotated in the $[1\overline{1}0]$ plane from the cubic axis to the $[110]$ direction. The EPR spectra have been explained by \emph{the existence of Yb-Yb pairs} which are coupled by the isotropic exchange but interact also with the other pairs by dipole and exchange coupling. The concentration of Yb$^{3+}$ ions in these pairs was estimated to be 0.2--0.5\%, and the origin of the EPR signal has been attributed to the effect of defects and/or vacancies stabilizing the valence of Yb ions~\cite{Slu_AltshulerGoryunov03}. Taking into account dipole-dipole splitting of the resonance lines, the authors have estimated the distance between the interacting Yb pairs, $a_\text{p} \approx 9.1$~\r{A}. It was suggested that $a_\text{p}$ is the average interpair distance, and pairs are distributed randomly so that their interaction energy fluctuates in space, contributing to fluctuations of the local CEF\index{crystal electric field} and resulting in an inhomogeneous broadening of the EPR line~\cite{Slu_AltshulerGoryunov03}. It was noted especially that the formal description of the EPR results points to the alignment of all Yb pairs, i.e., \textit{the appearance of a spontaneously chosen direction} in the otherwise cubic crystal, and this fact is equivalent to the existence of a sort of phase transition induced by the interpair coupling. The occurrence of a slight anisotropy in a cubic semiconductor should be attributed to a spontaneous symmetry breaking which, according to Altshuler \textit{et al.}~\cite{Slu_AltshulerGoryunov03}, is specific to the ground state of a Kondo insulator.\index{Kondo insulator} A strong temperature dependence of the EPR amplitude was found on cooling in the temperature range of 1.6--4.2~K~\cite{Slu_AltshulerGoryunov03}. The finding was interpreted in terms of the capture of electrons by Yb$^{3+}$ ions from electron traps with a binding energy of 18~K, which transforms the Yb centers into EPR-inactive ions with a fluctuating valence.

\paragraph{$\mu$SR spectroscopy}\index{YbB$_{12}$!$\mu$SR spectroscopy}
Positive-muon spectroscopy ($\mu$SR) and $^{170}$Yb M\"{o}ssbauer absorption measurements\index{M\"{o}ssbauer spectroscopy} have been performed on the cubic Kondo insulator YbB$_{12}$ down to 50~mK in temperature. Yaouanc \textit{et al.}~\cite{Slu_YaouancReotier99} observed a paramagnetic fluctuation mode\index{paramagnetic fluctuation mode} at low temperature, with a weak moment amplitude ($\sim$$10^{-2}\mu_\text{B}$) and a slow fluctuation frequency ($\sim$60~MHz) which remains constant between 0.04 and 4.2~K. No indication of a magnetic phase transition was found down to 0.04~K. It was argued~\cite{Slu_YaouancReotier99} that the smallness of the observed moments points to an itinerant spin picture within the strongly hybridized $4f$ band. The $\mu$SR data therefore imply that at a temperature of a few tens of mK, excitations with almost zero energy (100~neV~$\approx$~1~mK) are possible within the narrow band at the Fermi level, i.e. they suggest that the gap in YbB$_{12}$ (almost) closes along some direction $\mathbf{q}_\text{c}$ in $\mathbf{q}$-space, for which $\chi(\mathbf{q}_\text{c})$ remains paramagnetic~\cite{Slu_YaouancReotier99}.

$\mu$SR spectra of the single crystals of Yb$_{1-x}$Lu$_x$B$_{12}$ ($x = 0$, 0.125, 0.5, 1) and ErB$_{12}$ were measured between 1.8 and 300~K~\cite{Slu_KalviusNoakes02, Slu_KalviusNoakes03}. The authors found similar spectral shapes in all magnetic and nonmagnetic compounds studied over the whole temperature range, which excludes any contribution from Yb magnetic moments. In zero field the spectra alter their appearance around 20, 100 and 150~K. In a longitudinal field of 100~Oe, which largely suppresses the contribution from $^{11}$B nuclear moments, the $\mu$SR relaxation rate remained constant up to $\sim$150~K, where it suddenly peaks. It was concluded that the spectral shape is determined by the field of the $^{11}$B nuclear moments. These fields show dynamic behavior, but nuclear spin relaxation can be excluded, because its rate is below the $\mu$SR time window. It is suggested that the dynamical features arise from atomic motions within the B$_{12}$ clusters~\cite{Slu_KalviusNoakes02, Slu_KalviusNoakes03}.

\paragraph{Neutron scattering}\index{YbB$_{12}$!neutron scattering}
Magnetic excitations in YbB$_{12}$ have been studied by neutron scattering~\cite{Slu_AlekseevMignot04, Slu_NefeodovaAlekseev99, Slu_IgaBouvet99, Slu_MignotAlekseev05, Slu_NemkovskiMignot07}. According to the INS experiments~\cite{Slu_NemkovskiMignot07}, three peaks M1, M2, and M3 appear at low temperatures, $T < 60$~K, showing a significant dispersion. Both M1 (14.1~meV, $L$ point) and M2 (17.9~meV, $L$ point) correspond to an excitation from the Kondo singlet state to the magnetic state with $\Gamma_8$ symmetry~\cite{Slu_NemkovskiMignot07}. M1 is affected by the AFM interaction between Yb ions and is regarded as an intra-gap exciton-like mode~\cite{Slu_NemkovskiMignot07}. \textit{A significant 2D character of AFM short-range correlations} has been found in YbB$_{12}$~\cite{Slu_MignotAlekseev05, Slu_NemkovskiMignot07}. In particular, the correlation lengths that correspond to the couplings within ($\xi_{\parallel}$) and between ($\xi_{\perp}$) the (001) planes were estimated in Ref.~\cite{Slu_NemkovskiMignot07}, and the values $\xi_{\parallel} = 5.4 \pm 1.4$~\r{A} and $\xi_{\parallel} = 3.4 \pm 1.1$~\r{A} have been obtained. The spin gap was evaluated by the lower edge of the M2 peak, and it is close to the M1 peak position. It is worth noting that the spin gap in YbB$_{12}$ was found to be around 15~meV (for more details see Chapter~\ref{Chapter:Alekseev}), which is nearly the same as the charge gap.

\paragraph{Thermal conductivity and specific heat}\index{YbB$_{12}$!thermal properties}\index{thermal conductivity}
Thermal conductivity $\kappa(T)$ shows a pronounced increase below 60~K and exhibits a sharp peak at 15~K~\cite{Slu_IgaSuemitsu01}. This behavior below 60~K is a result of a strong increase in the phonon mean free path due to the gap formation in the electronic DOS. The electronic contribution to the thermal conductivity, $\kappa_\text{el}$, was estimated from the $\kappa(T)$ data by using the Wiedemann--Franz law. It was shown that $\kappa_\text{el}$ is less than 10\% of the total value even at 100~K~\cite{Slu_IgaSuemitsu01}. There is no second anomaly below the $\kappa(T)$ peak, which indicates that there is no thermal activation corresponding to $E_\text{p}$ observed in $\rho(T)$ below 15~K.

The heat capacity $C(T)$ of YbB$_{12}$ single crystals has been studied for the first time by Iga \textit{et al.}~\cite{Slu_IgaHiura99}. To estimate the magnetic contribution $C_\text{m}$, the authors used as a phonon heat-capacity reference the $C(T)$ dependence of the isostructural nonmagnetic compound LuB$_{12}$, taking into account a small difference in the Debye temperatures $\Theta_\text{D}$ between the two dodecaborides. The Schottky anomaly in $C_\text{m}(T)$\index{Schottky specific heat} was detected with an activation energy $E_\text{g}/k_\text{B} \approx 170$~K~\cite{Slu_IgaHiura99}. It was shown that the magnetic entropy in YbB$_{12}$ is close to the value of $R\ln 4$ (per Yb ion) near room temperature. It was mentioned that this result indicates the existence of four nearly generate states up to 300~K irrespective of the presence of the energy gap. If the fourfold degenerate state originates from the CEF\index{crystal electric field} splitting of Yb$^{3+}$ in a cubic symmetry, this state should be either the $\Gamma_8$ quartet or two closely located doublets $\Gamma_6$ and $\Gamma_7$~\cite{Slu_IgaHiura99}. It is worth noting that the finding contradicts the CEF splitting of the $J = 7/2$ multiplet of the Yb$^{3+}$ ion. Indeed, the overall splitting of 11.2~meV ($\sim$130~K) estimated previously from the INS experiments~\cite{Slu_AlekseevNemkovski14, Slu_AlekseevMignot04} suggests that both the $\Gamma_6$ and $\Gamma_7$ excited doublets should contribute also to the entropy at room temperature.

\begin{figure}[b!]
\centerline{\includegraphics[width=\textwidth]{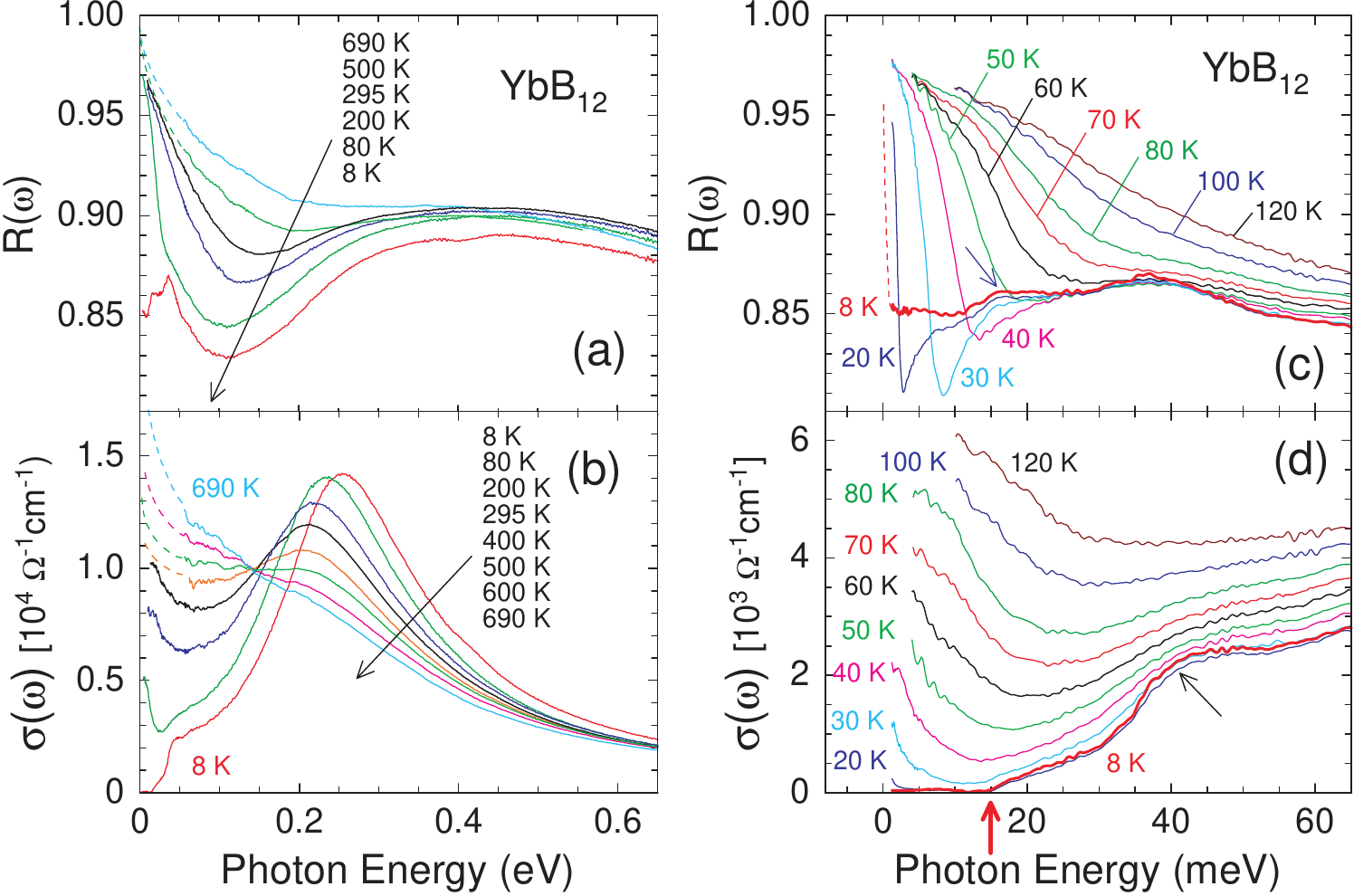}}
\caption{(a)~Optical reflectivity $R(\omega)$ and (b)~conductivity $\sigma(\omega)$ of YbB$_{12}$ between 8 and 690~K. (c,\,d) $R(\omega)$ and $\sigma(\omega)$, respectively, below 120~K in the low-energy region. The blue arrow in (c) indicates the hump in $R(\omega)$, and the black and red arrows in (d) indicate, respectively, the shoulder (at 40~meV) and the onset (at 15~meV) in $\sigma(\omega)$. The broken curves are the extrapolations. Reproduced from Ref.~\cite{Slu_OkamuraMichizawa05}.}
\label{Slu:Fig30}\index{YbB$_{12}$!optical properties!reflectivity}\index{YbB$_{12}$!optical properties!optical conductivity}
\end{figure}

\paragraph{Optical properties}\index{YbB$_{12}$!optical properties}
Measurements of the optical reflectivity, $R(\omega)$, were conducted on single crystals of YbB$_{12}$ at various temperatures \cite{Slu_OkamuraKimura98, Slu_OkamuraKimura99, Slu_OkamuraMichizawa05} in order to obtain their optical conductivity, $\sigma(\omega)$. Similar to LuB$_{12}$, the reflectivity spectra of YbB$_{12}$ have a clear plasma cutoff ($\omega_\text{p}$) near 1.6~eV and a sharp structure above 4~eV due to interband transitions~\cite{Slu_KimuraOkamura99}. The optical conductivity spectrum of YbB$_{12}$ clearly showed an energy gap formation below 80~K. The gap development involved a progressive depletion of $\sigma(\omega)$ below a shoulder at $\sim$40~meV. In addition, the authors observed a strong mid-infrared (mIR) absorption in $\sigma(\omega)$ peaked at $\sim$0.25~eV, which was also strongly temperature dependent~\cite{Slu_OkamuraKimura98}.

The temperature- and photon-energy ranges of the experiment have been extended from $T = 20$--290~K and $\geq 7$~meV~in Ref.~\cite{Slu_OkamuraKimura98} to $T = 8$--690~K and $\geq 1.3$~meV in Ref.~\cite{Slu_OkamuraMichizawa05}. The obtained $\sigma(\omega)$ reveals the entire evolution of the electronic structure from the metallic to semiconducting behavior with the temperature lowering, as illustrated in Fig.~\ref{Slu:Fig30}. Below 20~K, $\sigma(\omega)$ has revealed a clear onset at 15~meV, which was identified as the energy gap width. The energy of 15~meV agrees well with the gap widths obtained by other experimental techniques \cite{Slu_IgaShimizu98, Slu_TakabatakeIga98, Slu_TakedaArita04}. The authors concluded that the observed energy gap of 15~meV in $\sigma(\omega)$ arises from an indirect gap, predicted by the band model of the Kondo semiconductor; the feature at 0.2--0.25~eV was ascribed to a direct gap~\cite{Slu_OkamuraMichizawa05}.

\begin{figure}[t!]
\centerline{\includegraphics[width=0.8\textwidth]{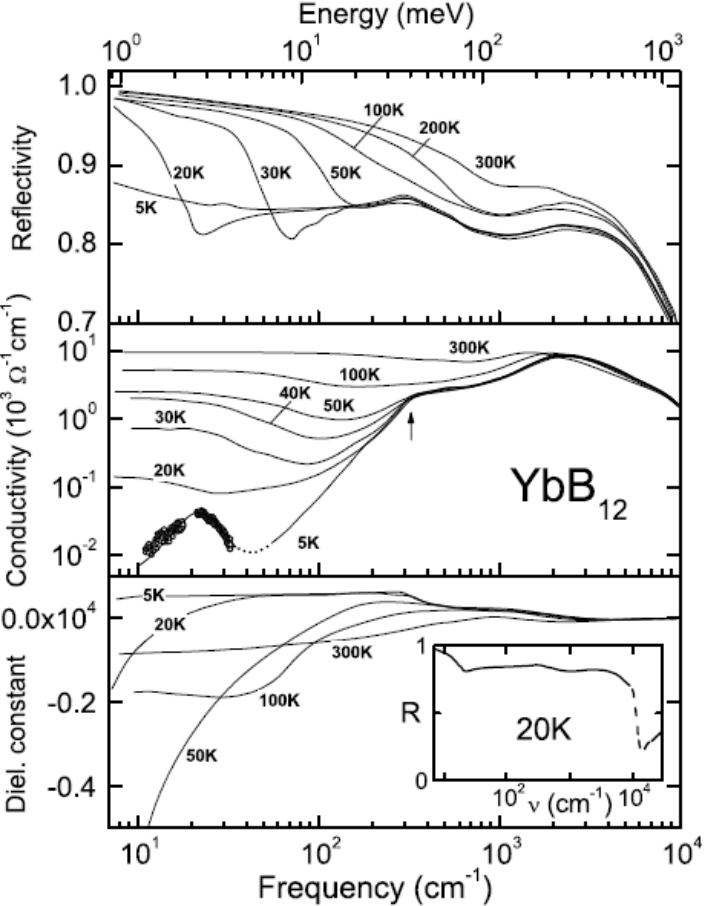}}
\caption{Spectra of reflectivity, conductivity and dielectric constant of YbB$_{12}$ at different temperatures, reproduced from Gorshunov \textit{et al.}~\cite{Slu_GorshunovHaas06}. The absorption peak at 22~cm$^{-1}$ seen at $T = 5$~K in the conductivity spectrum was recorded by measuring complex transmissivity of a thin ($\sim$20~$\mu$m) sample; the solid line shows a Lorentzian fit with the eigenfrequency $v_0 = 22$~cm$^{-1}$, dielectric contribution $\Delta\epsilon = 75$ and damping $\gamma = 15$~cm$^{-1}$. Inset: broadband reflectivity of YbB$_{12}$ showing two plasma edges at low (heavy quasiparticles) and high (unscreened electrons) frequencies. The dashed line corresponds to the data of Okamura \textit{et al.}~\cite{Slu_OkamuraKimura98}.}
\label{Slu:Fig31}\index{YbB$_{12}$!optical properties!reflectivity}\index{YbB$_{12}$!optical properties!optical conductivity}\index{YbB$_{12}$!optical properties!dielectric constant}
\end{figure}

\begin{figure}[b!]\vspace{-3pt}
\centerline{\includegraphics[width=0.8\textwidth]{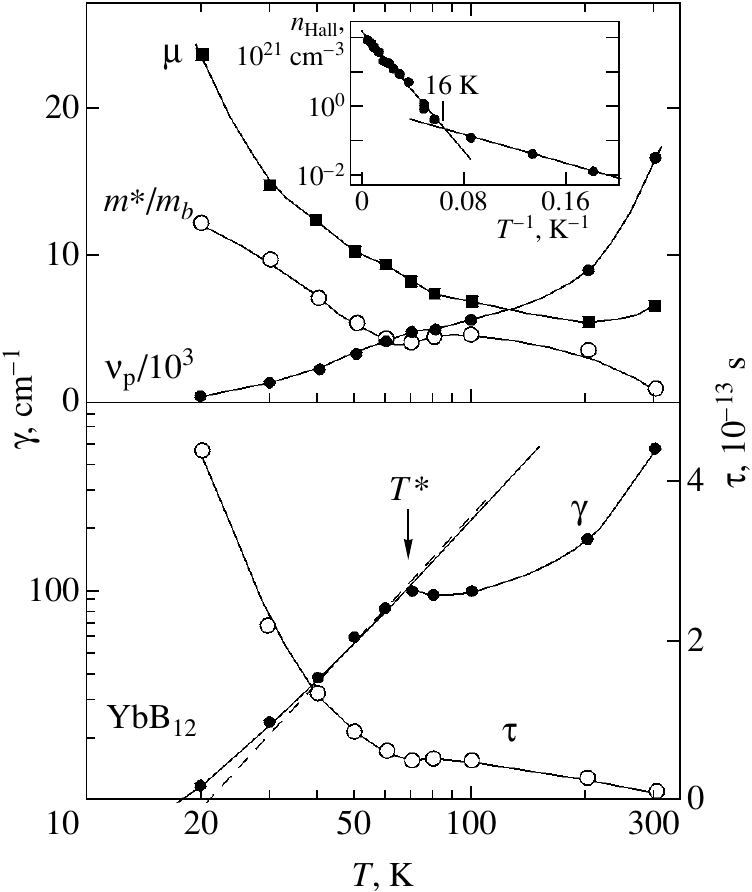}}
\caption{Temperature dependences of charge-carrier parameters in YbB$_{12}$. Upper panel: mobility $\mu$ in cm$^2$V$^{-1}$s$^{-1}$, effective mass $m^{\ast}/m_\text{b}$ ($m_\text{b} = 2.85 m_0$), plasma frequency $v_\text{pl}$ in cm$^{-1}$. The inset shows an Arrhenius plot of the Hall concentration of charge carriers. Lower panel: scattering rate $\gamma$ and relaxation time $\tau$. $T^{\ast} \approx 70$~K indicates the crossover temperature. Reproduced from Gorshunov \textit{et al.}~\cite{Slu_GorshunovProkhorov06}.\vspace{-3pt}}
\label{Slu:Fig32}\index{YbB$_{12}$!charge transport!charge-carrier mobility}\index{YbB$_{12}$!charge transport!effective electron mass}\index{YbB$_{12}$!plasma frequency}\index{YbB$_{12}$!charge transport!carrier concentration}\index{YbB$_{12}$!charge transport}\index{charge-carrier mobility}
\end{figure}

By employing the terahertz quasioptical technique~\cite{Slu_KozlovVolkov98}, Gorshunov~\textit{et~al.} performed reliable reflectivity measurements of YbB$_{12}$ down to the frequency of 8~cm$^{-1}$~\cite{Slu_GorshunovHaas06}, as shown in Fig.~\ref{Slu:Fig31}. In addition to the reflection experiments, direct measurements of the GHz and THz spectra of the dynamical conductivity $\sigma(\omega)$ and dielectric constant $\epsilon(\omega)$ have been performed at the lowest temperature of 5~K by measuring the complex transmissivity of a thin ($\sim$20~$\mu$m) crystal~\cite{Slu_GorshunovHaas06}. The progress in the lowest-frequency measurements at low temperatures allowed the authors to perform a quantitative analysis of the charge-carrier parameters in dependence on temperature in order to explain the electronic properties of YbB$_{12}$. Assuming a single band with one type of carriers, they used the Hall data obtained on crystals from the same batch in order to evaluate the carrier concentration $n = 1/(ecR_\text{H})$. As a result, the plasma frequency $\omega_\text{p} = 2\pi(ne^2/\pi m^{\ast})^{1/2}$, the effective mass $m^{\ast} = ne^2/(\omega_{\text p}/2\pi)^2$ and the mobility of the charge carriers $\mu = e/(2\pi m^{\ast}\gamma)$ (where $\gamma$ is the relaxation frequency) have been estimated~\cite{Slu_GorshunovHaas06}, see Fig.~\ref{Slu:Fig32}. It is seen from Fig.~\ref{Slu:Fig32} that upon cooling from 300 to 100~K, the effective mass $m^{\ast}(T)$ increases fivefold while the scattering rate $\gamma$ simultaneously decreases by a similar factor. This behavior is characteristic of a phonon-assisted scattering and contradicts the simple Kondo mechanism for which one would expect an increase in magnetic scattering upon cooling~\cite{Slu_VarmaYafet76}, which should suppress the mobility, while the latter is almost constant at these temperatures (Fig.~\ref{Slu:Fig32}). Also the temperature-dependent resistivity of YbB$_{12}$,\index{YbB$_{12}$!charge transport!resistivity} which shows an activated behavior rather than a Kondo-type dependence $\rho \propto -\log T$, is in contrast to the Kondo scenario. A possible reason for these observations is the presence of a gap in the DOS of YbB$_{12}$ at temperatures above 100~K. Indications of a gap at $T > 100$~K are also seen in photoemission experiments \cite{Slu_SusakiTakeda99}.\index{photoelectron spectroscopy}

At temperatures below $T^{\ast} \approx 70$~K, the gap opens in the DOS at $E_\text{F}$, which produces a shoulder in the $\sigma(v)$ spectrum at 40~meV (Figs.~\ref{Slu:Fig30}, \ref{Slu:Fig31}) and leads to the above-mentioned anomalies in the transport, magnetic, and thermodynamic properties. At $T^{\ast}$, an abrupt change in the charge-carrier characteristics is observed: the scattering rate starts to follow the Fermi-liquid-type dependence $\gamma(T) \propto T^2$~\cite{Slu_MillisLee87}, and the effective mass starts to increase (see Fig.~\ref{Slu:Fig32}). These findings indicate that below $T^{\ast}$ a transition occurs into a state with full coherence among the $f$ sites~\cite{Slu_VarmaYafet76, Slu_BickersCox87}, leading to the emergence of the heavy fermions. In other words, the low-frequency ($v < 100$~cm$^{-1}$) dispersion which the authors~\cite{Slu_GorshunovHaas06, Slu_GorshunovProkhorov06} observe in the spectra of YbB$_{12}$ at low temperatures ($T < T^{\ast} \approx 70$~K) is determined by the response of a Fermi liquid composed of heavy fermions, $m^{\ast}(\text{20~K}) \approx 34m_0$, with a strongly enhanced relaxation time $\tau(\text{20~K}) = 4\cdot10^{-13}$~s and mobility $\mu(\text{20~K}) = 24$~cm$^2$/(V$\cdot$s), as it is also shown in Fig.~\ref{Slu:Fig32}. Gorshunov \textit{et~al.} consider $T^{\ast} \approx 70$~K as the coherence temperature in YbB$_{12}$. It was noted in Refs.~\cite{Slu_GorshunovHaas06, Slu_GorshunovProkhorov06} that the findings depicted in Figs.~\ref{Slu:Fig30}\hspace{1.2pt}--\ref{Slu:Fig32} qualitatively resemble the optical response of heavy-fermion compounds where a Drude-like behavior of heavy quasiparticles is expected~\cite{Slu_MillisLee87} and observed~\cite{Slu_Degiorgi99, Slu_SchefflerDressel05} in the coherent state. Another feature of the heavy-fermion state in YbB$_{12}$ is the presence of two plasma edges in the reflectivity spectra that originate from plasma oscillations of heavy (renormalized) and light (unrenormalized) quasiparticles (Figs.~\ref{Slu:Fig30}, \ref{Slu:Fig31}).

One of the most important findings of the optical study~\cite{Slu_GorshunovHaas06, Slu_GorshunovProkhorov06} is a pronounced absorption peak which was observed at 22~cm$^{-1}$ ($\sim$2.7~meV) at the lowest temperature of 5~K. The origin of the low-frequency feature in the $\sigma(\omega)$ spectrum (see Fig.~\ref{Slu:Fig31}) has been connected with the formation of exciton-polaronic many-body intra-gap states close to the bottom of the conduction band. The authors~\cite{Slu_GorshunovHaas06} noted that the energy position of 2.7~meV of the low-temperature conductivity peak, which can be associated with the photon-assisted breaking of the intra-gap many-body states, is in accord with the activation energy of the charge transport characteristics. Indeed, at the temperature of $\sim$15~K, a crossover from $E_\text{g} \approx 20$~meV (excitations across the hybridization gap) to $E_\text{p} \approx 2.2$~meV (thermal activation of carriers from the ``impurity band'' into the conduction band) is observed \cite{Slu_IgaShimizu98, Slu_KasayaIga85}. It was suggested that the 22~cm$^{-1}$ peak in the conductivity spectrum of YbB$_{12}$ can correspond to the many-body states arising from the coupling of free electrons to soft valence fluctuations on Yb sites. Within this picture, the localization radius of these heavy fermions has been estimated to be $\sim$5~\r{A}~\cite{Slu_GorshunovHaas06, Slu_GorshunovProkhorov06}, close to the Yb-Yb distance ($\sim$5.3~\r{A}) in the YbB$_{12}$ lattice.\vspace{-1pt}

\paragraph{Tunneling spectroscopy}\enlargethispage{1pt}\index{YbB$_{12}$!tunneling spectroscopy}\index{tunneling spectroscopy!on YbB$_{12}$}
Measurements of the tunneling spectra have been carried out on YbB$_{12}$ single crystals using the break-junction technique~\cite{Slu_EkinoUmeda99}. It was found that the differential conductance shows very sharp gap-edge peaks with quite a low leakage at zero bias. The representative gap magnitude of $2\Delta \approx 220$--260~meV at 4.2~K was found to be unexpectedly larger than that from the transport measurements~\cite{Slu_EkinoUmeda99}. At the same time, the $2\Delta$ value agrees very well with the energy of the mid-infrared absorption peak in the $\sigma(\rho)$ spectra of YbB$_{12}$, i.e. $\sim$250~meV~\cite{Slu_GorshunovHaas06, Slu_OkamuraMichizawa05}.\vspace{-1pt}

\paragraph{Photoemission spectroscopy}\index{YbB$_{12}$!photoelectron spectroscopy}\index{photoelectron spectroscopy}
The temperature-dependent energy gap formation in both the valence bands and the Yb~$4f$ states of YbB$_{12}$ has been examined by means of high-resolution PES~\cite{Slu_TakedaArita06}. It was found that an energy gap ($<15$~meV) is gradually formed in the valence band upon cooling. Two characteristic temperatures have been detected at $T_1 \approx 150$~K and at $T^{\ast} \approx 60$~K. It was shown that the 55~meV peak at 250~K is shifted toward lower binding energy ($\sim$35~meV) upon cooling below $T_1$. The appearance of a 15~meV peak in the Yb~$4f$ and Yb~$5d$ states was clearly observed below $T^{\ast}$~\cite{Slu_TakedaArita06} and enhanced near the $L$ point. The authors concluded that the results give direct evidence that the coherent nature of the Yb~$4f$ state plays an important role in the energy gap formation via the $d$-$f$ hybridization below $T^{\ast} \approx 60$~K~\cite{Slu_TakedaArita06}. Then, the low-energy electronic structure of YbB$_{12}$ and its transient properties were investigated in more detail~\cite{Slu_OkawaIshida15} using ultra-high-resolution PES and time-resolved PES (TrPES). In the $T$-dependent laser-PES spectra, the authors found two different (pseudo)gaps with sizes of 25 and 15~meV, which were attributed to the single-site effect and the insulating hybridization-gap opening, respectively. The characteristic temperature $T_1$ was determined to be $\leq 150$~K, where the hybridization gap begins to open, although the Fermi edge remains as the in-gap state down to the lowest temperatures. In TrPES measurements, Okawa~\textit{et~al.} \cite{Slu_OkawaIshida15} found that the long-lived ($\geq$100~ps) component of the photoexcited electrons gradually develops upon cooling below 150~K, which was interpreted as a feature of the hybridization-gap evolution. Thus it was experimentally determined that the characteristic temperature $T_1 \approx 150$~K corresponds to a metal-to-insulator crossover in YbB$_{12}$.

\subsection{Pressure-induced insulator-to-metal transition in YbB$_\text{12}$}\index{YbB$_{12}$!metal-insulator transition!pressure-induced}
\index{metal-insulator transition}\index{insulator-to-metal transition}

In the case of YbB$_{12}$, the electrical resistivity has been first measured under pressure up to 8 and 20~GPa in Refs.~\cite{Slu_IgaKasaya93} and \cite{Slu_IgaLeBihan99}, respectively. However, it has shown semiconducting behavior over such a narrow pressure range. The pressure dependence of the activation energies $E_\text{g}$ and $E_\text{a}$ estimated from the electrical resistivity suggested that the semiconductor-to-metal transition may be expected at a much higher pressure of about 100~GPa~\cite{Slu_IgaLeBihan99, Slu_IgaKasaya93}.
\begin{figure}[b!]
\centerline{\includegraphics[width=0.8\textwidth]{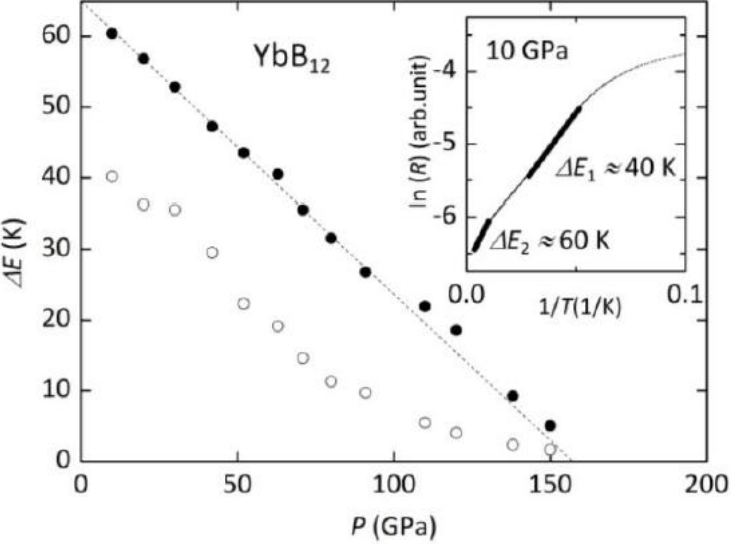}}
\caption{Pressure dependences of two activation energies of YbB$_{12}$ estimated from the resistivity measurements $R(T)$ up to 150~GPa. $\Delta E_1 = E_\text{g}$ and $\Delta E_2 = E_{\rm p}$ were detected in the intervals 18~K~<~$T$~<~33~K and 100~K~<~$T$~<~280~K, correspondingly. Reproduced from Kayama~\textit{et al.}~\cite{Slu_KayamaTanaka14}.}
\label{Slu:Fig33}\index{YbB$_{12}$!activation energy}
\end{figure}
More recent measurements of the electrical resistance $R(T)$ on YbB$_{12}$ single crystals were carried out under pressures up to 195~GPa~\cite{Slu_KayamaTanaka14}. It was then found that at pressures under 150~GPa the semiconducting behavior persists, but after that the low-temperature resistance is dramatically suppressed with increasing pressure. The pressure dependences of the two activation energies $E_\text{g}$ and $E_\text{p}$ estimated from the $R(T)$ dependences in the two temperature regions, 18~K~<~$T$~<~33~K (for $E_\text{g}$) and 100~K~<~$T$~<~280~K (for $E_\text{p}$) decrease monotonically with increasing pressure and are expected to vanish at $\sim$160~GPa by extrapolation, see Fig.~\ref{Slu:Fig33}. This marks a pressure-driven insulator-to-metal (I-M) transition. Moreover, metallic $R(T)$ behavior was observed at 164~GPa~\cite{Slu_KayamaTanaka14}, and at 195~GPa a drop in the resistance was observed below 0.8~K, which was attributed to the onset of superconductivity in YbB$_{12}$.

\subsection{Field-induced insulator-to-metal transition in YbB$_\text{12}$}\index{YbB$_{12}$!metal-insulator transition!field-induced}
\index{metal-insulator transition}\index{insulator-to-metal transition}

\begin{figure}[t!]
\centerline{\includegraphics[width=0.77\textwidth]{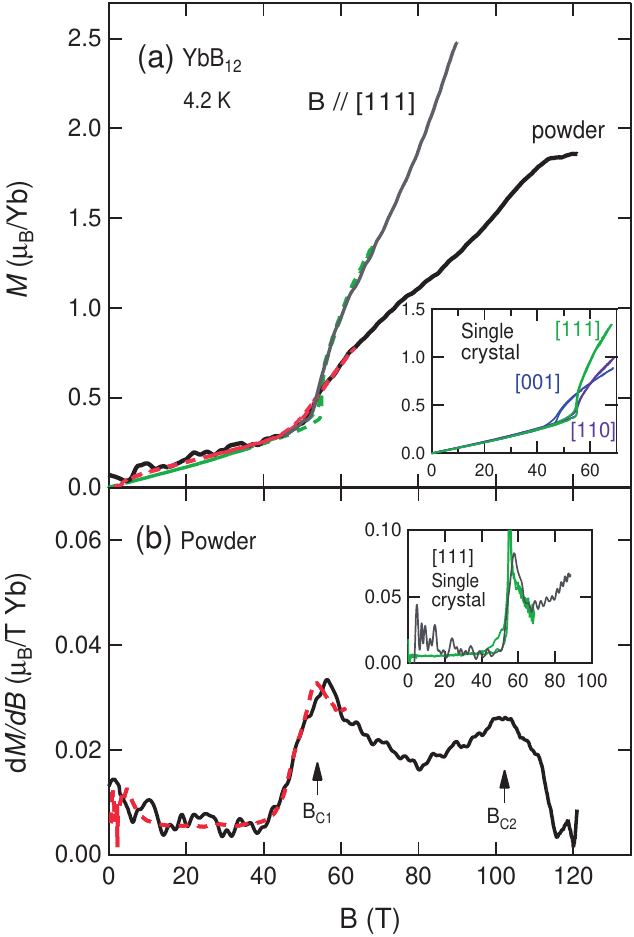}}
\caption{(a)~Magnetic-field dependence of the magnetization of YbB$_{12}$. The red dashed and black solid curves denote the magnetization in the powder sample measured by a nondestructive pulsed magnet and by the horizontal single-turn coil (HSTC), respectively. The green dashed and gray solid curves denote the magnetization in a single-crystal sample measured by a nondestructive pulsed magnet and by the HSTC, respectively. The inset shows the magnetization process of single crystals with different magnetic field directions. (b)~The field derivative of the magnetization (${\rm d}M/{\rm d}B$) curves for the results of the nondestructive pulsed magnet (red dashed curve) and the single-turn coil method (black solid curve). The inset shows ${\rm d}M/{\rm d}B$ for the single crystal in the $\mathbf{B}\parallel\langle111\rangle$ direction. The green solid curve was obtained with a nondestructive magnet and the gray solid curve was obtained with the HSTC. Reproduced from Terashima~\textit{et~al.}~\cite{Slu_TerashimaIkeda17}.\vspace{-3em}}
\label{Slu:Fig34}\index{YbB$_{12}$!magnetization!magnetic-field dependence}
\end{figure}

Strong negative MR is naturally expected for systems with field-induced metallization. The results of high-field magnetization and MR measurements indicate that the energy gap in YbB$_{12}$ collapses in a strong magnetic field by a metamagnetic transition~\cite{Slu_IgaLeBihan99, Slu_IgaSuga10}. An extreme drop in the resistivity $\rho(B)$ by two orders of magnitude was observed and considered as the manifestation of the first-order transition. At the critical field $B_\text{c1}$, the resistivity drops and then becomes constant, and the authors suggested that the electronic state above $B_\text{c}$ may be a metallic one. The critical field $B_\text{c1}$ shows an anisotropy with respect to the field direction: 47~T for $\mathbf{H}\parallel\langle100\rangle$ and 53~T for $\mathbf{H}\parallel\langle110\rangle$ and $\mathbf{H}\parallel\langle111\rangle$. The hysteresis on the magnetization and MR dependences also indicates this transition to be of the first-order type. The authors supposed that the anisotropy in the energy gap may originate from the $\mathbf{k}$-dependence of the mixing integral between the conduction and $4f$-electrons~\cite{Slu_IgaLeBihan99}.

\begin{figure}[b!]
\centerline{\includegraphics[width=\textwidth]{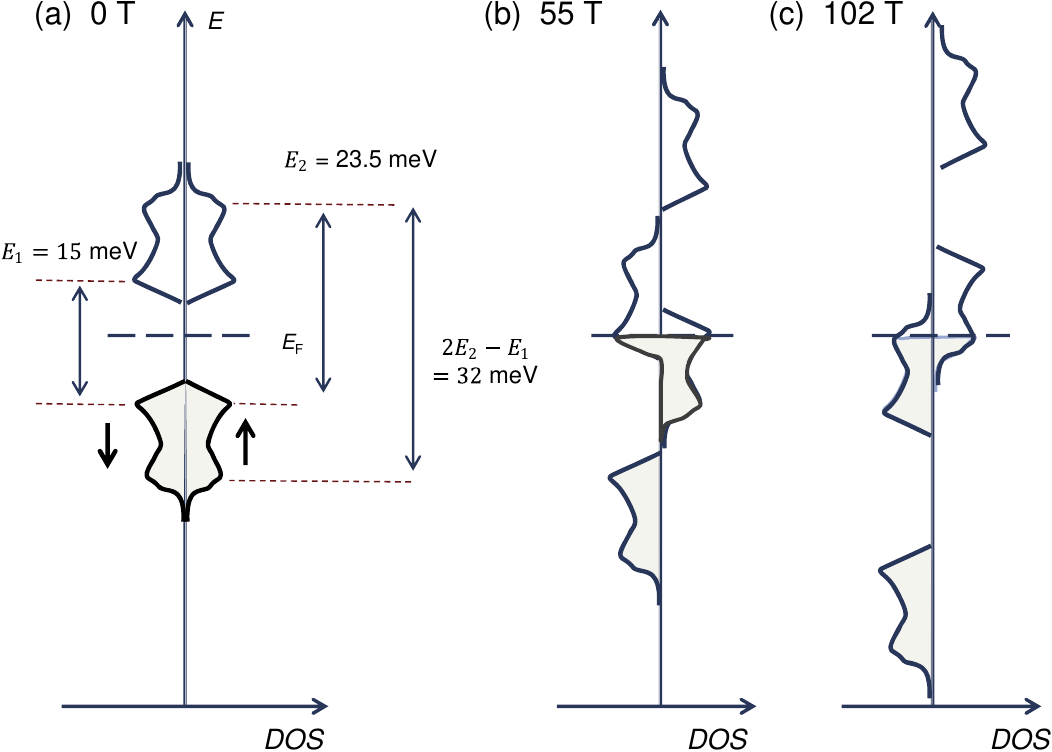}}
\caption{(a)~Schematic of the assumed density of states (DOS) at zero magnetic field. (b)~DOS expected in magnetic fields where the insulator-to-metal transition at $B_\text{c1}$ takes place. (c)~DOS expected in magnetic fields where the second metamagnetic transition takes place at $B_\text{c2}$. Reproduced from Ref.~\cite{Slu_TerashimaIkeda17}.}
\label{Slu:Fig35}\index{YbB$_{12}$!density of states}
\end{figure}

Recently the magnetization measurements were extended to ultrahigh magnetic fields up to 120~T with $\mathbf{B}\parallel\langle111\rangle$ to explore the possibility of additional magnetic transitions~\cite{Slu_TerashimaIkeda17}. Indeed, two clear metamagnetic transitions at $B_\text{c1}\approx 55$~T and $B_\text{c2} \approx 102$~T, and the tendency of the magnetization to saturate above 112~T have been observed (see Fig.~\ref{Slu:Fig34}). A two-gap structure in the DOS, probably due to an anisotropic hybridization of the $4f$ states with the conduction electrons (Fig.~\ref{Slu:Fig35}), has been suggested~\cite{Slu_TerashimaIkeda17} to explain the characteristic magnetization process. The authors have argued that the second metamagnetic transition at 102~T may correspond to the transition from the heavy-fermion metallic state to another metallic phase that either possesses a weaker Kondo effect than the phase below 102~T or conserves the strong AFM interaction between Yb magnetic moments and reduces the magnetization. One more very recent study of YbB$_{12}$ in the pulsed magnetic field may be also mentioned here to complete the section. Specific-heat measurements of a YbB$_{12}$ single crystal have been conducted in high magnetic fields of up to 60~T~\cite{Slu_TerashimaMatsuda18}, and the obtained data were used to deduce the magnetic field dependence of the Sommerfeld coefficient. It has been found that the linear coefficient of the electron heat capacity increases considerably at the I-M transition, being as large as 67~mJ/(mol$\cdot$K$^2$) at high fields, which is an indication of heavy quasiparticles~\cite{Slu_TerashimaMatsuda18}. Unfortunately, the simplest data analysis presented in~Ref.~\cite{Slu_TerashimaMatsuda18} did not consider the low-temperature magnetic components $C_\text{m}(T)$ that are typical for inhomogeneous systems with magnetic ions. But, at temperatures below $T^{\ast} \approx 60$~K in the cage-glass\index{cage glass} state of $R$B$_{12}$, the formation of AFM nanosize domains of Yb$^{3+}$ ions (short-range AFM order) should be expected, having a strong influence on the low-temperature heat capacity~\cite{Slu_GschneidnerTang90, Slu_Coles96}. Such an indication of ``false heavy fermions'' has been discussed in detail previously, for instance, by Gschneidner \textit{et al.}~\cite{Slu_GschneidnerTang90} and by Coles~\cite{Slu_Coles96}. On the other hand, the dramatic increase of the heat capacity in between 39 and 49~T may be certainly considered in terms of the metamagnetic transition in YbB$_{12}$~\cite{Slu_TerashimaMatsuda18}.
\index{YbB$_{12}$!metal-insulator transition|)}

\subsection[\mbox{Insulator-to-metal transition in} Yb$_x$\textit{R}$_{\text{1}-x}$B$_\text{12}$ (\textit{R}~=~Y, Lu, Sc, Ca, and Zr)]{Insulator-to-metal transition in Yb$_x$\textit{R}$_{\text{1}-x}$B$_\text{12}$ (\textit{R}~=~Y, Lu, Sc, Ca, and Zr)}\index{metal-insulator transition}\index{insulator-to-metal transition}
\index{Yb$_x$\textit{R}$_{1-x}$B$_{12}$!metal-insulator transition|(}

Another approach to investigate the I-M transition in YbB$_{12}$ is to study the substitutional solid solutions Yb$_{1-x}$R$_x$B$_{12}$, where $R$ is a nonmagnetic ion. In particular, the substitution of Lu$^{3+}$ for Yb$^{3+}$ is expected to remove the $4f$ magnetic moment at the corresponding RE site. At the same time, it does not change the number of conduction electrons, one per formula unit in an atomic picture~\cite{Slu_OkamuraMatsunami02}. In addition, the substitution does not change significantly the electronic structure beside the $4f$-related states. Hence, the replacement is expected to ``dilute'' the magnetic moments in this system, without changing the filling of the conduction band or influencing other electronic states. This dilution is expected to lower the overlap among the Yb orbitals at different Yb sites, which is believed to be essential in the hybridization gap scenario of the Kondo semiconductor \cite{Slu_OkamuraMatsunami02}.

On this way the combined results of resistivity~\cite{Slu_IgaHiura99, Slu_KasayaIga85, Slu_IgaKasaya88}, magnetic susceptibility and specific heat \cite{Slu_IgaHiura99, Slu_IgaKasaya88}, optical conductivity \cite{Slu_OkamuraMatsunami00, Slu_OkamuraMatsunami02}, tunneling spectroscopy \cite{Slu_EkinoUmeda99} of Yb$_{1-x}$Lu$_x$B$_{12}$ have suggested that the gap structure remains up to \mbox{$x = 0.5$}, and a mid-gap state develops near the Fermi level with increasing Lu concentration. From the analysis of specific heat, the authors have proposed that the ground state of Yb$_{1-x}$Lu$_x$B$_{12}$ ($x < 0.5$) is the $\Gamma_8$ quartet, but the magnetic entropy values ($\sim R\ln4/\text{mol\,Yb}$ at 300~K) contradict the CEF\index{crystal electric field} scheme with small enough total splitting ($\sim$11~meV) suggested in Ref.~\cite{Slu_AlekseevMignot04}. The transport experiments~\cite{Slu_IgaHiura99, Slu_KasayaIga85} on Yb$_{1-x}$Lu$_x$B$_{12}$ have shown that the electrical resistivity of YbB$_{12}$ at low temperatures is strongly reduced by substituting a small amount of Lu for Yb. These results indicate that the DOS at $E_\text{F}$ increases strongly with a small amount of Lu substitution. In addition, $\rho(T)$ shows thermally activated $T$ dependence only for $x \leq 1/2$, which indicates the absence of a transport gap in dilute Yb regime ($x>1/2$)~\cite{Slu_IgaHiura99, Slu_KasayaIga85}.

The magnetic field induced a collapse of the energy gap in the semiconducting solid solutions Yb$_{1-x}$Lu$_x$B$_{12}$, revealed in the measurements of magnetization and electrical resistivity in the pulsed magnetic fields up to 68~T~\cite{Slu_IgaSuga10}. It was found that for $x = 0.01$, positive MR appears in the range up to 20~T for the three principal directions and then the MR drops at almost the same field $B_\text{cs}$ (47~T for $\mathbf{B}\parallel\langle100\rangle$ and 53~T for $\mathbf{B}\parallel\langle110\rangle$ and $\langle111\rangle$) as in YbB$_{12}$~\cite{Slu_IgaSuga10}. It should be pointed out that the largest positive MR is observed for $\mathbf{B}\parallel\langle100\rangle$ both in the case of $x=0.01$ and especially for $x=0.05$~\cite{Slu_IgaSuga10}, and the finding is very similar to the MR anisotropy detected both for LuB$_{12}$ and magnetic dodecaborides $R$B$_{12}$, which is shown, for example, in Fig.~\ref{Slu:Fig28}\,(d).

\begin{figure}[t!]\vspace{-1pt}
\centerline{\includegraphics[width=0.72\textwidth]{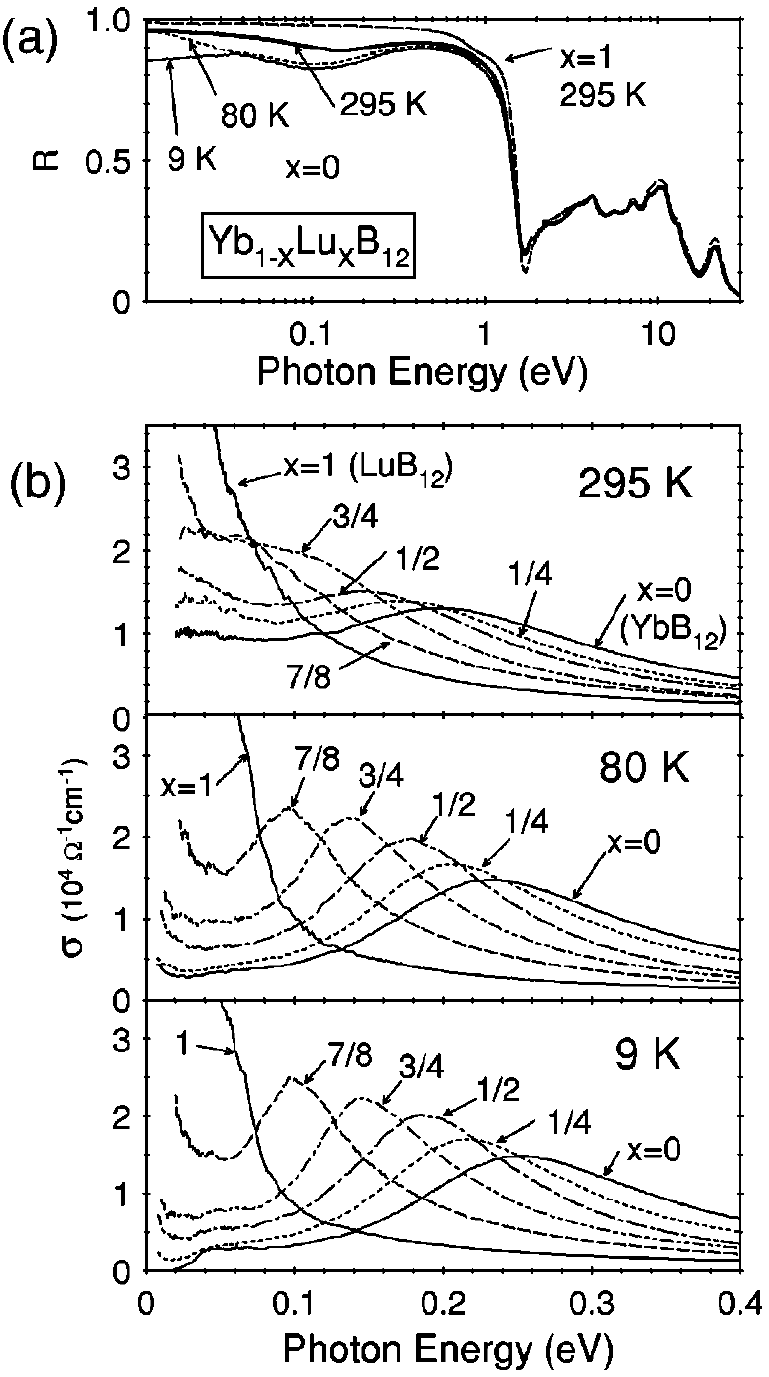}\vspace{-1pt}}
\caption{Optical reflectivity ($R$) and conductivity ($\sigma$) spectra of Yb$_{1-x}$Lu$_x$B$_{12}$. (a)~$R$ of $x=1$ (LuB$_{12}$) at 295~K and those of $x=0$ (YbB$_{12}$) at 295, 80, and 9~K. (b)~$\sigma$ for various values of $x$ at 295, 80, and 9~K. Reproduced from Okamura \textit{et al.}~\cite{Slu_OkamuraMatsunami00}.}
\label{Slu:Fig36}\index{Yb$_{1-x}$Lu$_x$B$_{12}$!optical properties}
\end{figure}

When discussing the evolution of the optical conductivity spectra of Yb$_{1-x}$Lu$_x$B$_{12}$, studied in detail by Okamura \textit{et~al.}~\cite{Slu_OkamuraMatsunami00, Slu_OkamuraMatsunami02}, it is worth noting that similar to YbB$_{12}$, in solid solutions with $x < 0.5$ a shoulder at $\sim$40~meV remains nearly unshifted in a wide range of temperatures (up to 70~K), but the energy gap in $\sigma(\omega)$ below 20~K is rapidly filled in from the bottom rather than by narrowing (Fig.~\ref{Slu:Fig36}). This result is consistent with the observation of the rapid filling of the gap in resistivity measurements at small $x$~\cite{Slu_IgaHiura99, Slu_KasayaIga85}. In contrast, the $\sigma(\omega)$ data~\cite{Slu_OkamuraMatsunami00, Slu_OkamuraMatsunami02} show that the DOS at $E_\text{F}$ is slightly reduced even at $x = 3/4$, as evidenced by the small depletion of spectral weight below $\sim$50~meV (Fig.~\ref{Slu:Fig36}). According to the conclusions of Okamura \textit{et al.}, the rapid filling of the gap in $\sigma(\omega)$ of Yb$_{1-x}$Lu$_x$B$_{12}$ at small $x$ clearly shows the importance of lattice effects in producing a well-developed gap. On the other hand, the shoulder position ($E^{\ast} \approx 40$~meV) is almost unchanged over a wide range of $x$, which strongly implies that $E^{\ast}$ is closely related to some single-site energy scale of Yb$^{3+}$ in YbB$_{12}$~\cite{Slu_OkamuraMatsunami00, Slu_OkamuraMatsunami02}.

Another important feature of the $\sigma(\omega)$ spectra is a broad mIR peak which was observed in YbB$_{12}$ at $\sim$250~meV \cite{Slu_OkamuraKimura98, Slu_OkamuraMatsunami00, Slu_OkamuraMatsunami02} (Fig.~\ref{Slu:Fig36}). It was clearly demonstrated in Refs.~\cite{Slu_OkamuraMatsunami00, Slu_OkamuraMatsunami02} that the peak position depends dramatically on the Lu concentration, changing monotonically to 120~meV for $x = 7/8$ (Fig.~\ref{Slu:Fig36}). In addition to the gradual shift of the mIR peak toward lower energy with increasing $x$, the displacement of the anomaly toward higher energy with decreasing temperature has been also observed \cite{Slu_OkamuraMatsunami02, Slu_OkamuraMatsunami00}.

\begin{figure}[!ht]
\centerline{\includegraphics[width=0.6\textwidth]{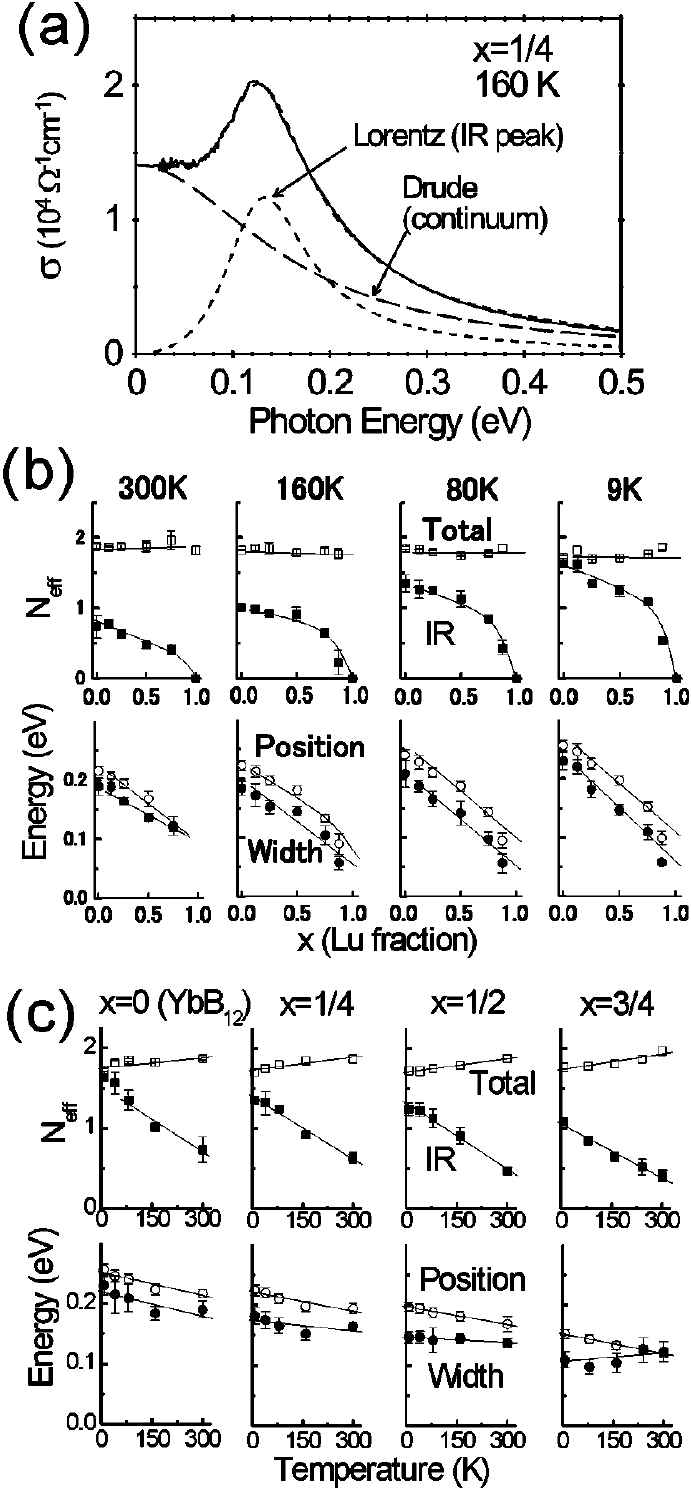}}
\caption{Fitting results of the mIR peak in the optical conductivity $\sigma(\omega)$ spectra of Yb$_{1-x}$Lu$_x$B$_{12}$. (a)~An example of the fitting for $x=1/4$. The solid line shows the measured spectrum, and dotted and dashed lines show the fitting. (b)~The effective carrier density per formula unit ($N_\text{eff}$) for the total (Drude + Lorentz) intensity and that for the mIR (Lorentz) peak, and the position and the width of mIR peak are plotted as a function of $x$ at four temperatures. (c)~The same data as in (b), but plotted as a function of temperature for four values of $x$. In (b) and (c), the solid lines are guided to the eye, and the error bars are derived from the uncertainty in the measured $R$. Reproduced from Okamura \textit{et al.}~\cite{Slu_OkamuraMatsunami00}.\vspace{-3em}}
\label{Slu:Fig37}\index{Yb$_{1-x}$Lu$_x$B$_{12}$!optical properties}
\end{figure}

To perform quantitative analyses of the $\sigma(\omega)$ spectra, the authors have fitted the spectral shape of the mIR peak for Yb$_{1-x}$Lu$_x$B$_{12}$ using the classical Lorentz oscillator model, and the broad continuum toward $\omega = 0$ using the Drude model. Figure~\ref{Slu:Fig37}\,(a) shows an example of the fitting. Using the optical sum rule, an effective carrier density $N_\text{eff}$ contributing to $\sigma(\omega)$ below $\omega=\omega_\text{p}$ was obtained as
\begin{equation}\label{Slu:Eq7}
N_\text{eff}=\frac{n}{m^\ast}=\frac{2m_0}{\pi e^2} \int_0^{\omega_\text{p}} \sigma(\omega){\rm d}\omega,
\end{equation}
where $n$ is the carrier density, and $m^\ast$ is the effective mass in units of the free electron mass $m_0$. In Figs.~\ref{Slu:Fig37}\,(b,\,c), $N_\text{eff}$ for the total $\sigma(\omega)$ (the sum of fitted Drude and Lorentz contributions) and that for the mIR peak (fitted Lorentz) are plotted as functions of $x$ and $T$, together with the position and the width of the mIR peak. Figure~\ref{Slu:Fig37}\,(b) shows that $N_\text{eff}$ contributing to the mIR peak is strongly nonlinear in 1\,--\,$x$, and hence it does not scale with the number of Yb~$4f$ electrons. Then, the peak energy and the width of the mIR feature show large, linear decreases with increasing $x$. Taking into account that a similar mIR anomaly has been found for the nonmagnetic reference compound LuB$_{12}$ at room temperature (see Fig.~\ref{Slu:Fig14}), and that the extrapolation of the peak width and position in Yb$_{1-x}$Lu$_x$B$_{12}$ [Fig.~\ref{Slu:Fig37}\,(b)] to $x = 1$ provides the same values $\sim$800~cm$^{-1}$ ($\sim$100~meV) that were measured in LuB$_{12}$~\cite{Slu_GorshunovZhukova18}, it is natural to conclude about the \emph{common nature of this collective mode} in YbB$_{12}$ and LuB$_{12}$. Thus, the origin of the mIR peak in Yb$_{1-x}$Lu$_x$B$_{12}$ can be attributed to the collective JT dynamics of the boron clusters (B$_{12}$, ferrodistortive effect, see Chapter~\ref{Chapter:Bolotina} for details) in the RE dodecaborides which is renormalized additionally by the fast charge and spin fluctuations on the Yb ions.

Okamura \textit{et al.}~\cite{Slu_OkamuraMatsunami00, Slu_OkamuraMatsunami02} noted that not only the mIR peak but also the Drude component shows considerable $x$- and $T$-dependencies, while the combined intensity (Drude\,+\,mIR peak) is kept almost constant, $N_\text{eff} \approx 1.8$ upon varying either $x$ and $T$. Thus, a spectral weight transfer exists between the Drude component and the mIR peak, which demonstrates that the dynamics of both the Drude free carriers and the conduction electrons involved in the collective mode are strongly connected to each other. In other words, despite of the appearance of the energy gap in YbB$_{12}$ only below 70~K, while the mIR peak is already observed at room temperature~\cite{Slu_OkamuraMatsunami02}, one can conclude that (i)~the depleted spectral weight in the gap region is transferred to the wide energy range that contains the mIR peak and (ii)~both the gap and the mIR peak show similar responses to $T$ and $x$. It will be shown below for Tm$_{1-x}$Yb$_x$B$_{12}$ that the cage-glass\index{cage glass} transition at $T^{\ast} = 60$--70~K to the disordered state of RE ions in the rigid boron sub-lattice may be responsible for the emergence of vibrationally coupled Yb-Yb pairs randomly distributed in the dodecaboride matrix.

Let us note at the end of the section that in Ref.~\cite{Slu_IgaYokomichi18}, the nonmagnetic-element substitution effects have been reported for YbB$_{12}$. It has been found that YbB$_{12}$ in a virtual gapless state would have a peak at about 25~K from magnetic susceptibility of Yb$_{1-x}R_x$B$_{12}$ ($R$ = Y, Lu, Sc, and Zr). The peak temperature $T\chi_\text{max}$ can be extrapolated to 25~K from the high-$x$ (Yb-dilute) region. The linear variations in the $x$-dependence of $T\chi_\text{max}$ show kinks at around $x = 0.5$ for all substituted element $R$ due to the development of an energy gap $E_\text{g}$. Furthermore, upturns shown in the magnetic susceptibility below 20~K in Yb$_{1-x}R_x$B$_{12}$ may be also related to the formation of the second gap $E_\text{p}$~\cite{Slu_IgaYokomichi18}. It is worth noting also the results of Matsuhra \textit{et al.}~\cite{Slu_MatsuhraYokomichi18} who succeeded in the synthesis of Yb$_{1-x}$Ca$_x$B$_{12}$ solid solutions up to $x = 0.15$ by using high-pressure synthesis. In these substituted alloys, a low-temperature increase in both $\chi$ and $C/T$ according to $-\log T$ are remarkably larger than those in powdered YbB$_{12}$, which, as concluded in Ref.~\cite{Slu_MatsuhraYokomichi18}, may originate from a giant increase in the DOS at $E_\text{F}$ due to an in-gap state.
\index{Yb$_x$\textit{R}$_{1-x}$B$_{12}$!metal-insulator transition|)}

\subsection{Metal-insulator transition in Tm$_{\text{1}-x}$Yb$_x$B$_\text{12}$}\label{Slu_SubSec:TmYbB12}
\index{Tm$_{1-x}$Yb$_x$B$_{12}$!metal-insulator transition|(}\index{metal-insulator transition}\index{insulator-to-metal transition}

As the concentration $x$ varies, the properties of the substitutional solid solutions Tm$_{1-x}$Yb$_x$B$_{12}$ continuously transform from the AFM metal TmB$_{12}$ ($x = 0$, $T_\text{N} \approx 3.2$~K)~\cite{Slu_SluchankoBogach09} to the paramagnetic insulator YbB$_{12}$ ($x = 1$) with strong charge and spin fluctuations~\cite{Slu_IgaShimizu98, Slu_GorshunovHaas06, Slu_AeppliFisk92}. Thus, the study of the charge transport, magnetic characteristics, heat capacity, and fine details of the crystal structure were undertaken in Refs.~\cite{Slu_SluchankoBogach09, Slu_SluchankoAzarevich12, Slu_BogachSluchanko13, Slu_SluchankoAzarevich13, Slu_SluchankoDudka18} to reveal the singularities of both the MIT and the AF-P phase transition in these compounds with an unstable electron configuration of the Yb ion. In strongly correlated electron systems,\index{strongly correlated electrons} the most significant changes in the physical characteristics take place near QCPs~\cite{Slu_Stewart06, Slu_LoehneysenRosch07, Slu_SachdevKeimer11, Slu_Stishov04, Slu_GantmakherDolgopolov10}\index{HoB$_{12}$!quantum criticality} and at the MITs~\cite{Slu_TakabatakeIga98, Slu_AeppliFisk92, Slu_GantmakherDolgopolov10, Slu_Dagotto05, Slu_GantmakherDolgopolov08}. Near QCPs corresponding to zero-temperature phase transitions, such changes are due to both thermal and quantum fluctuations. This leads to an enhancement of correlation effects and to the formation of complicated many-body states~\cite{Slu_Stewart06, Slu_LoehneysenRosch07, Slu_SachdevKeimer11, Slu_Stishov04, Slu_GantmakherDolgopolov10}. Similarly, insulator-type features arising at the MITs are usually closely related to the formation of bound many-body states in the systems of itinerant charge carriers, resulting to a drastic drop in the density/mobility of conduction electrons~\cite{Slu_TakabatakeIga98, Slu_AeppliFisk92, Slu_GantmakherDolgopolov10, Slu_Dagotto05, Slu_GantmakherDolgopolov08}.

\begin{figure}[t!]
\centerline{\includegraphics[width=0.82\textwidth]{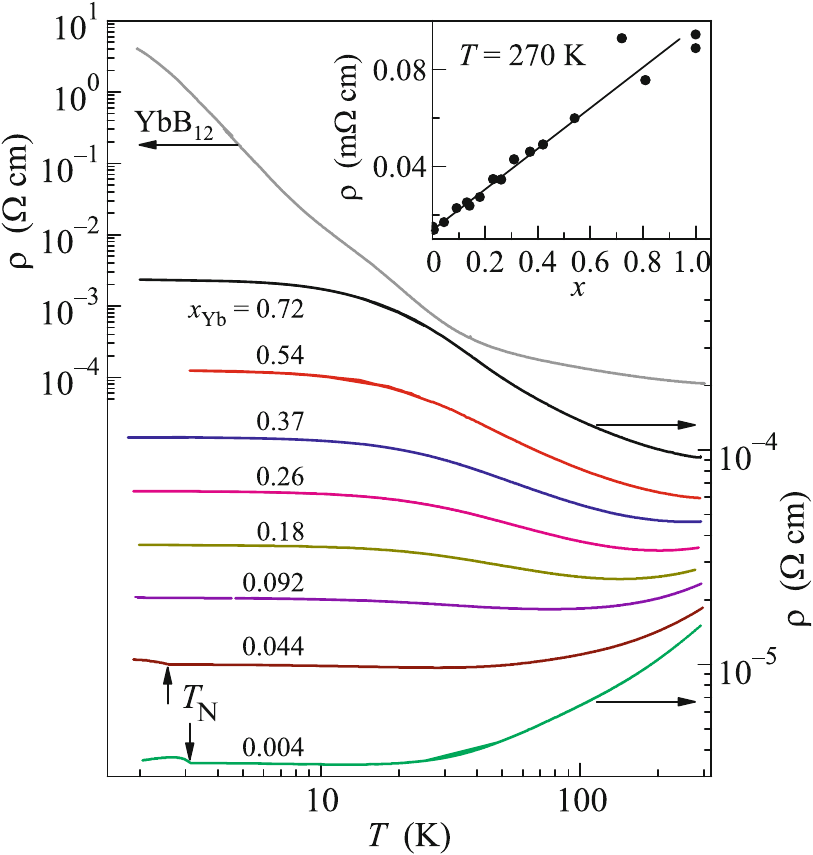}}
\caption{Temperature dependence of the electrical resistivity for Tm$_{1-x}$Yb$_x$B$_{12}$. The inset shows the dependence of resistivity on the ytterbium concentration at $T=270$~K. Reproduced from Ref.~\cite{Slu_SluchankoDudka18}.}
\label{Slu:Fig38}\index{Tm$_{1-x}$Yb$_x$B$_{12}$!charge transport!resistivity}
\end{figure}

\begin{figure}[t!]
\centerline{\includegraphics[width=0.8\textwidth]{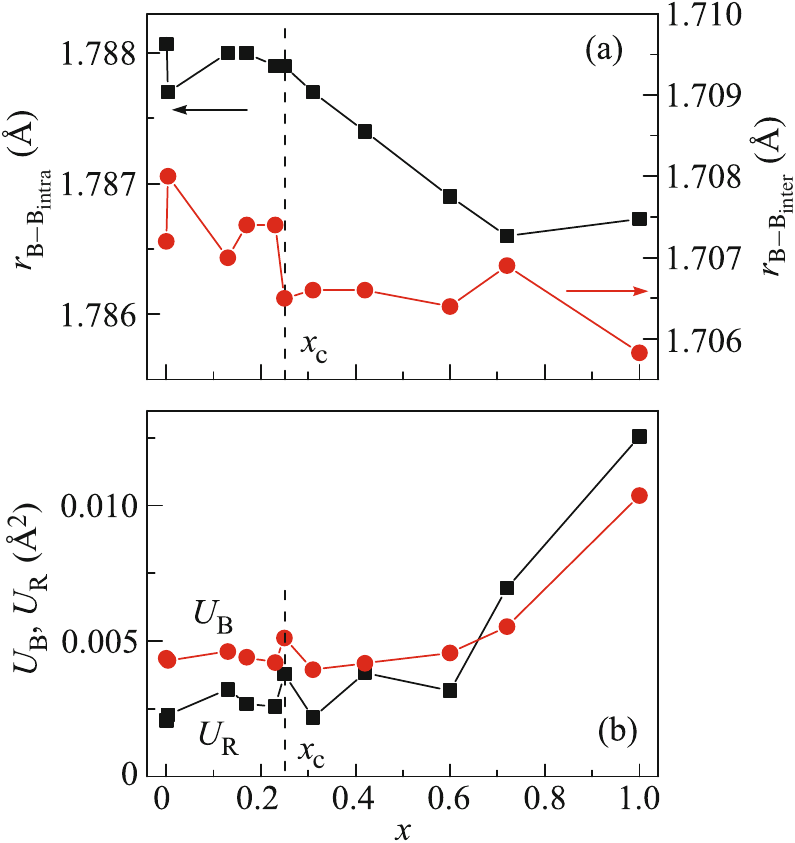}}
\caption{(a)~Boron-boron distances in the B$_{12}$ clusters, $r_{\text{(B-B)\,intra}}$, and between the clusters, $r_{\text{(B-B)\,inter}}$, and (b)~the equivalent atomic displacement parameters $U_{\rm B}$ and $U_R$ of the boron and cation atoms, respectively, in Tm$_{1-x}$Yb$_x$B$_{12}$ versus the ytterbium content $x$. Reproduced from Ref.~\cite{Slu_SluchankoDudka18}.}
\label{Slu:Fig39}\index{Tm$_{1-x}$Yb$_x$B$_{12}$!interatomic distances}\index{Tm$_{1-x}$Yb$_x$B$_{12}$!atomic displacement parameters}\index{atomic displacement parameters}
\end{figure}

\paragraph{Resistivity and Hall effect}\index{Tm$_{1-x}$Yb$_x$B$_{12}$!charge transport!resistivity}\index{Tm$_{1-x}$Yb$_x$B$_{12}$!charge transport!Hall effect}
The resistivity of Tm$_{1-x}$Yb$_x$B$_{12}$ crystals on cooling at small $x$ values exhibits a decrease typical for metals in combination with low-temperature features near the N\'{e}el temperature $T_\text{N}$, see Fig.~\ref{Slu:Fig38}. With an increasing Yb content, the AFM phase transition becomes suppressed at the QCP, $x_{\rm c} \approx 0.25$, the $\rho(T)$ dependences demonstrate a low-temperature increase, and for $x>0.5$ the plots correspond to the semiconducting behavior on cooling below 300~K. As a result, the resistivity in the Tm$_{1-x}$Yb$_x$B$_{12}$ series with MIT increases by a factor of about 2$\cdot$10$^6$ at low temperatures and by a factor of about 7 even at 270~K (inset in Fig.~\ref{Slu:Fig38}). The results of room-temperature Hall effect measurements allowed us to detect a variation in the charge carrier density $n(x) = (R_{\rm H}e)^{-1}$ versus the ytterbium content $x$ in the Tm$_{1-x}$Yb$_x$B$_{12}$ series~\cite{Slu_SluchankoDudka18}. It was shown that the single-band carrier concentration $n$ decreases by more than a factor of 2, and the Hall mobility $\mu_{\rm H}(x) = R_{\rm H}/\rho$ also exhibits a significant decrease (by a factor of about 3)~\cite{Slu_SluchankoDudka18}. The room-temperature behavior of $n(x)$ and $\mu_{\rm H}(x)$ near the QCP at $x_\text{c} \sim 0.25$ was found to be monotonic within the experimental error. The band gap in the Tm$_{1-x}$Yb$_x$B$_{12}$ for $x \geq 0.5$ was estimated to be $E_\text{g}/k_\text{B} \approx 200$~K \cite{Slu_SluchankoAzarevich12}. Thus, a metallic behavior can be expected for all $x$(Yb) compositions at $T>E_\text{g}/k_\text{B}$. It was concluded in Ref.~\cite{Slu_SluchankoDudka18} that strong electron correlations and many-body effects are apparently responsible for the renormalization of the charge-carrier density and mobility observed at room temperature. Then, to determine the mechanism responsible for the increase in $\rho(x)$ at room temperature, the authors have analyzed the equivalent atomic displacement parameters in the boron $U_{\rm B}$ and cation $U_{\rm R}$ subsystems, as determined from the x-ray diffraction data.\index{x-ray diffraction} Figure~\ref{Slu:Fig39} shows the $U_{\rm B}(x)$ and $U_{\rm R}(x)$ dependences, demonstrating a considerable growth of the amplitude of atomic displacements with the increase in $x$ both in the boron sublattice (more than by a factor of 2) and for the RE ions (by a factor of about 6)~\cite{Slu_SluchankoDudka18}. It was pointed out in Ref.~\cite{Slu_SluchankoDudka18} that the situation in Tm$_{1-x}$Yb$_x$B$_{12}$ at $x \geq 0.7$ is very unusual, because the displacement amplitude of heavy RE ions with $m_{\rm A} \approx 170$ exceeds by far that for the light boron atoms ($m_{\rm A} \approx 11$), as shown in Fig.~\ref{Slu:Fig39}. It was suggested in Ref.~\cite{Slu_SluchankoDudka18} that the increase in the amplitude of atomic displacements by a factor of 2--6 must be responsible for the enhancement of charge-carrier scattering in the crystal lattice of dodecaborides and for the corresponding (by a factor of about~7) decrease in the conductivity (inset in Fig.~\ref{Slu:Fig38}).

\begin{figure}[t!]\vspace{-1pt}
\centerline{\includegraphics[width=\textwidth]{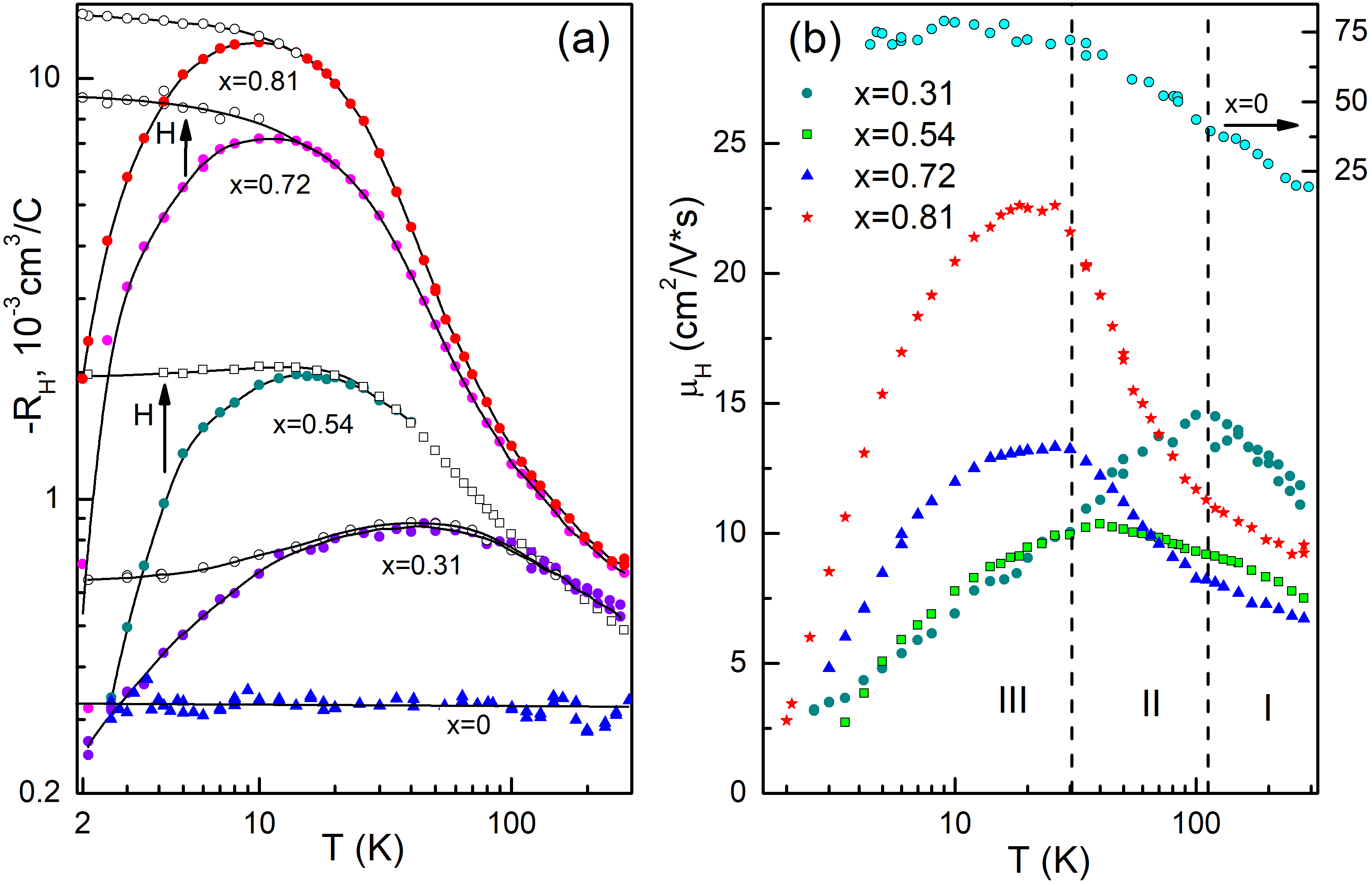}\vspace{-2pt}}
\caption{Temperature dependences of (a)~the Hall coefficient $R_{\rm H}(T, H_0)$ at low ($H_0 = 15$~kOe, solid symbols) and high ($H_0 = 70$~kOe, open symbols) magnetic fields and (b)~mobility $\mu_{\rm H}$($T$, $H_0 = 15$~kOe) of charge carriers for Tm$_{1-x}$Yb$_x$B$_{12}$ with $x = 0.31$, 0.54, 0.72, and 0.81. For comparison the $R_{\rm H}(T)$ and $\mu_{\rm H}(T)$ dependences for TmB$_{12}$ at $H_0 = 3.7$~kOe are shown in panels (a) and (b), respectively. Reproduced from Ref.~\cite{Slu_SluchankoAzarevich12}.\vspace{-1pt}}
\label{Slu:Fig40}
\end{figure}

\begin{figure}[t!]
\centerline{\includegraphics[width=\textwidth]{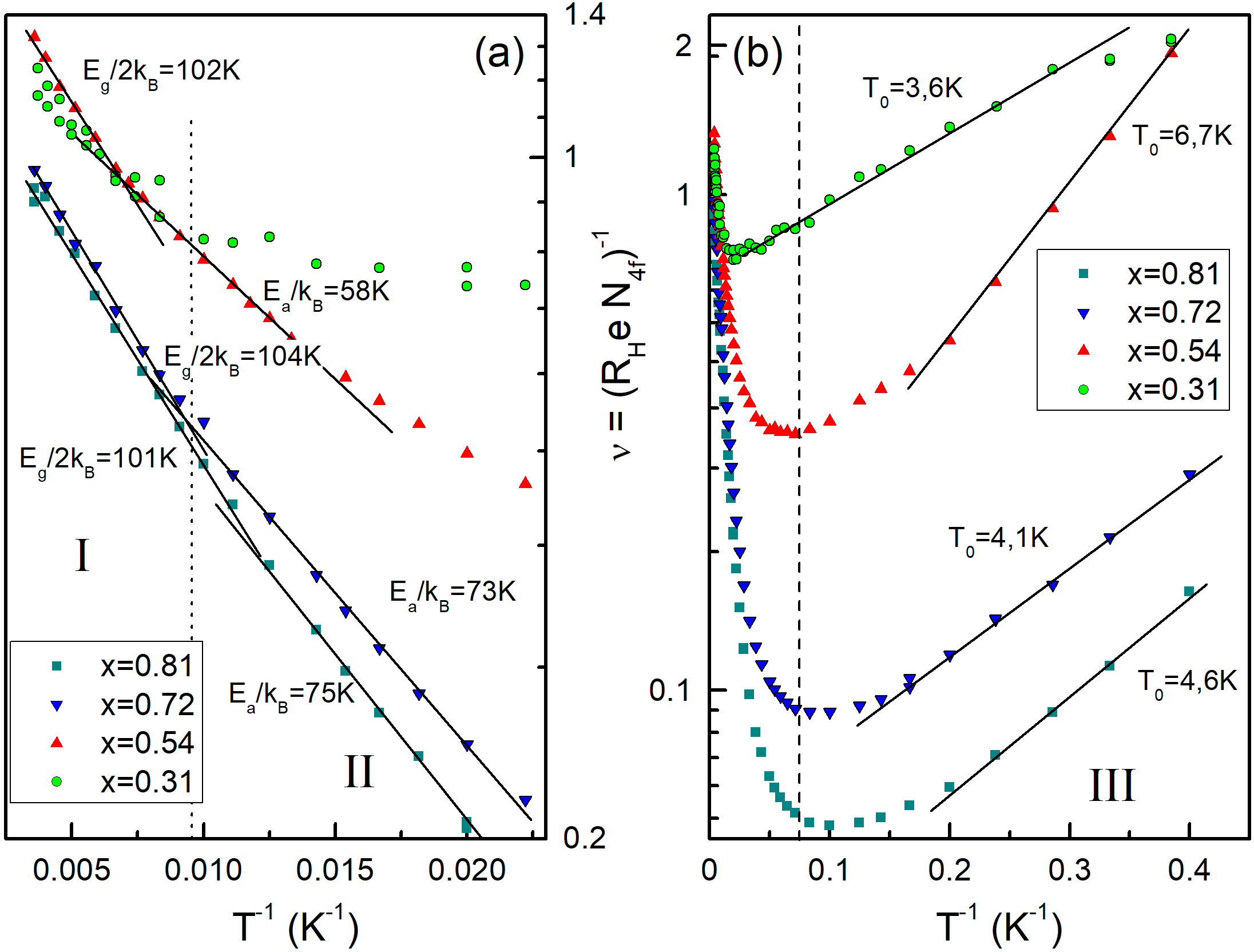}}
\caption{(a--b)~Arrhenius-type dependences of the reduced concentration $v = (R_\text{H}eN_{4f})^{-1}$ of charge carriers for Tm$_{1-x}$Yb$_x$B$_{12}$ compounds with $x=0.31$, 0.54, 0.72, and 0.81 in the different temperature ranges I--III. Reproduced from Ref.~\cite{Slu_SluchankoAzarevich12}.}
\label{Slu:Fig41}
\end{figure}

The detailed measurements of the resistivity in the Hall geometry in Tm$_{1-x}$Yb$_x$B$_{12}$ have been carried out in Ref.~\cite{Slu_SluchankoAzarevich12}. The comprehensive analysis of the angular, temperature, and field dependences of the resistivity from the Hall probes' allowed separating and classifying two components: the Hall and transverse even (TE) effects in the charge transport of these strongly correlated electron systems.\index{strongly correlated electrons} It was found that the second-harmonic contribution $\rho_\text{TE}$ (TE effect) appears in the angular dependences of the Hall probe's resistivity of the Tm$_{1-x}$Yb$_x$B$_{12}$ dodecaborides near the QCP ($x_{\rm c} \approx 0.25$), and it enhances drastically as the concentration $x$ increases in the interval $x \geq x_\text{c}$. A pronounced negative maximum in the temperature dependence of the Hall coefficient for $x \geq x_{\rm c}$ and the sign reversal of $R_\text{H}(T)$ at LHe temperatures for Tm$_{1-x}$Yb$_x$B$_{12}$ with $x \geq 0.5$ have been revealed~\cite{Slu_SluchankoAzarevich12} (see Fig.~\ref{Slu:Fig40}), and a decrease in $R_\text{H}(T)$ at $T < 30$~K has been identified as a signature of the coherent regime of charge transport. It was shown that at temperatures below the negative maximum of the Hall coefficient (interval III in Fig.~\ref{Slu:Fig40}), the temperature dependence $R_\text{H}(T) \propto \exp(-T_0/T)$ is observed with the estimated values $T_0 \approx 3.5$--7~K. A similar behavior of $R_\text{H}(T)$ was previously observed for heavy-fermion compounds CeAl$_2$~\cite{Slu_SluchankoBogach04} and CeAl$_3$~\cite{Slu_SluchankoGlushkov06}, and it was interpreted in terms of the dependence predicted in Ref.~\cite{Slu_YeKim99} for the Hall coefficient in a system with Berry-phase effects, where the carrier moves by hopping in a topologically nontrivial spin background. According to Refs.~\cite{Slu_YeKim99, Slu_KimMajumdar98}, the Hall effect is modified in such a situation because of the appearance of the internal magnetic field $H_\text{int} \sim (1/k_\text{B}T)\exp(-T_0/T)$, which is added to the external field $H$. It has been shown that the external magnetic field drastically enforces the TE effect~\cite{Slu_SluchankoAzarevich12}, also suppressing the coherent regime of the spin-flip scattering of charge carriers on the magnetic moments of the RE ions, see Fig.~\ref{Slu:Fig40}\,(a).

For the Tm$_{1-x}$Yb$_x$B$_{12}$ compounds with $x \geq 0.5$ in the intervals 120--300~K (I) and 50--120~K (II), the authors~\cite{Slu_SluchankoAzarevich12} found an Arrhenius-type behavior of the Hall coefficient with the activation energies $E_\text{g}/k_\text{B} \approx 200$~K and $E_\text{p}/k_\text{B} \approx 58$--75~K (Fig.~\ref{Slu:Fig41}) and the following microscopic parameters: effective masses $m^{\ast} \approx 20m_0$ and localization radii of the heavy-fermion many-body states $\sim$5~\r{A} (120--300~K) and $\sim$9~\r{A} (50--120~K). We have argued in Ref.~\cite{Slu_SluchankoAzarevich12} that the MIT, which develops both in YbB$_{12}$ as the temperature decreases and in Tm$_{1-x}$Yb$_x$B$_{12}$ solid solutions as the Yb concentration increases in the range $0 < x \leq 1$, is induced by a formation of Yb-Yb pairs randomly distributed in the $R$B$_{12}$ matrix. In the framework of the spin-polaron approach, the appearance of the odd $R_\text{H}(T, H)$ and even $\rho_\text{TE}(T,H)$ anomalous components in the resistivity detected in the Hall-effect geometry in the Tm$_{1-x}$Yb$_x$B$_{12}$ series has been discussed in terms of the interference effects between local $4f$-$5d$ and long-range spin fluctuations, leading to the formation of a filamentary structure (network) of the interconnected many-body complexes in the dielectric matrix of these compounds.

\begin{figure}[t!]
\centerline{\includegraphics[width=\textwidth]{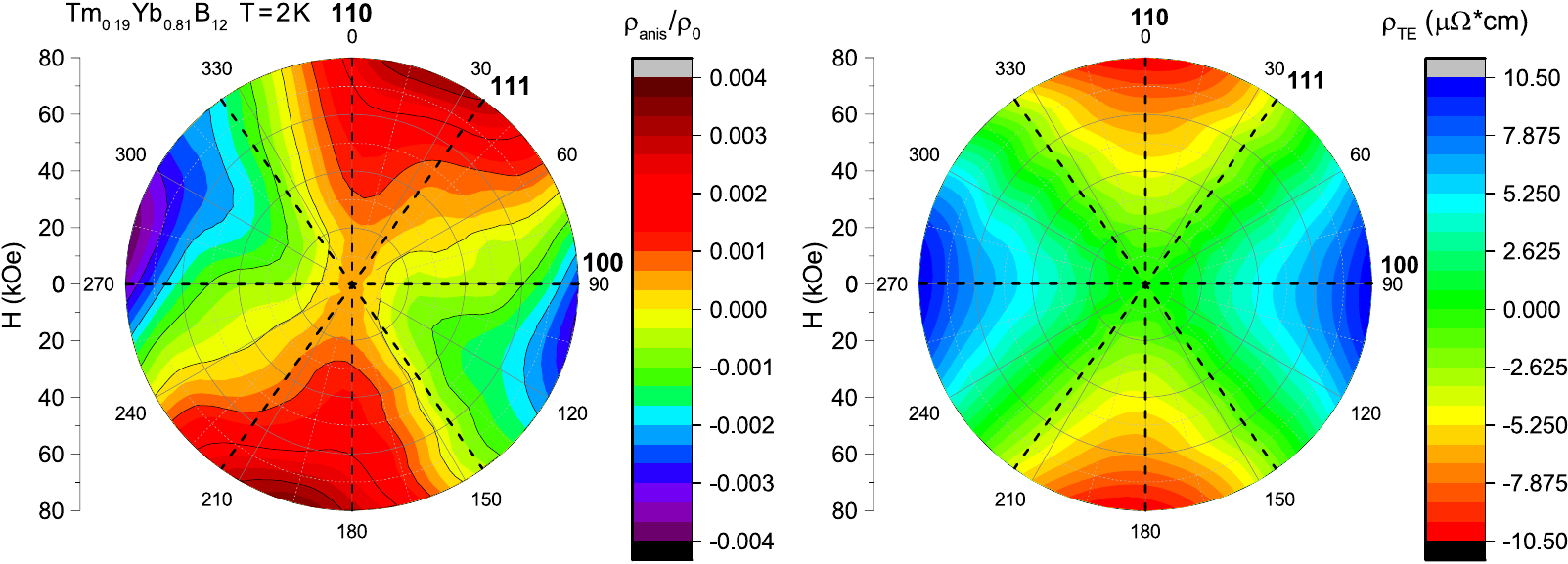}}
\caption{Transverse magnetoresistance $\rho_\text{anis}/\rho_0(\phi, H_0)$ (left) and the transverse even effect $\rho_\text{TE}(\phi, H_0)$ (right) for Tm$_{0.19}$Yb$_{0.81}$B$_{12}$ as a function of magnetic-field magnitude and direction in polar coordinates. Reproduced from Ref.~\cite{Slu_SluchankoAzarevich19}.\vspace{-10pt}}
\label{Slu:Fig42}\index{Tm$_{1-x}$Yb$_x$B$_{12}$!transverse magnetoresistance}\index{transverse magnetoresistance}\index{Tm$_{1-x}$Yb$_x$B$_{12}$!transverse even effect}
\end{figure}

\paragraph{Charge stripes and magnetotransport anisotropy}\index{Tm$_{1-x}$Yb$_x$B$_{12}$!magnetotransport anisotropy}\index{Tm$_{1-x}$Yb$_x$B$_{12}$!charge stripes}
To testify fine details of the crystal and electron structure, accurate x-ray diffraction\index{x-ray diffraction} experiments and structural data analysis were developed for single crystals of the semiconducting Tm$_{0.19}$Yb$_{0.81}$B$_{12}$ solid solutions~\cite{Slu_SluchankoAzarevich19}. Combined with the reconstruction of difference Fourier maps of residual electron density and applying the maximum entropy method to deduce the normal electron density, the approach allowed us to visualize the dynamic charge stripes\index{dynamic charge stripes}\index{charge stripes} oriented predominantly along a face diagonal of the unit cell in Tm$_{1-x}$Yb$_x$B$_{12}$ (for more details see Chapter~\ref{Chapter:Bolotina}). For single-domain crystals of Tm$_{0.19}$Yb$_{0.81}$B$_{12}$, measurements of the charge transport anisotropy at temperatures in a coherent regime (interval III in Fig.~\ref{Slu:Fig40}) were performed in a scheme corresponding to the transverse MR and Hall resistance, recorded as the crystal was rotated around the current axis $\mathbf{I}$ parallel to one of the $\langle110\rangle\perp\mathbf{H}$. Both the transverse even effect $\rho_\text{TE}(\phi, H_0)$ and the transverse MR have been separated (see Fig.~\ref{Slu:Fig42}), demonstrating uniaxial anisotropy with the anisotropy axis predominantly oriented along the $\mathbf{H} \parallel [110]$ direction~\cite{Slu_SluchankoAzarevich19}.

\begin{figure}[t!]
\centerline{\includegraphics[width=\textwidth]{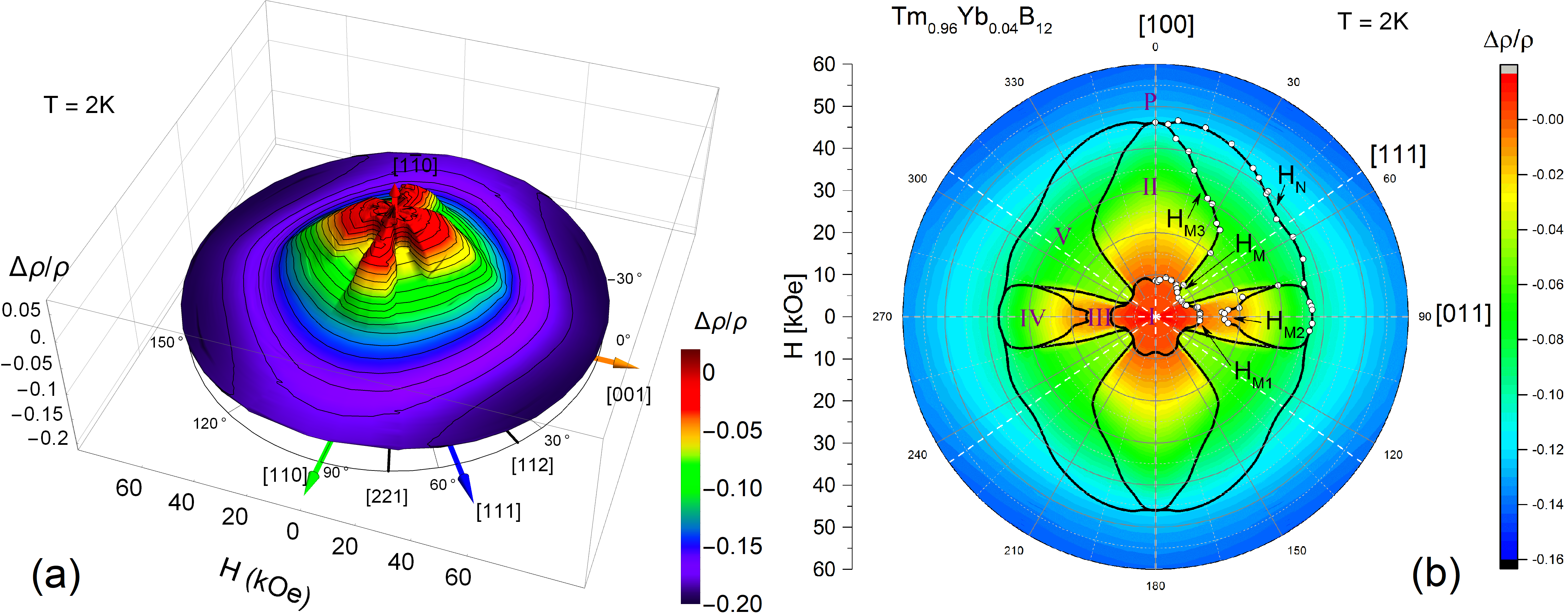}}
\caption{(a)~The magnetoresistance of Tm$_{0.96}$Yb$_{0.04}$B$_{12}$ as a function of magnetic field and its direction at $T=2$~K. The rotation axis is $\mathbf{I} \parallel [1\overline{1}0]$, three principal cubic directions in the $[1\overline{1}0]$ plane are shown with arrows. (b)~The $H$-$\phi$ magnetic phase diagram at $T = 2$~K. The phase boundaries are derived from the magnetoresistance measurements. Roman numerals denote various AFM phases. Reproduced from Ref.~\cite{Slu_AzarevichBogach}.}
\label{Slu:Fig43}\index{Tm$_{1-x}$Yb$_x$B$_{12}$!magneresistance anisotropy}\index{anisotropic magnetoresistance!in Tm$_{1-x}$Yb$_x$B$_{12}$}\index{field-angular anisotropy!in magnetoresistance}
\end{figure}

Besides, on the metal side of the MIT, single-domain crystals of Tm$_{0.96}$Yb$_{0.04}$B$_{12}$ with an AFM ground state ($T_{\rm N} \approx 2.6$~K~\cite{Slu_SluchankoAzarevich12, Slu_SluchankoAzarevich13}) have been studied in detail by low-temperature MR, magnetization, and heat capacity measurements. The angular $H-\phi-T_0$ AFM phase diagram in the form of a Maltese cross\index{Maltese cross anisotropy} has been deduced for this AFM metal from precise MR measurements, reproduced here in Fig.~\ref{Slu:Fig43}~\cite{Slu_AzarevichBogach}. The authors argued that the observed dramatic symmetry lowering is a consequence of strong renormalization of the indirect exchange interaction (RKKY mechanism)\index{RKKY interaction} due to the presence of dynamic charge stripes\index{dynamic charge stripes}\index{charge stripes} in the matrix of this AFM metal. They proposed that additionally to the development of the fcc lattice instability (cooperative dynamic JT effect\index{dynamic Jahn-Teller effect}\index{Jahn-Teller effect!cooperative} in the boron sub-lattice) and appearance of the nanometer-size electronic phase separation (dynamic charge stripes), the electron density fluctuations on Yb sites also play an essential role both in the suppression of the AFM state and in driving the changes in orientation of the stripes and the phase-diagram boundaries in Tm$_{1-x}$Yb$_x$B$_{12}$ antiferromagnets~\cite{Slu_AzarevichBogach, Slu_SluchankoAzarevich14a}.

\begin{figure}[t!]
\centerline{\includegraphics[width=\textwidth]{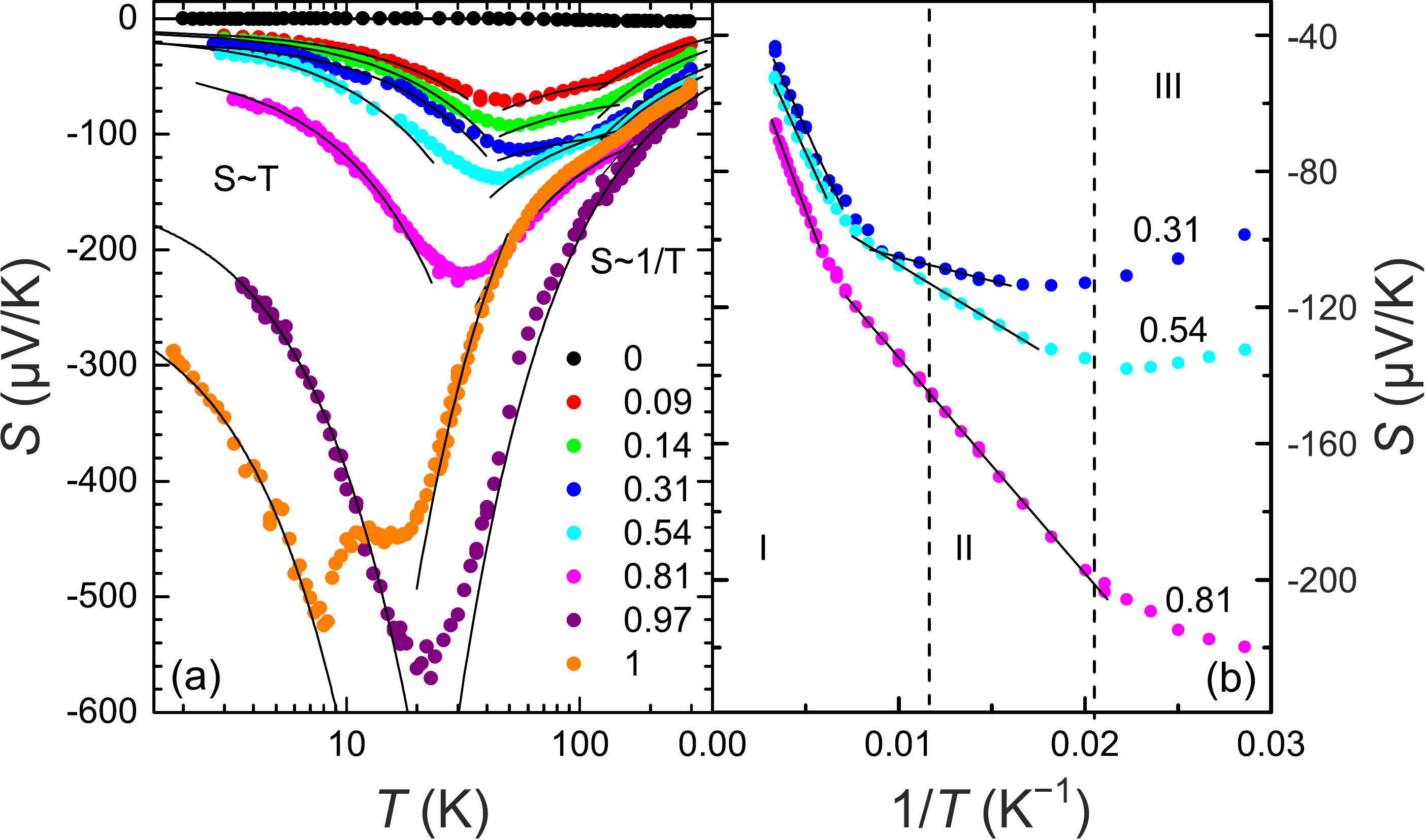}}
\caption{(a)~Temperature dependences of the Seebeck coefficient $S(T)$ of the Tm$_{1-x}$Yb$_x$B$_{12}$ solid solutions for the orientation of the temperature gradient ${\nabla T}\parallel [110]$. Solid lines show the approximations of the S($T$) curves by the activation ($T>50$~K, $S \propto 1/T$) and linear ($T<30$~K, $S(T) \propto A(x)T$) dependences. (b)~The $S(T)$ curves in the Arrhenius plot. The temperature ranges of the activation behavior of the thermopower are denoted as I and II. Adapted from Ref.~\cite{Slu_SluchankoBogach09}.}
\label{Slu:Fig44}\index{Tm$_{1-x}$Yb$_x$B$_{12}$!Seebeck coefficient}\index{Seebeck coefficient!in Tm$_{1-x}$Yb$_x$B$_{12}$}\index{thermopower}
\end{figure}

\paragraph{Seebeck coefficient}\index{Tm$_{1-x}$Yb$_x$B$_{12}$!Seebeck coefficient}\index{Seebeck coefficient!in Tm$_{1-x}$Yb$_x$B$_{12}$}
According to Ref.~\cite{Slu_SluchankoBogach09}, strong variations in the charge-transport characteristics of Tm$_{1-x}$Yb$_x$B$_{12}$ have been observed in the temperature dependences of the Seebeck coefficient $S(T)$. Indeed, the low magnitudes of the thermopower\index{thermopower} for TmB$_{12}$ ($|S(T)| < 3$~$\mu$V/K, see Figs.~\ref{Slu:Fig44} and \ref{Slu:Fig20}), typical of a compound with metallic conductivity, change to $|S(T)| > 15$~$\mu$V/K already for the Tm$_{0.96}$Yb$_{0.04}$B$_{12}$ solid solution. Further substitution of Yb for Tm increases the Seebeck coefficient magnitude up to 580~$\mu$V/K (Fig.~\ref{Slu:Fig44}), and the negative minimum on the $S(T)$ curves shifts down in temperature with increasing $x$. The authors \cite{Slu_SluchankoBogach09} noted that the $S(T)$ dependences for all solid solutions with $x > 0.01$ in the intermediate temperature range $T \geq 50$~K are characterized by a complex activation behavior (Fig.~\ref{Slu:Fig44}) and $S(1/T)$ representation allows estimating both the gap $E_\text{g}/k_\text{B} = 100$--160~K (interval~I) and the binding energy of intra-gap many-body states $E_\text{p}$ (interval II)~\cite{Slu_SluchankoBogach09}. It is worth noting that the estimated values $E_{\rm g}/k_{\rm B} = 100$--160~K are close to the estimate of the gap value found for YbB$_{12}$ in the measurements of the Hall effect and resistivity (180 and 134~K, respectively \cite{Slu_IgaShimizu98}) as well as NMR on the Yb ions and specific heat ($\sim$170~K)~\cite{Slu_IkushimaKato00, Slu_IgaHiura99}. They are also comparable to the spin-gap value $\sim$14~meV determined by neutron spectroscopy~\cite{Slu_MignotAlekseev05, Slu_NemkovskiMignot07}\index{inelastic neutron scattering} and EPR~\cite{Slu_AltshulerBresler02}. With a decrease below 30~K, the $S(T)$ curves approach about linear asymptotic behavior (Fig.~\ref{Slu:Fig44}), which points to a metallic conductivity via the band of many-body states in Tm$_{1-x}$Yb$_x$B$_{12}$. It has been argued~\cite{Slu_SluchankoBogach09} that a considerable increase (about 5 times) in the $A(x)$ slope in the $S(T) \propto A(x)T$ curves when Yb is substituted for Tm in the Tm$_{1-x}$Yb$_x$B$_{12}$ series may be attributed to the Mott term (diffusion thermopower)\index{thermopower}\index{Mott thermopower}\index{diffusive thermopower} that points to the DOS renormalization near $E_\text{F}$. Similar effects were detected in the investigation of the electronic specific heat $C/T = \gamma(T)$ of the Yb$_x$Lu$_{1-x}$B$_{12}$ compounds in Ref.~\cite{Slu_IgaKasaya88}, where an increase by about an order of magnitude in the Sommerfeld coefficient has been observed upon varying the Yb content between 10 and 80~\mbox{at.\,\%}.

\paragraph{Magnetic properties}\index{Tm$_{1-x}$Yb$_x$B$_{12}$!magnetic properties}
In the temperature range 40--300~K, Tm$_{1-x}$Yb$_x$B$_{12}$ compounds with $x < 0.9$ are paramagnets whose magnetic susceptibility $\chi(T)$ is well described by the Curie-Weiss law with a negative paramagnetic Curie temperature $\Theta_{\rm p}$ corresponding to the AFM exchange interaction~\cite{Slu_SluchankoBogach09}. The estimation of the effective magnetic moment $\mu_\text{eff}$ per unit cell was done by considering the additive contributions from the localized magnetic moments of the Tm$^{3+}$ ($\mu$(Tm)~=~$7.5\mu_{\rm B}$) and Yb$^{3+}$ ($\mu$(Yb)~=~$4.5\mu_{\rm B}$) ions present in the Tm$_{1-x}$Yb$_x$B$_{12}$ solid solutions. Analysis of $\chi(T)$ dependences~\cite{Slu_SluchankoBogach09, Slu_BogachSluchanko13} indicates that $\Theta_{\rm p}(x)$ increases monotonically, and in addition to the Curie-Weiss contribution, a magnetic-fluctuation component appears in the low-temperature magnetic response of Tm$_{1-x}$Yb$_x$B$_{12}$. Both to reconstruct the $H$-$T$-$x$ magnetic phase diagrams and to clarify the nature of the additional paramagnetic response, the authors~\cite{Slu_BogachSluchanko13} have investigated the permanent (up to 11~T) and pulsed (up to 50~T) field magnetization of the Tm$_{1-x}$Yb$_x$B$_{12}$ compounds in the composition range $0 < x \leq 0.81$. It was found that at low temperatures there are two paramagnetic contributions including a Pauli-type component, which corresponds to the response of the heavy-fermion many-body states that appear in the energy gap in the vicinity of the Fermi level (DOS of (3--4)$\cdot$10$^{21}$ cm$^{-3}$meV$^{-1}$), and a contribution which saturates in high magnetic fields and may be attributed to the localized magnetic moments (0.8--3.7~$\mu_{\rm B}$ per unit cell) of the nanosize clusters of Tm(Yb) ions with AFM exchange (short-range order nanodomains)~\cite{Slu_BogachSluchanko13}. The last finding may be interpreted in terms of the Griffiths-phase formation \cite{Slu_Bray87} and the spin-polaron effect, which are the two dominant factors responsible for the low-temperature changes in the magnetic properties of Tm$_{1-x}$Yb$_x$B$_{12}$.\vspace{-1pt}

\begin{figure}[t]\vspace{-2pt}
\centerline{\includegraphics[width=0.85\textwidth]{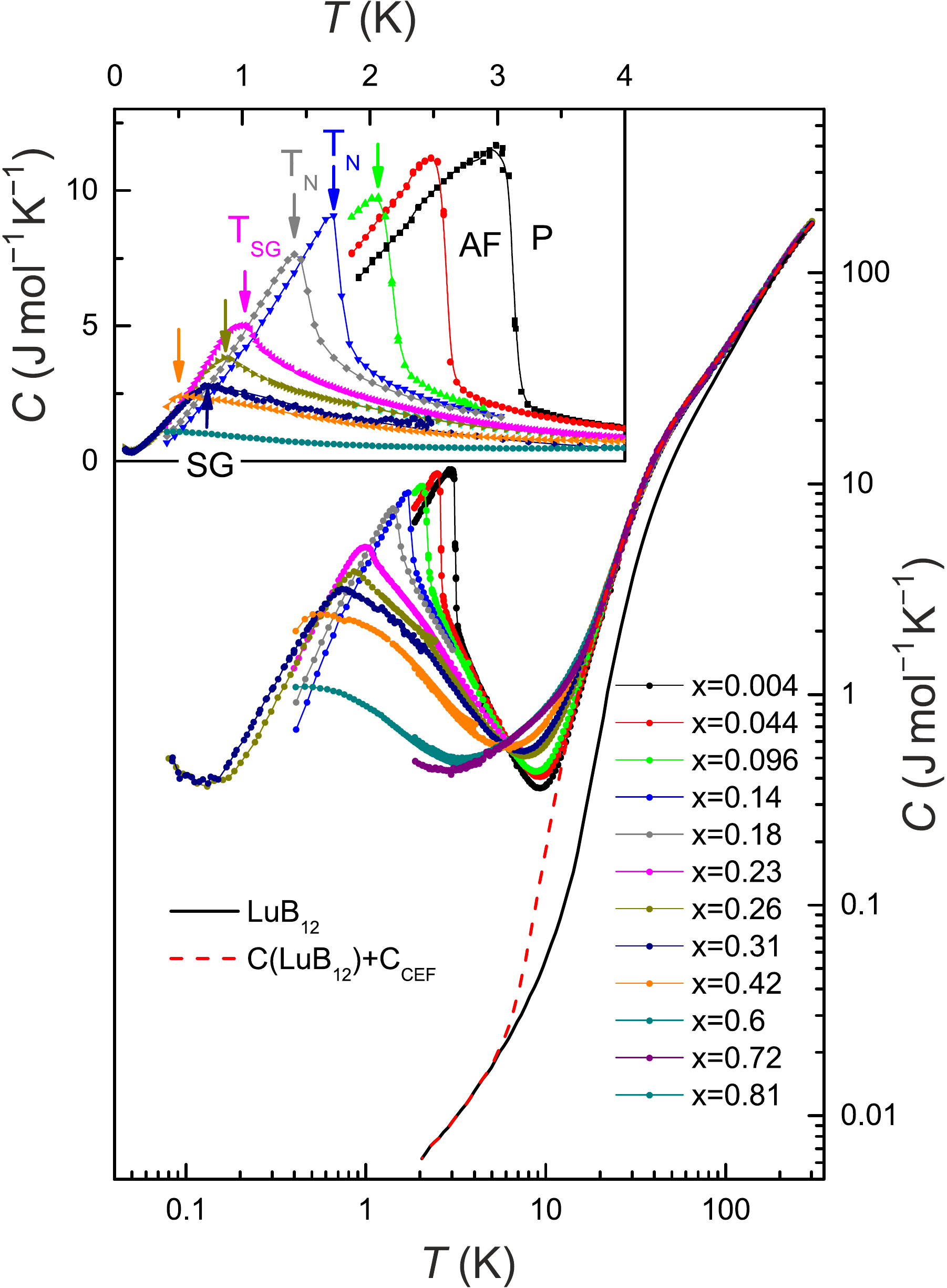}}\vspace{-2pt}
\caption{Temperature dependence of the specific heat for Tm$_{1-x}$Yb$_x$B$_{12}$, nonmagnetic reference compound LuB$_{12}$ (solid line), and the sum of the magnetic contribution from Tm$^{3+}$ ions ($C_\text{CEF}$) and specific heat of LuB$_{12}$ (dashed line) \cite{Slu_Azarevich14}. The inset depicts the low-temperature range corresponding to the magnetic phase transitions (indicated by arrows). The antiferromagnetic, paramagnetic, and spin glass phases are denoted as AF, P, and SG, respectively \cite{Slu_SluchankoDudka18}.}
\label{Slu:Fig45}
\end{figure}

\begin{figure}[t]
\centerline{\includegraphics[width=0.8\textwidth]{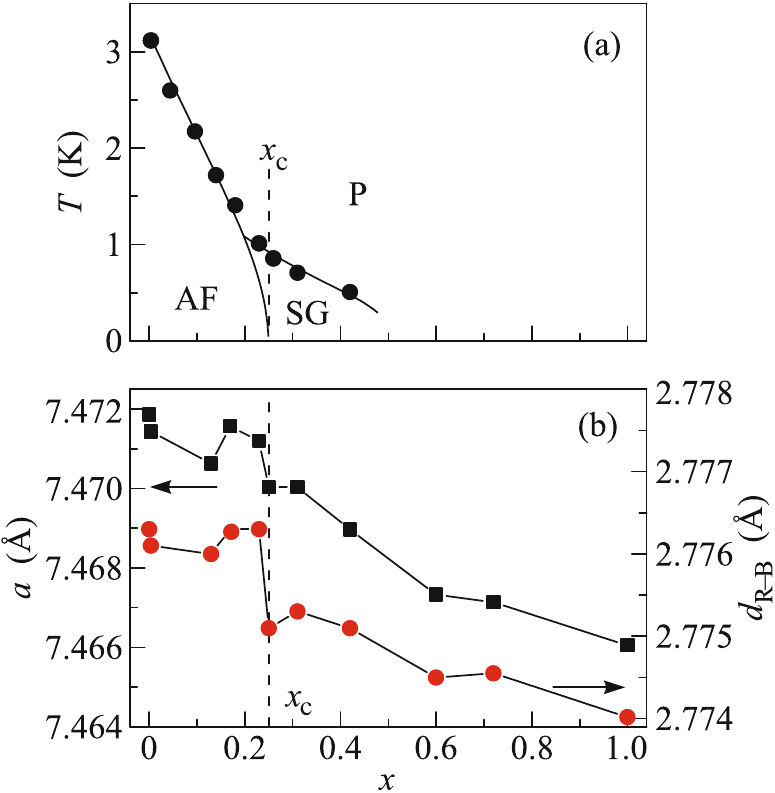}}
\caption{(a)~The magnetic phase diagram in the $T$-$x$ plane with the QCP at $x_\text{c} \approx 0.25$. (b)~Average lattice parameter for the cubic unit cell and boron-cation distance versus the ytterbium content for Tm$_{1-x}$Yb$_x$B$_{12}$ solid solutions. Reproduced from Ref.~\cite{Slu_SluchankoDudka18}.}
\label{Slu:Fig46}
\end{figure}

\paragraph{Heat capacity}\index{Tm$_{1-x}$Yb$_x$B$_{12}$!specific heat}\index{heat capacity!in Tm$_{1-x}$Yb$_x$B$_{12}$}\index{specific heat!in Tm$_{1-x}$Yb$_x$B$_{12}$}
For the refinement of the magnetic $T-x$ phase diagram of Tm$_{1-x}$Yb$_x$B$_{12}$, the temperature dependences of the specific heat have been investigated in detail, as shown in Fig.~\ref{Slu:Fig45} \cite{Slu_SluchankoDudka18}. Surprisingly, the log--log plots nearly coincide for different Yb concentrations. These plots exhibit a steep decrease below 40~K, which is fitted with a good accuracy by the Einstein model for the specific heat related to the nearly universal quasilocal vibrations of both Yb and Tm ions in the rigid boron cage~\cite{Slu_SluchankoBogach10}. Near $T_\text{N}$, the $C(T)$ curves exhibit a sharp increase related to the formation of the long-range AFM order~\cite{Slu_SluchankoBogach10}, and with the growth of the Yb content in Tm$_{1-x}$Yb$_x$B$_{12}$ the step-like $C(T)$ anomaly changes to a peak for $x > x_{\rm c} \approx 0.25$. Taking into account that at low temperatures the Yb ion becomes nonmagnetic, we have concluded \cite{Slu_SluchankoDudka18} that the substitution of ytterbium for magnetic Tm$^{3+}$ results in disordering and the formation of a mictomagnetic state (spin glass). Thus, it was argued \cite{Slu_SluchankoDudka18} that the spin glass phase exists above the AF-P transition, and a hidden QCP appears in the $T-x$ plane of Tm$_{1-x}$Yb$_x$B$_{12}$ compounds, see Fig.~\ref{Slu:Fig46}\,(a).

Then, an isosbestic point was observed at $T^{\ast} \approx 6$~K on the $C$($T$, $H = 0$) curves \cite{Slu_SluchankoDudka18, Slu_Azarevich14} (Fig.~\ref{Slu:Fig45}), and it was found that the temperature dependence of the specific heat for the entire composition range $0.004 \leq x \leq 0.81$ linearly scales with respect to the Yb concentration. It was shown that at low temperatures a linear in $x$ decrease in $C(T_0,x)$ should be attributed to the formation of a charge gap, which leads to a decrease in the electron DOS at $E_\text{F}$. Within the proposed approach, the magnetic contribution to the specific heat of Tm$_{1-x}$Yb$_x$B$_{12}$, consisting of several additive components, can be distinguished~\cite{Slu_Azarevich14}. It was shown that for all Tm$_{1-x}$Yb$_x$B$_{12}$ compositions, the magneto-vibronic contribution to the heat capacity with a maximum near $T^{\ast} \approx 60$~K, which is practically independent of the Yb concentration, is dominant (Fig.~\ref{Slu:Fig45})~\cite{Slu_Azarevich14}. An analysis of the low-temperature contributions to the specific heat allows one to distinguish the Zeeman component from Tm$^{3+}$ ions with the triplet ground state $\Gamma_5$ of the $^3H_6$ multiplet, as well as to find the $g$-factors ($g_1 \approx 2.5$, $g_2 \approx 5$) and their changes in the series Tm$_{1-x}$Yb$_x$B$_{12}$. It was found that one more contribution to $C(T)$ with a maximum near $T\approx10$~K is independent of the magnetic field and apparently should be attributed to the effects of a nanosize clustering of Yb ions~\cite{Slu_Azarevich14}.

\begin{figure}[t!]\vspace{-3pt}
\centerline{\includegraphics[width=0.85\textwidth]{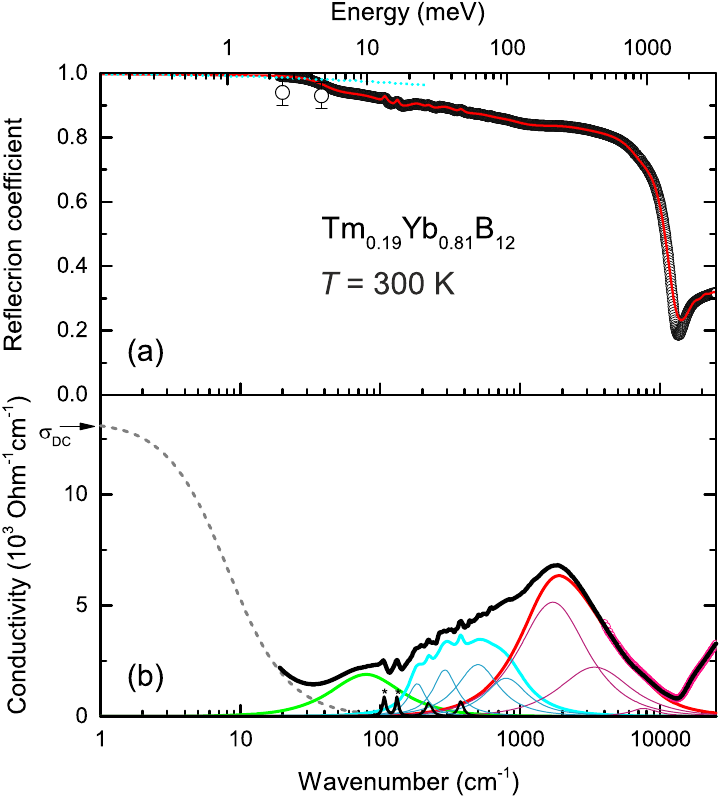}}\vspace{-2pt}
\caption{(a)~Room-temperature spectra of the reflection coefficient for a Tm$_{0.19}$Yb$_{0.81}$B$_{12}$ single crystal. The data points at 20 and 38~cm$^{-1}$ show measurements with the backward-wave oscillator spectrometer. The dotted line is the Hagen-Rubens reflectivity $R=1-\sqrt{4\nu/\sigma_\text{DC}}$ calculated with the measured $\sigma_\text{DC} = 13\,000$~$\Omega^{-1}$cm$^{-1}$. The solid line is the result of fitting the spectrum with a set of Lorentzians. Corresponding excitations are shown separately in panel~(b) together with the optical conductivity spectrum (black solid line) obtained by Kramers-Kronig analysis of the reflectivity. Red dots above 4000~cm$^{-1}$ correspond to a direct measurement on an ellipsometer. The dashed line models the mismatch between the THz-FIR and DC conductivity with the Drude expression $\sigma(v)=\sigma_\text{DC}(1-i\nu/\gamma)^{-1}$. Asterisks denote quasi-local vibrations of heavy $R$ ions at 107 and 132~cm$^{-1}$ with the damping $\gamma \approx 8$~cm$^{-1}$. Reproduced from Ref.~\cite{Slu_SluchankoAzarevich19}.\vspace{-3pt}}
\label{Slu:Fig47}\index{Tm$_{1-x}$Yb$_x$B$_{12}$!optical properties!reflection coefficient}\index{Tm$_{1-x}$Yb$_x$B$_{12}$!optical properties!optical conductivity}
\end{figure}

\paragraph{Optical spectra}\index{Tm$_{1-x}$Yb$_x$B$_{12}$!optical properties}
To shed more light on the characteristics of the dynamic charge stripes\index{dynamic charge stripes}\index{charge stripes} detected in the x-ray diffraction\index{x-ray diffraction} experiment on the semiconducting Tm$_{0.19}$Yb$_{0.81}$B$_{12}$ solid solution (see Chapter~\ref{Chapter:Bolotina}), broadband reflectivity measurements have been carried out at room temperature~\cite{Slu_SluchankoAzarevich19}. Then, using the Kramers-Kronig analysis, the spectrum of dynamical conductivity was calculated, which is shown here in Fig.~\ref{Slu:Fig47}. Among the most important results, two issues have been mentioned in Ref.~\cite{Slu_SluchankoAzarevich19}: (i)~the conductivity at terahertz frequencies, $\sigma$(30--40~cm$^{-1}$)~$\approx$~1500~$\Omega^{-1}$cm$^{-1}$, is about an order of magnitude below the measured DC conductivity $\sigma_\text{DC} \approx 13\,000~\Omega^{-1}$cm$^{-1}$ and (ii)~modeling the mismatch with the Drude conductivity term (dashed line in Fig.~\ref{Slu:Fig47}) provides the scattering rate of carriers $\gamma \approx 8$~cm$^{-1}$, which coincides very well with the damping of two quasilocal vibrations of the heavy $R$ ions (Tm and Yb) located at 107~cm$^{-1}$ ($\sim$154~K) and 132~cm$^{-1}$ ($\sim$190~K), marked by two asterisks in Fig.~\ref{Slu:Fig47}\,(b). For comparison, similar values of the Einstein temperature $\Theta_\text{E} = 160$--206~K have been detected previously in EXAFS~\cite{Slu_MenushenkovYaroslavtsev13},\index{EXAFS} heat capacity~\cite{Slu_SluchankoAzarevich11} and INS~\cite{Slu_RybinaNemkovski10} studies of the dodecaborides.

Taking into account that these rattling modes\index{rattling phonon mode} induce a variation in the $5d$-$2p$ hybridization in $R$B$_{12}$ \cite{Slu_SluchankoBogach18}, the authors \cite{Slu_SluchankoAzarevich19} have suggested that this kind of ``modulation'' of the conduction band is responsible for the dynamic charge stripes\index{dynamic charge stripes}\index{charge stripes} formation in the dodecaboride matrix. Besides, the frequency $\sim$2.4$\cdot$10$^{11}$~Hz (8~cm$^{-1}$) can be deduced from the conductivity spectra as the characteristic of the quantum motion of charges in the dynamic stripes. As a result, the room-temperature DC conductivity in Tm$_{0.19}$Yb$_{0.81}$B$_{12}$ should be attributed to the charge transport in dynamic stripes that are more conductive than the surrounding semiconducting matrix. Consequently, the value of the DC conductivity is determined by the stripes that percolate through the crystal while the relatively smaller ac conductivity is provided by the THz-FIR reflectivity of the ``whole'' sample (conducting stripes + semiconducting matrix).

\paragraph{Quantum critical behavior}\index{Tm$_{1-x}$Yb$_x$B$_{12}$!quantum criticality}\index{quantum criticality}
It has been predicted \cite{Slu_KoppChakravarty05, Slu_Lonzarich05} that the AFM instability arising near the QCP at $T = 0$ should substantially modify the characteristics of strongly correlated electron systems\index{strongly correlated electrons} in a wide temperature range, including room temperature. To examine these predictions, we have studied fine details of the crystal structure of the system Tm$_{1-x}$Yb$_x$B$_{12}$ in a wide vicinity of the QCP at $x_{\rm c} \approx 0.25$~\cite{Slu_SluchankoDudka18}. It was found that in addition to the decrease in the lattice constant $a$ upon substitution of ytterbium for thulium (lanthanide compression), the $a(x)$ plot near the QCP at $x_{\rm c} \approx 0.25$ exhibits an anomaly, which evidently results from the abrupt decrease in the boron-cation distance near $x_{\rm c}$, see Fig.~\ref{Slu:Fig46}\,(b). The high accuracy in determining the structural parameters allowed the authors to determine also the variation in the distance between boron atoms within the B$_{12}$ clusters, $r_{\text{(B-B)\,intra}}$, and between the neighboring B$_{12}$ clusters, r$_{\text{(B-B)\,inter}}$. In Fig.~\ref{Slu:Fig39}\,(a), it is seen that the distance between the B$_{12}$ clusters in Tm$_{1-x}$Yb$_x$B$_{12}$ with $x$ near $x_{\rm c}$ decreases stepwise, whereas the size of the B$_{12}$ clusters remains the same. In contrast, above the QCP, the distance between the boron clusters tends to a constant value, whereas their size decreases gradually in the composition range of $x = 0.25$--0.7. It has been emphasized in Ref.~\cite{Slu_SluchankoDudka18} that the amplitude of changes in the structural parameters $a$, $d_{\text{\textit{R}-B}}$, $r_{\text{(B-B)\,intra}}$, and $r_{\text{(B-B)\,inter}}$ is too small to explain the nature of the MIT in the composition range $x > 0.5$, so, a pronounced (by a factor of about~7) increase in resistivity in the Tm$_{1-x}$Yb$_x$B$_{12}$ series at $T=270$~K (Fig.~\ref{Slu:Fig38}) should be attributed to the development of the fcc lattice instability in combination with the unstable electron configuration of the~Yb~ion.

To test the quantum-critical behavior in the Tm$_{1-x}$Yb$_x$B$_{12}$ series near $x_{\rm c} \approx 0.25$ at low temperatures, we have also studied the temperature dependence of the magnetic contribution to the specific heat $C_\text{m}$ of the Tm$_{0.74}$Yb$_{0.26}$B$_{12}$ single crystal~\cite{Slu_SluchankoAzarevich10a, Slu_SluchankoBogach10}. A logarithmic divergence of the renormalized Sommerfeld coefficient $C_\text{m}/T \propto -\ln T$ was found at the temperature below 4~K which is typical for a system at QCP~\cite{Slu_Loehneysen96}, and it is attributed usually to the dramatic renormalization of the quasiparticles' effective mass and to the issue of a non-Fermi-liquid behavior of heat capacity. It was noted in Refs.~\cite{Slu_SluchankoAzarevich10a, Slu_SluchankoBogach10} that the magnetic field $H>30$~kOe suppresses completely the quantum critical regime, and at low temperatures the specific-heat curves $C_{\rm m}(T)$ demonstrate the Schottky anomaly.\index{Schottky specific heat}\vspace{-5pt}
\index{Tm$_{1-x}$Yb$_x$B$_{12}$!metal-insulator transition|)}\index{metal-insulator transition|)}

\vspace{-3pt}\section{Conclusions}\vspace{-3pt}\enlargethispage{1pt}
\label{Sec:Slu_Conclusions}

Until recently, it was commonly believed that the filling of the $4f$ shell in the RE dodecaborides $R$B$_{12}$ ($R$~=~Tb, Dy, Ho, Er, Tm, Yb and Lu) in the range from TbB$_{12}$ ($n_{4f} = 8$) to LuB$_{12}$ ($n_{4f} = 14$) has only a single singularity at YbB$_{12}$ due to the Yb-ion valence instability ($n_{4f} \approx 13.05$). The results presented in this chapter clearly demonstrate the complexity of all the $R$B$_{12}$ compounds including the AFM (TbB$_{12}$--TmB$_{12}$) and nonmagnetic (LuB$_{12}$) metals.

The reason for this complexity is the cooperative dynamic JT instability\index{dynamic Jahn-Teller effect} of the rigid boron covalent network, which produces trigonal distortions and results in the symmetry lowering of the fcc lattice. The ferrodistortive dynamics in the boron sublattice (see also Chapter~\ref{Chapter:Bolotina}) generates both the collective modes (overdamped oscillators in the frequency range 250--1000~cm$^{-1}$) that can be seen in the dynamic conductivity spectra of all the metallic $R$B$_{12}$ compounds~\cite{Slu_GorshunovPrivate} and the rattling modes\,---\,quasilocal vibrations\index{rattling phonon mode} of the heavy RE ions inside the oversized B$_{24}$ cavities (the radius $\sim$1.2~\r{A} of the B$_{24}$ complex considerably exceeds the ionic radius of the RE ions $\sim$0.9--0.95~\r{A}). The large amplitude displacements of the $R$-ions (Einstein oscillators, $k_\text{B}\Theta_\text{E} = 14$--18~meV) immediately cause (i)~the development of vibrational instability at intermediate temperatures $T_\text{E} \approx 150~K$ on reaching the Ioffe-Regel limit\index{Ioffe-Regel limit} and (ii)~strong changes in the hybridization of the $R$~$5d$ and B~$2p$ atomic orbitals by varying their overlap. Accordingly, these overlap oscillations along the $[110]$ axis (the direction of the shortest $R$-$R$ distance in the fcc lattice) are responsible for the modulation of the conduction band constructed from these $R$~$5d$ and B~$2p$ electron states. The modulation of the conduction-electron density with the frequency $\sim$2$\cdot$10$^{11}$~Hz~\cite{Slu_SluchankoAzarevich19} should be discussed in terms of the emergence of dynamic charge stripes\index{dynamic charge stripes}\index{charge stripes} which are the feature of a nanoscale electron instability and electronic phase separation. These nonequilibrium charge carriers dominate in the RE dodecaborides, taking 50--70\% from the total number of the conduction electrons, and these \textit{hot electrons} can no longer participate in the indirect exchange (RKKY interaction)\index{RKKY interaction} between the RE magnetic moments which are located at the distance of $\sim$5.3~\r{A} from each other. As a result, the suppression of the magnetic exchange interaction along the [110] direction of the nanosize filamentary structure should be considered as the main consequence of the electron instability, providing a magnetic symmetry lowering and leading to the field-angular phase diagrams of the $R$B$_{12}$ antiferromagnets in the form of the Maltese cross.\index{Maltese cross anisotropy} Because of the loosely bound state of RE ions in the rigid boron cage, there is also an order-disorder phase transition to the cage-glass\index{cage glass} state at the temperature $T^{\ast} \approx 60$~K resulting in the random displacements of the $R$-ions from their positions in the fcc lattice. The disordering induces immediately a formation of the nanosize clusters of RE ions (couples, triples, etc.) with AFM exchange, and these nanodomains should be apparently taken into account to explain the nature of the short-range magnetic order detected in the $R$B$_{12}$ antiferromagnets at temperatures far above $T_\text{N}$. It was certainly established that these AFM nanosize clusters, for example, in HoB$_{12}$ have a cigarlike form, elongated parallel to the $\langle111\rangle$ axis~\cite{Slu_SiemensmeyerHabicht07}.

The situation becomes much more complicated in the dodecaborides with Yb ions. Indeed, in this case, in addition to the JT instability of the boron cage, the instability of Yb~$4f$-electron configuration appears, providing one more mechanism of the charge and spin fluctuations in $R$B$_{12}$. When this additional charge degree of freedom is switched on, a strong and monotonic shift of the collective modes to high frequencies is observed in the Yb$_x$Lu$_{1-x}$B$_{12}$ series, until the two Lorentzians appear in the dynamic conductivity spectra of YbB$_{12}$ at 40~meV ($\sim$320~cm$^{-1}$) and 250~meV ($\sim$2000~cm$^{-1}$)~\cite{Slu_OkamuraMichizawa05}.

An appearance of Yb-Yb pairs with the temperature lowering can be probably considered the main factor which is responsible for the charge- and spin-gap formation in YbB$_{12}$, and the many-body states are characterized by the localization radius $\sim$5~\r{A}~\cite{Slu_SluchankoAzarevich12, Slu_GorshunovHaas06}, which is about equal to the Yb-Yb distance in the fcc lattice. Taking into account that this indirect gap opening is a phonon-assisted process~\cite{Slu_GorshunovHaas06, Slu_OkamuraMichizawa05}, and there is a coupling between the intensities of the anomalous acoustic phonons\index{acoustic phonons} and the M1, M2 magnetic excitations in YbB$_{12}$~\cite{Slu_AlekseevMignot12}, one has to conclude that the Yb dimers should be vibrationally coupled complexes. The process of the Yb pair formation develops gradually below the temperature $T^{\ast} \approx 60$~K, in the cage-glass\index{cage glass} state where static displacements of Yb ions occur in the double-well potentials~\cite{Slu_SluchankoAzarevich12}.\index{double-well potential}

It is worth emphasizing here that (i)~the mechanism underlying the formation of heavy fermions in the vicinity of Yb ions is different from the Kondo one, (ii)~the localization radius for these many-body states (heavy fermions) turns out to be nearly equal to the $R$-$R$ distance ($\sim$5~\r{A}), being much smaller than that of the ``Kondo cloud'' ($>20$~\r{A}), and (iii)~the strong local spin fluctuations near the RE ions lead to the formation of both \emph{heavy fermions} and the \emph{spin-polarized nanodomains}~\cite{Slu_KaganKugel01, Slu_Nagaev67}\index{spin-polarized nanodomains} which represent the $5d$ component of the magnetic structure in the $R$B$_{12}$ antiferromagnets~\cite{Slu_KhoroshilovKrasnorussky19, Slu_SluchankoKhoroshilov15}. Taking into account that the anisotropic Kondo Hamiltonian is a special case of the ``spin-boson'' Hamiltonian describing the properties of a dissipative two-level system~\cite{Slu_LeggettChakravarty87}, it is natural to argue~\cite{Slu_Sluchanko15} that the origin of spin fluctuations in YbB$_{12}$ is related to the quantum oscillations of a heavy particle (magnetic ion) between the states in the DWP,\index{double-well potential} and the Kondo physics is invalid in this case.

\begin{figure}[t!]\vspace{-3pt}
\centerline{\includegraphics[width=0.8\textwidth]{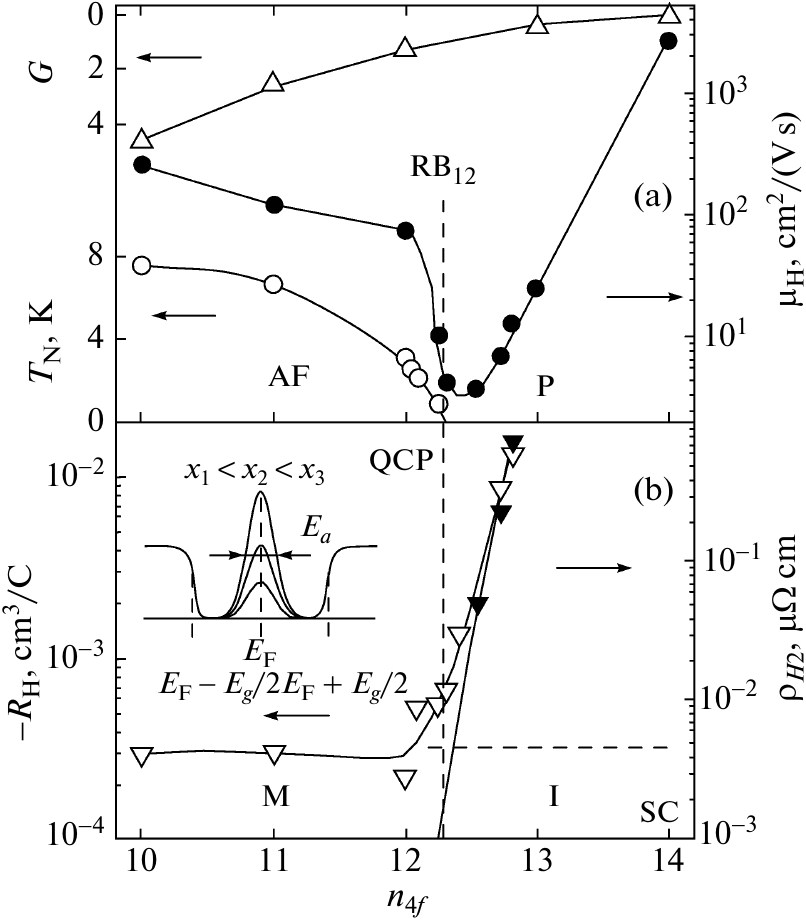}}
\caption{(a)~The de~Gennes factor $G = (g_J - 1)^2\cdot J(J + 1)$, the Hall mobility $\mu_{\rm H}(n_{4f}$) at the LHe temperature (the data for HoB$_{12}$ ($n_{4f} = 10$), ErB$_{12}$ ($n_{4f}$ = 11), TmB$_{12}$ ($n_{4f}$ = 12), and LuB$_{12}$ ($n_{4f}$ = 14) are from Refs.~\cite{Slu_SluchankoBogach09, Slu_SluchankoBogomolov06, Slu_SluchankoAzarevich12}, and the data for YbB$_{12}$ ($n_{4f} \approx 13.05$) are from Refs.~\cite{Slu_GorshunovHaas06, Slu_GorshunovProkhorov06}) and the N\'{e}el temperature $T_\text{N}(n_{4f})$ versus the filling of the $4f$ shell of the RE ions in the $R$B$_{12}$ compounds; AF and P denote the antiferromagnetic and paramagnetic phases; M, I, and SC stand for the metal, insulator, and superconductor. (b)~The Hall coefficient $R_\text{H}(n_{4f})$ and the even harmonic contribution $\rho_\text{H2}(n_{4f})$ for $H = 15$ and 80~kOe. The inset shows the renormalization effect in the intra-gap density of states in the series of Tm$_{1-x}$Yb$_x$B$_{12}$ compounds. Reproduced from Ref.~\cite{Slu_SluchankoAzarevich12}.}
\label{Slu:Fig48}\index{rare-earth dodecaborides!de Gennes factor}\index{de Gennes factor}\index{rare-earth dodecaborides!charge transport!Hall mobility}\index{rare-earth dodecaborides!charge transport!Hall coefficient}\index{rare-earth dodecaborides!transverse even effect}\index{even harmonic contribution}\index{transverse even effect}
\end{figure}

In the framework of the common scenario for the $R$B$_{12}$ family, the symmetry lowering which develops at low temperatures, $T < 30$~K, should be attributed to the formation in YbB$_{12}$ of a filamentary structure of conduction channels\,---\,dynamic charge stripes.\index{dynamic charge stripes}\index{charge stripes}\index{conduction channels} The intra-gap states with a binding energy of 2.7--5~meV are characterized by a localization radius $a_{\rm p} \approx 9$~\r{A}~\cite{Slu_SluchankoAzarevich12, Slu_AltshulerGoryunov03}. The $a_\text{p}$ value exceeds the lattice constant ($\sim$7.5~\r{A}), and the arrangement of both the Yb-pairs and the charge stripes leads to the \textit{spontaneous selection of a special direction} in the formally cubic crystal. Thus, the coherent regime of charge transport can be attributed to the percolation through the network of the many-body states and charge stripes in YbB$_{12}$ and Yb-based dodecaborides~\cite{Slu_SluchankoAzarevich12}.

To summarize, when discussing the mechanisms responsible for the strong charge-carrier scattering in the $R$B$_{12}$ compounds, it is worth emphasizing that the local on-site spin fluctuations play the dominant role in the charge transport of all magnetic RE dodecaborides. Indeed, when $n_{4f}$ increases from HoB$_{12}$ to LuB$_{12}$ in the range $10 \leq n_{4f} \leq 14$, one should expect a monotonic increase in the mobility of charge carriers $\mu_{\rm H}(n_{4f}$), which follows from a decrease in the de~Gennes factor $G = (g_J - 1)^2\cdot J(J + 1)$\index{rare-earth dodecaborides!de Gennes factor}\index{de Gennes factor} characterizing the magnetic scattering intensity in the $R$B$_{12}$ series. Instead, below the QCP in the range $10 \leq n_{4f} \leq 12.3$, a decrease in $\mu_{\rm H}(n_{4f}$) is observed (Fig.~\ref{Slu:Fig48}), and the mobility minimum near $x = 0.3$--0.5 in Tm$_{1-x}$Yb$_x$B$_{12}$ corresponds to the antiferromagnetic-paramagnetic transition with the QCP near $x_\text{c} \approx 0.25$. Furthermore, the substitution of Yb for Tm in the range $x \geq 0.5$ ($n_{4f} \geq 12.5$ in Fig.~\ref{Slu:Fig48}) leads to an increase in $\mu_\text{H}(x)$, while a monotonic and sharp increase in the resistivity $\rho(x)$ (Fig.~\ref{Slu:Fig38}) and the Hall coefficient $R_{\rm H}(x)$ (Fig.~\ref{Slu:Fig48}) at low temperatures is observed in the whole range $0 < x < 1$~\cite{Slu_SluchankoAzarevich12}. The metal-insulator transition is accompanied by rather low mobility values $\mu_\text{H}(T) \approx 27$~cm$^2$/(V$\cdot$s) in YbB$_{12}$, which are obviously caused by the scattering of carriers on strong spin and charge $4f$-$5d$ fluctuations. Evidently, these fast fluctuations of electron density should be discussed among the factors responsible for the suppression of the AFM ground state in the Yb-based dodecaborides. And finally, in passing to the nonmagnetic superconducting metal LuB$_{12}$ ($n_{4f} = 14$), the Hall mobility increases strongly ($\sim$100 times, see Fig.~\ref{Slu:Fig48}), but the nonmagnetic reference compound also demonstrates the same kind of structural and electron instabilities.\vspace{-2pt}

\section*{Acknowledgments}
\addcontentsline{toc}{section}{Acknowledgments}

This work was supported by the Russian Science Foundation, project No.~17-12-01426. The author is grateful to S.~Demishev, V.~Glushkov, K.~Flachbart, G.~Grechnev, N.~Shitsevalova, B.~Gorshunov, S.~Gabani, K.~Siemenmeyer, N.~Bolotina, A.~Dudka, V.~Mironov, T.~Mori, and V.~Moshchalkov for helpful discussions and to K.~Krasikov for his assistance in preparing the manuscript. Many of the presented results would be impossible without high-quality
samples provided by N.~Shitsevalova and V.~Filipov.

%BIBLIOGRAPHY
%This is the right way to include the bibliography to ensure every chapter has its own one. Requires bibunits package.

\putbib[Chapter_Sluchanko]

\end{bibunit} 

%\input{Chapter_Ponosov.tex}

%\input{Chapter_Alekseev.tex}

%\input{Chapter_Matsumura.tex}

%\input{Chapter_Thalmeier.tex}

%\input{Chapter_Inosov.tex}

%\input{Chapter_Schlottmann.tex}

%\input{Chapter_Fisk.tex}

% for BibTeX users
%\bibliographystyle{psp-book-van}   % Bibliography: Author-Date system
%\bibliography{psp-book-sample}     % pls. call your database here

% for non-BibTeX users
%\input bibliography.tex

%\printindex

\end{document}